\documentclass[letterpaper,11pt]{article}
\usepackage[margin=1in]{geometry}  

\usepackage{centernot}
\usepackage{listings}
\usepackage{amsmath}
\usepackage{amsthm}
\usepackage{tikz}
\usepackage{caption,subcaption}
\usepackage{array}
\usepackage{mdwmath}
\usepackage{multirow}
\usepackage{mdwtab}
\usepackage{eqparbox}
\usepackage{multicol}
\usepackage{amsfonts}
\usepackage{tikz}
\usepackage{multirow,bigstrut,threeparttable}
\usepackage{amsthm}
\usepackage{array}
\usepackage{bbm}
\usepackage{epstopdf}
\usepackage{mdwmath}
\usepackage{mdwtab}
\usepackage{eqparbox}
\usepackage{tikz}
\usepackage{latexsym}
\usepackage{cite}
\usepackage{amssymb}
\usepackage{bm}
\usepackage{amssymb}
\usepackage{graphicx}
\usepackage{mathrsfs}
\usepackage{epsfig}
\usepackage{psfrag}
\usepackage{setspace}
\usepackage[
            CJKbookmarks=true,
            bookmarksnumbered=true,
            bookmarksopen=true,
            colorlinks=true,
            citecolor=blue, 
            linkcolor=blue,
            anchorcolor=blue, 
            urlcolor=blue
            ]{hyperref}
\usepackage{algorithm}
\usepackage{algpseudocode}
\usepackage{stfloats}

\input xy
\xyoption{all}

\theoremstyle{plain}
\newtheorem{theorem}{Theorem}
\newtheorem{lemma}{Lemma}

\newtheorem{corollary}{Corollary}

\theoremstyle{definition}

\def\NN{\mathbb{N}}

\def\calM{\mathcal{M}}

\def\calR{\mathcal{R}}
\def\calS{\mathcal{S}}

\def\bE{\mathbf{E}}

\def\bP{\mathbf{P}}

\def\bR{\mathbf{R}}

\def\1{\mathbbm{1}}
\def\var{\mathsf{Var}}

\usepackage{booktabs}
\usepackage{adjustbox}
\usepackage{multirow}
\usepackage{listings}

\theoremstyle{plain}

\def \bP {\mathbb{P}}
\def \bE {\mathbb{E}}
\def \bR {\mathbb{R}}

\def \var {\mathsf{Var}}

\usepackage{xspace}

\newcommand{\stepa}[1]{\overset{\rm (a)}{#1}}
\newcommand{\stepb}[1]{\overset{\rm (b)}{#1}}
\newcommand{\stepc}[1]{\overset{\rm (c)}{#1}}
\newcommand{\stepd}[1]{\overset{\rm (d)}{#1}}

\newcommand{\TV}{{\sf TV}}

\definecolor{myblue}{rgb}{.8, .8, 1}
\definecolor{mathblue}{rgb}{0.2472, 0.24, 0.6} 
\definecolor{mathred}{rgb}{0.6, 0.24, 0.442893}
\definecolor{mathyellow}{rgb}{0.6, 0.547014, 0.24}

\begin{document}

\title{Minimax Estimation of Divergences between Discrete Distributions}
\author{Yanjun Han, Jiantao Jiao, and Tsachy Weissman\thanks{Yanjun Han and Tsachy Weissman are with the Department of Electrical Engineering, Stanford University, email: \url{ {yjhan,tsachy}@stanford.edu}. Yanjun Han and Tsachy Weissman were partially supported by
NSF Grant 1140567-11-QCAND. 
Jiantao Jiao is with the Department of Electrical Engineering and Computer Sciences and Department of Statistics at University of California, Berkeley, email: \url{jiantao@eecs.berkeley.edu}. Jiantao Jiao was partially supported by NSF Grants IIS-1901252 and CCF-1909499. A preliminary version \cite{han2016minimax_isita} of this paper appeared in the International Symposium on Information Theory and Its Applications (ISITA) in Monterey, CA, USA, 2016. An earlier arXiv version \cite{han2016minimax} of this work is also available, while the current work made substantial changes.}}

\maketitle

\begin{abstract}
We study the minimax estimation of $\alpha$-divergences between discrete distributions for integer $\alpha\ge 1$, which include the Kullback--Leibler divergence and the $\chi^2$-divergences as special examples. Dropping the usual theoretical tricks to acquire independence, we construct the first minimax rate-optimal estimator which does not require any Poissonization, sample splitting, or explicit construction of approximating polynomials. The estimator uses a hybrid approach which solves a problem-independent linear program based on moment matching in the non-smooth regime, and applies a problem-dependent bias-corrected plug-in estimator in the smooth regime, with a soft decision boundary between these regimes. 
\end{abstract}

\tableofcontents

\section{Introduction}
Divergences, as fundamental measures of the discrepancy between different probability distributions, are key quantities in information theory and statistics and arise in various disciplines. For example, the Kullback--Leibler (KL) divergence \cite{kullback1951information} is an information-theoretic measure arising naturally in data compression \cite{catoni2004statistical}, communications \cite{csiszar2011information}, probability theory \cite{sanov1958probability}, statistics \cite{kullback1997information}, optimization \cite{dempster1977maximum}, machine learning \cite{bishop2006pattern, kingma2013auto}, and many other disciplines. Besides the KL divergence, the Hellinger distance \cite{hellinger1909neue} plays key roles in the classical asymptotic theory \cite{Hajek1970characterization,Hajek1972local} and pairwise tests \cite{birge1983approximation,LeCam1986asymptotic} in statistics, and the Pearson $\chi^2$-divergence is an important measure in goodness-of-fit tests \cite{ingster2012nonparametric,acharya2020inference}.  

In this paper, we consider the problem of estimating divergences based on observations from both distributions. Given jointly independent $m$ samples from distribution $P=(p_1,\cdots,p_k)$ and $n$ samples from distribution $Q=(q_1,\cdots,q_k)$ over some common alphabet $[k]$, the target is to estimate the following $\alpha$-divergence \cite{amari2012differential}: 
\begin{align}\label{eq:D_alpha}
D_\alpha(P \|Q) \triangleq \frac{1}{\alpha(\alpha-1)}\left(\sum_{i=1}^k \frac{p_i^\alpha}{q_i^{\alpha-1}} - 1 \right), \quad \alpha\in \bR \backslash \{0, 1\}, 
\end{align}
where the cases $\alpha\in \{0,1\}$ are understood as the limit as $\alpha\to 0$ and $\alpha\to 1$, respectively. Specifically, $D_0(P\|Q)=D_{\text{KL}}(Q\|P)$, $D_1(P\|Q) = D_{\text{KL}}(P\|Q)$, where $D_{\text{KL}}$ is the well-known KL divergence
\begin{align}\label{eq:KL}
D_{\text{KL}}(P \| Q) = \sum_{i=1}^k p_i\log \frac{p_i}{q_i}. 
\end{align}
Note that $D_{\alpha}(P\|Q) = \infty$ if $P \centernot\ll Q$ when $\alpha\ge 1$ and $Q \centernot\ll P$ when $\alpha\le 0$. The $\alpha$-divergence belongs to the general class of $f$-divergences \cite{csiszar1967information} with $f_\alpha(t) = (t^{\alpha}-1)/(\alpha(\alpha-1))$ if $\alpha\notin (0,1)$, and is the negated $f$-divergence with $f_\alpha(t) = (1-t^{\alpha})/(\alpha(\alpha-1))$ if $\alpha\in (0,1)$. 

The family of $\alpha$-divergences includes multiple well-known divergences widely used in information theory and statistics, where the choices of $\alpha=1/2,1,2$ give (possibly up to scalings) the squared Hellinger distance, KL divergence, and $\chi^2$-divergence, respectively. For clarity of presentation, throughout the paper we assume that $\alpha\ge 1$ is an integer, which is enough to reflect our ideas. The case of non-integral $\alpha$ is more complicated as it involves some two-dimensional polynomial approximation of $p^\alpha/q^{\alpha-1}$, but such a complication is mostly irrelevant to the efficacy of our proposed idea and thus left for future work.

Estimating functionals (or properties) of discrete distributions has attracted a recent line of research in the past decade, such as \cite{Lepski--Nemirovski--Spokoiny1999estimation,Paninski2003,Paninski2004,Cai--Low2011,Valiant--Valiant2011,Valiant--Valiant2011power,Valiant--Valiant2013estimating, Jiao--Venkat--Han--Weissman2015minimax,wu2016minimax,han2016minimax,orlitsky2016optimal,ZVVKCSLSDM16,bu2018estimation,jiao2018minimax,wu2019chebyshev,hao2019broad,hao2019unified,han2020estimation,han2020optimal}. We refer to the survey paper \cite{verdu2019empirical} for an overview. The main aim of this paper is not simply to apply the previously known methodology to a new functional; instead, we intend to improve the previous methodologies and construct minimax rate-optimal estimators satisfying the following properties: 
\begin{enumerate}
	\item No Poissonization: Most of the previous estimators are constructed under the Poissonized model where the frequency counts of symbols are mutually independent, while they cannot directly work for the original i.i.d. sampling model and there is no direct reduction between the optimal estimators. In this paper, we aim to construct explicit minimax rate-optimal estimators under the i.i.d. sampling model, the exact model we are considering. 
	\item No sample splitting: In addition to Poissonization, most previous estimators also require a sample splitting to acquire further independence structure. Specifically, the observations are split into two or more parts, where the first part is to locate the probability of each symbol for regime classification, and the other parts are used for estimation (see, e.g. \cite{Jiao--Venkat--Han--Weissman2015minimax}). However, sample splitting does not make full use of all observations, and the hard decisions in regime classification make the final estimator unstable to slight changes in input and therefore lead to poor concentration properties \cite{han2021competitive}. Therefore, it is practically desirable to propose an estimator with soft decisions and no sample splitting. 
	\item No explicit polynomial construction: Many of the above work rely on an explicit polynomial approximation in both the estimator construction and analysis. An important and well-known approach is the best polynomial approximation under the $\ell_\infty$ norm, while other approximations with sound pointwise bounds could also be useful (e.g. \cite{jiao2018minimax}). This leads to scenarios where constructing an explicit polynomial is undesirable or difficult. First, the learner might not know which approximation norm to use when performing the best polynomial approximation. Second, even if the learner knows the type of pointwise bound to aim for, it could be challenging to find a polynomial with the desired pointwise bound. To address these concerns, it will be desirable if there is a single approach which is as good as the best polynomial approximation under any norm (i.e. \emph{adapts} to different approximation norms), and does not need to construct the polynomial explicitly. 
	
\end{enumerate}

In this paper, we construct minimax rate-optimal estimators satisfying all the above properties for the $\alpha$-divergences with integer $\alpha\ge 1$, while the ideas are generalizable to other functionals. Also as a byproduct, we obtain the sharp minimax estimation rates for the above $\alpha$-divergences, which could be of independent interest. 

\subsection{Notations}
Let $\NN$ be the set of all positive integers, and for $n\in \NN$, let $[n]\triangleq \{1,2,\ldots,n\}$. For $k\in\NN$, let $\calM_k$ be the set of all discrete distributions supported on $[k]$. Let $\mathsf{B}(n,p)$ be the Binomial distribution with number of trials $n$ and success probability $p\in [0,1]$, and $\mathsf{HG}(N,K,k)$ be the hypergeometric distribution with $k$ draws without replacement from a bin of $N$ total elements and $K\le N$ desired elements. For non-negative sequences $\{a_n\}$ and $\{b_n\}$, we write $a_n\ll b_n$ (or $a_n = o(b_n)$) to denote that $\lim_{n\to\infty}a_n/b_n = 0$, and $a_n\lesssim b_n$ (or $a_n = O(b_n)$) to denote that $\limsup_{n\to\infty}a_n/b_n<\infty$, and $a_n\gg b_n$ (or $a_n = \omega(b_n)$) to denote $b_n \ll a_n$, and $a_n \gtrsim b_n$ (or $a_n = \Omega(b_n)$) to denote $b_n\lesssim a_n$, and $a_n\asymp b_n$ (or $a_n = \Theta(b_n)$) to denote both $a_n\lesssim b_n$ and $b_n\lesssim a_n$. We also use $\lesssim_\alpha$ (resp. $\gtrsim_\alpha, \asymp_\alpha$) and $O_\alpha$ (resp. $\Omega_\alpha, \Theta_\alpha$) to denote the respective meanings with hidden constants depending only on $\alpha$. 

\subsection{Main Results}
Note that for $\alpha\ge 1$, the $\alpha$-divergence $D_{\alpha}(P\|Q)$ is unbounded if $P \centernot\ll Q$. Hence, we restrict our attention to the following set of distribution pairs $(P,Q)$: 
\begin{align}\label{eq:parameter_set}
\calM_k(U) \triangleq \left\{(P,Q)\in \calM_k \times \calM_k: \frac{p_i}{q_i} \le U, \forall i\in [k]  \right\}, 
\end{align}
where $U\ge 1$ is a prespecified upper bound on the likelihood ratio $dP/dQ$. Restricting to the set $\calM_k(U)$, it is clear that the $\alpha$-divergence $D_\alpha(P\|Q)$ is upper bounded by some quantity depending only on $U$. Our first result characterizes the minimax estimation error of $D_{\alpha}(P\|Q)$ for integer $\alpha\ge 2$. 

\begin{theorem}\label{thm.alpha_div}
	Let $\alpha\ge 2$ be an integer. Then for any $m\gtrsim_\alpha U^{2(\alpha-1)}$, $n\gtrsim_\alpha kU^\alpha / \log k + U^{2\alpha-1}$, $U\gtrsim_\alpha (\log k)^2$ and $\log k\gtrsim_\alpha \log n$,  
	\begin{align*}
	\inf_{\widehat{D}}\sup_{(P,Q)\in \calM_k(U)} \bE_{(P,Q)} |\widehat{D} - D_\alpha(P\|Q)| \asymp_\alpha \frac{kU^{\alpha}}{n\log n} + \frac{U^{\alpha-1}}{\sqrt{m}} + \frac{U^{\alpha-1/2}}{\sqrt{n}},
	\end{align*}
	where the infimum is taken over all possible estimators $\widehat{D}$ depending only on $m$ observations from $P$ and $n$ observations from $Q$. In particular, the upper bound is attained by an estimator without any Poissonization, sample splitting, or explicit polynomial construction. 
\end{theorem} 
The following corollary on the sample complexity is then immediate. 
\begin{corollary}
	For integer $\alpha\ge 2$ and $U\gtrsim_\alpha (\log k)^2$, consistent estimation of $D_\alpha(P\|Q)$ is possible over $\calM_k(U)$ iff $m\gg U^{2(\alpha-1)}$ and $n\gg kU^\alpha / \log k + U^{2\alpha-1}$. 
\end{corollary}

Hence, given a small upper bound on the likelihood ratio, it only requires few samples from $P$ and a nearly linear sample size from $Q$. However, when the upper bound $U$ becomes larger, the number of samples required for consistent estimation grows polynomially in $U$ and may become enormous for large $\alpha$, indicating the impossibility of learning in such scenarios. Also note that in the theorem statement $U$ cannot be too small, as $U$ close to $1$ makes the divergence $D_\alpha(P\|Q)$ close to zero. 

As for $\alpha=1$, the $\alpha$-divergence becomes the KL divergence $D_{\text{\rm KL}}(P\|Q)$, and the following theorem summarizes the minimax risk of estimating $D_{\text{\rm KL}}(P\|Q)$. 
\begin{theorem}\label{thm.KL}
	If $m\gtrsim k/\log k, n\gtrsim kU/\log k, U\gtrsim (\log k)^2$ and $\log k\gtrsim \log(m+n)$, we have
	\begin{align*}
	\inf_{\widehat{D}}\sup_{(P,Q)\in \calM_k(U)} \bE_{(P,Q)} |\widehat{D} - D_{\text{\rm KL}}(P\|Q)| \asymp \frac{k}{m\log m} + \frac{kU}{n\log n} + \frac{\log U}{\sqrt{m}} + \sqrt{\frac{U}{n}}, 
	\end{align*}
	where the infimum is taken over all possible estimators $\widehat{D}$ depending only on $m$ observations from $P$ and $n$ observations from $Q$. In particular, the upper bound is attained by an estimator without any Poissonization, sample splitting, or explicit polynomial construction. 
\end{theorem} 
Similarly, we have the following sample complexity on estimating the KL divergence. 
\begin{corollary}
	For $U\gtrsim_\alpha (\log k)^2$, consistent estimation of $D_{\text{\rm KL}}(P\|Q)$ is possible over $\calM_k(U)$ iff $m\gg k/\log k$ and $n\gg kU/ \log k$. 
\end{corollary}
Consequently, Theorems \ref{thm.alpha_div} and \ref{thm.KL} provide a complete characterization of the minimax rates of estimating $\alpha$-divergences for all $\alpha\in \NN$. 

\subsection{Related Work}
There have been several attempts to estimate the KL divergence for the continuous case, see~\cite{Wang--Kulkarni--Verdu2005divergence,Lee--Park2006estimation,Gretton--Borgwardt--Rasch--Scholkopf--Smola2006kernel,Perez2008kullback,Wang--Kulkarni--Verdu2009divergence,Nguyen--Wainwright--Jordan2010estimating} and references therein. These approaches usually do not operate in the minimax framework, and focus on consistency but not rates of convergence, unless strong smoothness conditions on the densities are imposed to achieve the parametric rate (i.e., $\Theta(n^{-1})$ in mean squared error). In the discrete setting, \cite{Cai--Kulkarni--Verdu2006universal} and \cite{Zhang--Grabchak2014nonparametric} proved consistency of some specific estimators without arguing minimax optimality. 

As for the minimax analysis of general functional estimation, there are three major approaches developed in the recent literature. The most prominent approach is via the best polynomial approximation, which operates in an ad-hoc manner and typically leads to tight minimax rates. Specifically, the minimax-optimal estimation procedures have been found for many non-smooth 1D functionals, including the entropy \cite{Paninski2003,Paninski2004,Valiant--Valiant2011,Jiao--Venkat--Han--Weissman2015minimax,wu2016minimax}, $L_1$ norm of the mean vector \cite{Cai--Low2011}, support size \cite{Valiant--Valiant2013estimating, wu2019chebyshev}, support coverage \cite{orlitsky2016optimal,ZVVKCSLSDM16}, distance to uniformity \cite{Valiant--Valiant2011power,jiao2018minimax}, and general 1-Lipschitz functions \cite{hao2019broad,hao2019unified}. Similar procedures can also be generalized to 2D functionals: for example, the optimal estimation of $L_1$ or total variation distance between discrete distributions was studied in \cite{jiao2018minimax}, and for the KL divergence considerd in this paper, \cite{bu2018estimation} and an earlier version of this work \cite{han2016minimax} independently obtained the minimax rate following an explicit polynomial approximation approach. Generalizations to nonparametric functionals are also available \cite{Lepski--Nemirovski--Spokoiny1999estimation,han2020estimation,han2020optimal}. We refer to the survey paper \cite{verdu2019empirical} for an overview of the results. However, this line of research typically satisfies none of the three properties mentioned in the introduction: explicit polynomials are necessary by definition of this approach, and nearly all works adopt sample splitting and Poissonization for analytical simplicity. We also point out some important exceptions. The work \cite{han2020optimal} gets rid of the Poissonization by applying a careful Efron-Stein-Steele inequality to the i.i.d. sampling model, but it still heavily relies on sample splitting. The sample splitting could also be avoided via explicit construction of linear estimators \cite{orlitsky2016optimal,wu2019chebyshev}, while a general methodology to interpolate between different estimation regimes is still underexplored. 

The second approach is based on linear programming, where the final estimator is the solution to an appropriate linear program. Early applications of linear programs to functional estimation include \cite{Paninski2003,Valiant--Valiant2011,Valiant--Valiant2011power,Valiant--Valiant2013estimating} which directly involve the functional of interest, but there were only suboptimal error guarantees. Later, new moment matching based methods independent of the target functional appeared in \cite{han2018local} for estimating sorted distributions and \cite{rigollet2019uncoupled} for estimating the multiset of the mean in Gaussian location model, both of which have optimal error guarantees. This idea of moment matching, or method of moments, dates back to \cite{pearson1894contributions}, and was also used in other problems such as \cite{hardt2015tight,wu2018optimal} for learning Gaussian mixtures and \cite{kong2017spectrum,tian2017learning} for learning a population of parameters. Although the moment matching approach does not require explicit polynomial approximation, it only achieves the minimax rate in a small parameter regime for most functionals \cite{han2018local} and it is unknown how to adapt this approach to a given functional. Moreover, for the \emph{local} moment matching idea which is necessary in most functional estimation problems where the regime classification is required, both Poissonization and sample splitting were still necessary in \cite{han2018local}. A very recent work \cite{han2021competitive} was able to remove the sample splitting in the 1D case and prove the near-optimal exponential concentration property, but it still applied the Poissonization and did not involve the target functional. Hence, for linear programming based approaches, it is open to find a linear program which gives a fully minimax rate-optimal estimator satisfying the three target properties in the introduction. 

The third approach is based on the profile maximum likelihood (PML) distribution, a concept first proposed in \cite{orlitsky2004modeling} on estimating the probability multiset. It was later shown in \cite{acharya2017unified} that plugging the PML distribution into general symmetric functionals attains the optimal sample complexity by a union bound over all possible profiles, and a better statistical guarantee was obtained recently in \cite{han2021competitive} using an involved chaining argument. There were also modifications of PML \cite{charikar2019general,hao2019broad} tailored for specific functionals, and its success was shown for a number of problems including the estimation of KL divergence \cite{acharya2018profile,hao2019broad}. However, the PML approach is designed to only work for symmetric functionals, and there is a provable non-negligible gap between the performance of PML and that of the minimax optimal estimator \cite{han2020high}. In addition, exact or approximate PML distributions are very hard to compute \cite{charikar2019efficient,ACSS20}, which further makes it not suitable for our problem. 

\subsection{Organization}
The rest of the paper is organized as follows. In Section \ref{sec:construction}, a hybrid estimator combining the idea of functional-independent moment matching and the functional-dependent bias-corrected plug-in approach is constructed for both the $\alpha$-divergences and the KL divergence, with a soft boundary between these two regimes. Section \ref{sec:analysis} presents the roadmap of the analysis of the above hybrid estimator for the $\alpha$-divergences with $\alpha\ge 2$. In particular, it presents the idea of implicit polynomial approximations achieved by moment matching, as well as key technical insights to overcome the dependence incurred by removing Poissonization and sample splitting. The lower bounds in this paper involve very similar techniques to the past thread on functional estimation, e.g. \cite{Cai--Low2011,Jiao--Venkat--Han--Weissman2015minimax,wu2016minimax,bu2018estimation}, so we defer it to the supplementary materials.


The Appendix \ref{appendix:auxiliary} lists some necessary auxiliary lemmas for this paper, while we refer the upper bound analysis for the KL divergence estimator, the minimax lower bounds, and the proofs of all main and auxiliary lemmas, to the appendices \ref{appendix:KL}, \ref{appendix:lower_bound}, \ref{appendix:main_proof}, and \ref{appendix:auxiliary_proof}, respectively. 

\section{Estimator Construction}\label{sec:construction}
In this section, we construct minimax optimal estimators of $\alpha$-divergences for integer $\alpha\ge 2$ and $\alpha=1$ separately. 

\subsection{Construction for $\alpha$-divergences with integer $\alpha\ge 2$}\label{subsec:alpha_div}
The high-level description of the estimator construction is as follows. First, following the general recipe in \cite{Jiao--Venkat--Han--Weissman2015minimax}, we split the entire set $[0,1]^2$ of probability pairs $(p,q)$ into the smooth regime $\calR_{\text{s}}$ and the non-smooth regime $\calR_{\text{ns}}$. In the non-smooth regime $\calR_{\text{ns}}$, we construct a 2D measure $\widehat{\mu}$ supported on a set slightly larger than $\calR_{\text{ns}}$ such that the moments $\int p^\alpha q^d\widehat{\mu}(dp,dq)$ of $\widehat{\mu}$ for all $d = 0,1,\cdots,O(\log n)$ is close to the true moments $\sum_{i=1}^k p_i^\alpha q_i^d \1((p_i,q_i)\in \calR_{\text{ns}})$ restricted to $\calR_{\text{ns}}$. Then we plug the measure $\widehat{\mu}$ into the target functional as the estimate in the non-smooth regime. As for the smooth regime, we apply an appropriate bias-corrected plug-in estimator of each $(p_i,q_i)$. However, one crucial difference from the previous approaches is that we do \emph{not} attribute each symbol $i\in [k]$ to smooth/non-smooth regimes, either deterministically or randomly. Instead, each symbol $i\in [k]$ contains both non-smooth and smooth components of contributions towards the final functional which can be computed separately to construct the overall estimator. In the detailed implementation, some smoothing operation will be applied to both the true moments and the plug-in estimators to account for different components, which is the reason why we no longer require the sample splitting. 

Fix constants $c_1,c_2>0$ to be chosen later, and assume that $n$ is an even integer. Moreover, for each $i\in [k]$, let $m\widehat{p}_i, n\widehat{q}_i$ be the number of occurrences of symbol $i$ in the $m$ observations from $P$ and the $n$ observations from $Q$, respectively. The detailed estimator construction for the $\alpha$-divergence with integer $\alpha\ge 2$ is then as follows. 
\begin{enumerate}
	\item Estimate the non-smooth component: 
	\begin{itemize}
		\item Estimators for smoothed moments: for $d\in \NN$, define the following function $g_d: \NN \to \bR$ with $g_0(x) \equiv 1$, and 
		\begin{align}\label{eq:moment_est}
		g_d(x) = \prod_{d' = 0}^{d-1} \frac{x-d'}{n/2 - d'}. 
		\end{align}
		We also define a modified version $\widetilde{g}_d$ of $g_d$ as 
		\begin{align}\label{eq:moment_est_trunc}
		\widetilde{g}_d(x) = \min\left\{g_d(x), g_d(\lceil 2c_1\log n \rceil) \right\}. 
		\end{align}
		Now for each $d\in \NN$, we define our estimator of the smoothed $(\alpha,d)$ moments as follows: 
		\begin{align}\label{eq:smoothed_moments}
		\widehat{M}_{\alpha,d} = \sum_{i=1}^k \left(\prod_{\ell=0}^{\alpha-1}\frac{m\widehat{p}_i - \ell}{m-\ell}\cdot \sum_{0\le s\le c_1\log n} \bP\left(\mathsf{HG}\left(n,n\widehat{q}_i, \frac{n}{2}\right) = s\right) \widetilde{g}_d(n\widehat{q}_i - s)\right),
		\end{align}
		where we recall that $\mathsf{HG}(N,K,k)$ denotes the hypergeometric distribution with $k$ draws without replacement from a bin of $N$ total elements and $K$ desired elements. The definitions of $g_d$ and $\widetilde{g}_d$ are motivated by the later identity \eqref{eq:unbiased_moments}, and $\widehat{M}_{\alpha,d}$ is motivated by \eqref{eq:unbiased_smoothed_moments}.
		\item The linear program: define $G = \lceil c_2\log n\rceil$ and the (non-probability) measure $\widehat{\mu}$ as the solution to the following linear program: 
		\begin{align}\label{eq:LP}
		\begin{split}
		\text{minimize} & \qquad  L(\widehat{\mu}, \widehat{M}) \triangleq \sum_{d=0}^G \left(\frac{3c_1\log n}{n} \right)^{-d}\cdot \left|\int p^\alpha q^d\widehat{\mu}(dp,dq) - \widehat{M}_{\alpha,d} \right| \\
		\text{subject to} &\qquad \text{supp}(\widehat{\mu}) \subseteq \mathcal{S}\triangleq \left\{(p,q)\in [0,1]^2: q\le \frac{3c_1\log n}{n}, p\le Uq \right\}, \\
		&\qquad \widehat{\mu}(\bR^2) \le k, 
		\end{split}
		\end{align}
		where $\text{supp}(\mu)$ denotes the support of the measure $\mu$. The specific form of the linear program is motivated by \eqref{eq:deterministic_ineq}.
		\item Final estimator of the non-smooth component: compute 
		\begin{align}\label{eq:nonsmooth_final}
		\widehat{D}_{\alpha,\text{ns}} = \int \frac{p^\alpha}{q^{\alpha-1}} \widehat{\mu}(dp,dq). 
		\end{align}
	\end{itemize}
	\item Estimate the smooth component: define the function $h_\alpha: \NN\to \bR$ with 
	\begin{align}\label{eq:bias_correction}
	h_\alpha(x) = \1(x\neq 0)\cdot \left(\frac{1}{(2x/n)^{\alpha-1}} -  \frac{\alpha(\alpha-1)}{n}\cdot \frac{1 - 2x/n}{(2x/n)^\alpha} \right),
	\end{align}
	which is a bias-corrected plug-in estimator for $q^{1-\alpha}$. The final estimator of the smooth component is 
	\begin{align}\label{eq:smooth_final}
	\widehat{D}_{\alpha,\text{s}} = \sum_{i=1}^k \left(\prod_{\ell=0}^{\alpha-1}\frac{m\widehat{p}_i - \ell}{m-\ell}\cdot \sum_{s>c_1\log n}\bP\left(\mathsf{HG}\left(n,n\widehat{q}_i, \frac{n}{2}\right) = s\right) h_\alpha(n\widehat{q}_i - s) \right). 
	\end{align}
	\item The final estimator $\widehat{D}_\alpha$ is then defined as
	\begin{align}\label{eq:final}
	\widehat{D}_\alpha = \frac{1}{\alpha(\alpha-1)}\left( \widehat{D}_{\alpha,\text{ns}} +  \widehat{D}_{\alpha,\text{s}} -1 \right). 
	\end{align}
\end{enumerate}

A few remarks of the estimator $\widehat{D}_\alpha$ are in order: 
\begin{itemize}
	\item The high-level idea: In the above estimator construction, the final estimator $\widehat{D}_\alpha$ consists of the non-smooth component $\widehat{D}_{\alpha,\text{ns}}$ in \eqref{eq:nonsmooth_final} and the smooth component $\widehat{D}_{\alpha,\text{s}}$ in \eqref{eq:smooth_final}. These two components aim to estimate the respective components in the following decomposition of the target functional
	\begin{align*}
	\sum_{i=1}^k \frac{p_i^\alpha}{q_i^{\alpha-1}} = D_{\alpha,\text{ns}} + D_{\alpha, \text{s}},
	\end{align*}
	with
	\begin{align}
	D_{\alpha,\text{ns}} &\triangleq \sum_{i=1}^k \frac{p_i^\alpha}{q_i^{\alpha-1}} \cdot \bP\left(\mathsf{B}\left(\frac{n}{2},q_i\right) \le c_1\log n \right), \label{eq:nonsmooth_component} \\
	D_{\alpha,\text{s}} &\triangleq \sum_{i=1}^k \frac{p_i^\alpha}{q_i^{\alpha-1}} \cdot \bP\left(\mathsf{B}\left(\frac{n}{2},q_i\right) > c_1\log n \right). \label{eq:smooth_component} 
	\end{align}
	Recall that $\mathsf{B}(n,p)$ denotes the Binomial distribution with number of trials $n$ and success probability $p$, thus \eqref{eq:nonsmooth_component} and \eqref{eq:smooth_component} provide a \emph{soft} regime classification in the following sense: for each symbol $i\in [k]$, if $q_i$ is large then it mainly contributes to the smooth component in \eqref{eq:smooth_component}, and if $q_i$ is small then it mostly contributes to the non-smooth component in \eqref{eq:nonsmooth_component}. Hence, each symbol contributes to both components with weights determined by a suitable Binomial probability, and our estimator aims to estimate these two components separately. The reason why we choose Binomial probabilities as the weights in \eqref{eq:nonsmooth_component} and \eqref{eq:smooth_component}, as well as use the parameter $n/2$ instead of $n$ in the Binomial probability, is that the smoothed moments with the current weights admit unbiased estimators, which will be discussed in details below. 
	\item Regime classification: As explained in the high-level idea, the above estimator construction does not involve a hard regime classification as in previous works. However, the support constraint of $\widehat{\mu}$ in the linear program \eqref{eq:LP} and the thresholds $c_1\log n$ in \eqref{eq:nonsmooth_component}, \eqref{eq:smooth_component} indicate that a soft regime classification is still performed with $q = 2c_1\log n/n$ being the boundary of the non-smooth (where $q$ is smaller than the threshold) and smooth regimes (where $q$ is greater than the threshold). The choice ensures that symbols with probability far away from the boundary have weight $\bP(\mathsf{B}(n/2,q)\le c_1\log n)$ close to either zero or one, as shown in Lemma \ref{lemma.localization} in the Appendix.
	\item The function $g_d(x)$ and the modification $\widetilde{g}_d(x)$: The choice of the function $g_d(x)$ in \eqref{eq:moment_est} is to satisfy the following identity: for $h\sim \mathsf{B}(n/2,q)$, it holds that
	\begin{align}\label{eq:unbiased_moments}
	\bE\left[g_d(h) \right] &= \sum_{x=0}^{n/2} \binom{n/2}{x}q^x(1-q)^{n/2-x} \prod_{d'=0}^{d-1} \frac{x-d'}{n/2 - d'} \nonumber\\
	& = \sum_{x=d}^{n/2} \binom{n/2 - d}{x-d} q^x(1-q)^{n/2 - x} = q^d.
	\end{align}
	In other words, the quantity $g_d(h)$ is an unbiased estimator of $q^d$. Therefore, the function $g_d(x)$ is used for the unbiased estimation of monomials of $q$. We cut off the function $g_{d}(x)$ outside the interval $[0,2c_1\log n]$ to ensure a uniform small difference $|\widetilde{g}_{d}(x) - \widetilde{g}_{d}(x+1)|$ for all $x\in \NN$, which helps to reduce the variance of the smoothed moment estimator $\widehat{M}_{\alpha,d}$ discussed below. 
	\item The smoothed moment estimator $\widehat{M}_{\alpha,d}$: Ideally we would like to estimate the local moments $\sum_{i=1}^k p_i^\alpha q_i^d\cdot \1(q_i\le 2c_1\log n/n)$ for each $d\in \NN$, if a hard regime classification could work. However, the above quantity does not admit an unbiased estimator, and is information-theoretically hard to estimate if some $q_i$'s are close to the boundary. Most previous works overcome the above difficulty via the \emph{sample splitting}, where they split the empirical probability $\widehat{q}_i$ into two independent halves $\widehat{q}_{i,1}$ and $\widehat{q}_{i,2}$, and the modified moments $\sum_{j=1}^k p_i^\alpha q_{i}^d\cdot \1(\widehat{q}_{i,1}\le 2c_1\log n/n)$ admit a simple unbiased estimator based on $\widehat{p}_i$ and $\widehat{q}_{i,2}$. 
	
	Instead of sample splitting, we consider the following \emph{smoothed moments} defined as
	\begin{align}\label{eq:smoothed_moments_true}
	M_{\alpha,d} \triangleq \sum_{i=1}^k p_i^\alpha q_i^d\cdot \bP\left(\mathsf{B}\left(\frac{n}{2}, q_i\right) \le c_1\log n \right). 
	\end{align}
	By Lemma \ref{lemma.localization}, we see that the Binomial probability in \eqref{eq:smoothed_moments_true} roughly corresponds to the indicator $\1(q_i\le 2c_1\log n/n)$, while \eqref{eq:smoothed_moments_true} makes a soft decision. The key advantage of introducing the smoothed moments taking the form of \eqref{eq:smoothed_moments_true} is that $M_{\alpha,d}$ now admits an unbiased estimator: 
	\begin{align}\label{eq:unbiased_smoothed_moments}
	\bE\left[\sum_{i=1}^k \left(\prod_{\ell=0}^{\alpha-1}\frac{m\widehat{p}_i - \ell}{m-\ell}\cdot \sum_{0\le s\le c_1\log n} \bP\left(\mathsf{HG}\left(n,n\widehat{q}_i, \frac{n}{2}\right) = s\right) g_d(n\widehat{q}_i - s)\right)\right] = M_{\alpha,d}. 
	\end{align}
	To see \eqref{eq:unbiased_smoothed_moments}, note that for $s\sim \mathsf{HG}(n,n\widehat{q}, n/2)$ and $t = n\widehat{q} - s$, they are independent random variables each of which follows a Binomial distribution $\mathsf{B}(n/2,q)$ \emph{unconditionally} on $n\widehat{q}\sim \mathsf{B}(n,q)$. Hence, using the independence of $\widehat{p}_i$ and $\widehat{q}_i$, the unbiased condition \eqref{eq:unbiased_moments} leads to \eqref{eq:unbiased_smoothed_moments}. An alternative view of \eqref{eq:unbiased_smoothed_moments} is via the reduction by sufficiency of the estimator using sample splitting; \cite{han2021competitive} discussed this idea concerning Poisson distributions. 
	\item The linear program: The high-level intuition behind the specific objective function $L(\widehat{\mu}, \widehat{M})$ in the linear program \eqref{eq:LP} is that after applying the duality between moment matching and polynomial approximation, the following \emph{deterministic} inequality holds for the estimation error of $\widehat{D}_{\alpha,\text{ns}}$ in estimating $D_{\alpha,\text{ns}}$: 
	\begin{align*}
	|\widehat{D}_{\alpha,\text{ns}} - D_{\alpha,\text{ns}} | \lesssim_{\alpha,c_1,c_2} \frac{kU^\alpha}{n\log n} + n^{O(c_2) + \alpha - 1}L(\widehat{\mu}, \widehat{M}).
	\end{align*}
	In other words, the random estimation error is controlled exactly by the objective function $L(\widehat{\mu}, \widehat{M})$. Now comes the key intuition: there must be an \emph{unknown} feasible solution $\mu$ to \eqref{eq:LP} which depends on the unknown distributions $(P,Q)$, with moments close to the smoothed moments $M_{\alpha,d}$ in \eqref{eq:smoothed_moments_true}. Hence, the fact that $\widehat{\mu}$ is a minimizer gives $L(\widehat{\mu}, \widehat{M}) \le L(\mu, \widehat{M})$, where the last quantity is a linear combination of the differences $|\widehat{M}_{\alpha,d} - M_{\alpha,d}|$ and is thus much easier to upper bound the expectation. 
	
	In summary, the form of the linear program \eqref{eq:LP} ensures that as long as $\widehat{M}_{\alpha,d}$ are accurate estimators of the true smoothed moments in \eqref{eq:smoothed_moments_true}, the plug-in approach \eqref{eq:nonsmooth_component} of the resulting minimizer $\widehat{\mu}$ will also be accurate for the target quantity $D_{\alpha,\text{ns}}$. Moreover, the bivariate polynomial we will use to approximate $p^\alpha/q^{\alpha-1}$ on the set $\calS$ will take the form $p^\alpha Q(q)$ for some univariate polynomial $Q$ with degree at most $G$, so the degree of $p$ is always $\alpha$ and $\widehat{\mu}$ is required to be supported on $\mathcal{S}$. The last constraint $\widehat{\mu}(\bR^2)\le k$ is mostly a technical condition and satisfied by the target measure $\mu$ in \eqref{eq:measure_true}.
	
	\item Comparison with explicit polynomial approximation: We also compare the linear program \eqref{eq:LP} with another linear program which computes the best polynomial approximation under the $\ell_\infty$ norm. On the surface level, the former linear program has infinitely many variables and the latter has infinitely many constraints. The main difference is that the linear program for explicit polynomial approximation depends on the target norm of approximation (e.g. $\ell_\infty$ norm, weighted $\ell_\infty$ norm, or $\ell_2$ norm), while our linear program performs an implicit polynomial approximation which adapts to different approximation norms. In other words, the usage of different approximation norms requires different linear programs for explicit polynomial approximation, but only needs a single linear program for implicit approximation. However, the linear program for explicit approximation could be pre-solved independent of the data, while the linear program for implicit approximation needs to be solved every time for new data, giving a higher computational complexity as a price of adaptation and implicit approximation. 
	
	In addition, it might be tempted to think that the above linear programs are dual to each other. While this is intuitively true, we note that the adaptivity nature of our linear program implies that it is not the dual formulation for the best polynomial approximation under any \emph{fixed} norm. Moreover, the objective value for explicit approximation is typically the polynomial approximation error and corresponds to the bias, while \eqref{eq:deterministic_ineq}, \eqref{eq:objective_upper_bound} show that the expected objective value of \eqref{eq:LP} actually corresponds to the variance in the non-smooth regime. Therefore, we do not expect a direct and explicit duality relation to hold.

	\item The smooth component $\widehat{D}_{\alpha,\text{s}}$: For symbols with a significant contribution to $D_{\alpha,s}$ in \eqref{eq:smooth_component}, Lemma \ref{lemma.localization} shows that we must have $q > c_1\log n/n$. Consequently, the target functional $p^{\alpha}/q^{\alpha-1}$ becomes smooth on both arguments $(p,q)$, and the plug-in approach (possibly with bias correction) is expected to work well. To this end, the function $h_\alpha(x)$ in \eqref{eq:bias_correction} serves as a plug-in estimator with order-one bias correction in the sense that if $X\sim \mathsf{B}(n/2, q)$ with $q\ge c_1\log n/n$, it holds that $\bE[h_\alpha(X)]\approx q^{1-\alpha}$ (cf. Lemma \ref{lemma:smooth_bias}). Consequently, using the same insights of the smoothed moments and the identity \eqref{eq:unbiased_smoothed_moments}, it can be seen that the estimator $\widehat{D}_{\alpha,s}$ in \eqref{eq:smooth_final} is close to $D_{\alpha,s}$ in \eqref{eq:smooth_component} in expectation. 
	\item Dependence on $(k,U)$: A potential drawback of the above estimator construction is that the estimator requires the knowledge of both the support size $k$ and the upper bound $U$ which are used in the constraints of the linear program \eqref{eq:LP}. However, we remark that our main aim is to theoretically provide a minimax rate-optimal methodology which does not require any Poissonization, sample splitting, or explicit polynomial construction in general functional estimation problems, and therefore we focus more on the key theoretical insights than other possible improvements. Moreover, we also point out that if we are allowed to use an explicit polynomial approximation (still without Poissonization or sample splitting), the linear program \eqref{eq:LP} can be replaced by an explicit unbiased estimator of the above polynomial and become agnostic to both parameters $k$ and $U$. 
	\item Computational complexity: The main computation lies in the steps \eqref{eq:smoothed_moments}, \eqref{eq:LP}, \eqref{eq:nonsmooth_final} and \eqref{eq:smooth_final}, where other steps take $O(1)$ or $O(\log n)$ time for each evaluation. For steps \eqref{eq:smoothed_moments} and \eqref{eq:smooth_final}, the inner summation is over at most $n\widehat{q}_i$ different values of $s$ for each symbol $i\in [k]$, and therefore the total computational time is $O(\alpha k\sum_{i=1}^k n\widehat{q}_i) = O(\alpha nk)$. 
	
	As for the linear program \eqref{eq:LP} and integration \eqref{eq:nonsmooth_final}, we may quantize the support of $\widehat{\mu}$ for both tasks to perform in polynomial time. We claim that restricting the support of $\widehat{\mu}$ to the following grid
	\begin{align*}
	\mathcal{G} = \left[ \left(\frac{U\log n}{nM}\cdot \mathbb{N} \right) \times \left(\frac{\log n}{nM}\cdot \mathbb{N} \right) \right] \cap \mathcal{S}
	\end{align*}
	with $M = n^{O(c_2)}$ suffices to give an estimator with the same theoretical property. Note that there are $O(M^2)$ grid points in $\mathcal{G}$, which could grow with $n$ slowly as $c_2$ could be chosen arbitrarily close to zero. To see why this grid works, simple algebra gives that for every $d\ge 0$ and any $(p,q)\in \mathcal{S}$, we could find some $(p', q')\in \mathcal{G}$ such that $|p^\alpha q^d - (p')^\alpha (q')^d| \lesssim M^{-1}\cdot d(U\log n/n)^\alpha(\log n/n)^d$. Therefore, any feasible solution $\widehat{\mu}$ of \eqref{eq:LP} gives rise to a feasible solution $\widehat{\mu}'$ supported on $\mathcal{G}$ with the objective value differing by at most $(U/n)^\alpha kn^{O(c_2)}/M$. Once we obtain an approximate solution of \eqref{eq:LP} within an additive error $\varepsilon$, we will use the inequality $L(\widehat{\mu},\widehat{M}) \le L(\mu, \widehat{M}) + \varepsilon$ in the later analysis \eqref{eq:LP_triangle}, which in view of \eqref{eq:deterministic_ineq} yields the same theoretical rate as long as $\varepsilon \lesssim (U/n)^\alpha k/n^{O(c_2)}$. Finally, plugging in the previous expression of $\varepsilon$ shows that $M \asymp n^{Cc_2}$ for a suitable constant $C>0$ is sufficient, proving the desired claim. 
	
	Finally, we remark that our computational complexity is polynomial in parameters $(n,k)$, as opposed to the usual near-linear complexity in prior work based on explicit polynomial approximation. We think this is a price to pay for implicit polynomial approximation, and a similar phenomenon occurs in the local moment matching approach \cite{han2018local}. 
\end{itemize}

Based on the above discussions, we make a summary of how the constructed estimator $\widehat{D}_{\alpha}$ satisfies the three properties in the introduction. First, to get rid of the sample splitting, the key idea is to use the smoothed moments which admit unbiased estimators. Second, to avoid the explicit construction of the polynomial, we use a linear program based on moment matching in the non-smooth regime which is independent of the target functional (except for the final plug-in step in \eqref{eq:nonsmooth_final}). Finally, to handle the original sampling model without Poissonization, we use Binomial probabilities as the weights in the estimator construction, while the dependence across symbols is handled in Section \ref{sec:analysis} based on an application of the Efron--Stein--Steele inequality (cf. Lemma \ref{lemma:ES_multinomial} in the Appendix). 

The performance of the estimator $\widehat{D}_\alpha$ is summarized in the following theorem. 
\begin{theorem}\label{thm:alpha_upper}
	For integer $\alpha\ge 2$, constant $c_1>0$ large enough and $c_2\log n\ge 1$, it holds that
	\begin{align*}
	\sup_{(P,Q) \in \calM_k(U)} \bE_{(P,Q)} |\widehat{D}_\alpha - D_\alpha(P\|Q) | \lesssim_{\alpha,c_1,c_2} \frac{kU^\alpha}{n\log n} + \frac{\sqrt{k}U^\alpha}{n^{1-cc_2}} + \frac{U^{\alpha-1}}{\sqrt{m}} + \frac{U^{\alpha-1/2}}{\sqrt{n}},
	\end{align*}
	where $c>0$ is an absolute constant. 
\end{theorem}
Based on Theorem \ref{thm:alpha_upper} and the assumption $\log k\gtrsim_\alpha \log n$ in Theorem \ref{thm.alpha_div}, we conclude that by choosing $c_2>0$ small enough, Theorem \ref{thm:alpha_upper} implies the upper bound of Theorem \ref{thm.alpha_div}. 

\subsection{Construction for KL divergence}\label{subsec.KL}
The estimator construction for the KL divergence $D_{\text{KL}}(P\|Q)$ is entirely similar. Specifically, if we write the KL divergence $D_{\text{KL}}(P\|Q) = - H(P) - \sum_{i=1}^k p_i\log q_i$ as the negated sum of the Shannon entropy $H(P) = \sum_{i=1}^k -p_i\log p_i$ and the cross entropy $\sum_{i=1}^k p_i\log q_i$, the previous estimator construction can be applied to both entropies separately. The detailed construction is as follows (assuming that both $m$ and $n$ are even): 
\begin{enumerate}
	\item Estimate the non-smooth component of the cross entropy: first, following the same steps \eqref{eq:moment_est}--\eqref{eq:LP} with $\alpha=1$, we solve for the problem-independent measure $\widehat{\mu}^{(1)}$. Then we again apply the plug-in approach for the non-smooth component of the cross entropy: 
	\begin{align*}
	\widehat{D}_{\text{KL}, \text{ns}}^{(1)} = \int -p\log q \cdot \widehat{\mu}^{(1)}(dp,dq). 
	\end{align*}
	\item Estimate the non-smooth component of the Shannon entropy: 
	\begin{itemize}
		\item Construct the function
		\begin{align*}
		g_d'(x) = \prod_{d'=0}^{d-1}\frac{x-d'}{m/2-d'}
		\end{align*}
		and its modification $\widetilde{g}_d'(x) = \min\{g_d'(x), g_d'(\lceil 2c_1\log m\rceil)  \}$, as well as the estimators of smoothed moments: 
		\begin{align*}
		\widehat{M}_d^{(2)} = \sum_{i=1}^k \sum_{0\le s\le c_1\log m} \bP\left(\mathsf{HG}\left(m,m\widehat{p}_i, \frac{m}{2}\right) = s\right) \widetilde{g}_d'(m\widehat{p}_i - s). 
		\end{align*}
		\item Set $G' = \lceil c_2\log m\rceil$ and solve the following linear program for the optimal $\widehat{\mu}^{(2)}$: 
		\begin{align*}
		\text{minimize} & \qquad  L^{(2)}(\widehat{\mu}^{(2)}, \widehat{M}^{(2)}) \triangleq \sum_{d=0}^{G'} \left(\frac{3c_1\log m}{m} \right)^{-d}\cdot \left|\int p^d\widehat{\mu}^{(2)}(dp) - \widehat{M}_{d}^{(2)}  \right| \\
		\text{subject to} &\qquad \text{supp}(\widehat{\mu}^{(2)}) \subseteq  \left[0, \frac{3c_1\log m}{m} \right], \quad \widehat{\mu}^{(2)}(\bR) \le k. 
		\end{align*}
		\item Apply the plug-in approach to estimate the non-smooth component of the Shannon entropy as
		\begin{align*}
		\widehat{D}_{\text{KL}, \text{ns}}^{(2)} = \int p\log p \cdot \widehat{\mu}^{(2)}(dp). 
		\end{align*}
	\end{itemize}
	\item Estimate the smooth component of the cross entropy: define function $h: \NN \to\bR$ with
	\begin{align*}
	h^{(1)}(x) = \1(x\neq 0)\cdot \left( -\log \frac{2x}{n} - \frac{1-2x/n}{2x} \right),
	\end{align*}
	and compute the following estimator
	\begin{align*}
	\widehat{D}_{\text{KL}, \text{s}}^{(1)} = \sum_{i=1}^k \left(\widehat{p}_i\cdot \sum_{s>c_1\log n}\bP\left(\mathsf{HG}\left(n,n\widehat{q}_i, \frac{n}{2}\right) = s\right) h^{(1)}(n\widehat{q}_i - s) \right). 
	\end{align*}
	\item Estimate the smooth component of the Shannon entropy: define function $h': \NN\to \bR$ with
	\begin{align*}
	h^{(2)}(x) = \frac{2x}{m}\log \frac{2x}{m} - \frac{1-2x/m}{m},
	\end{align*}
	and compute the following estimator 
	\begin{align*}
	\widehat{D}_{\text{KL}, \text{s}}^{(2)} = \sum_{i=1}^k \sum_{s>c_1\log m}\bP\left(\mathsf{HG}\left(m,m\widehat{p}_i, \frac{m}{2}\right) = s\right) h^{(2)}(m\widehat{p}_i - s). 
	\end{align*}
	\item Final estimator: 
	\begin{align*}
	\widehat{D}_{\text{KL}} = \widehat{D}_{\text{KL}, \text{ns}}^{(1)} + \widehat{D}_{\text{KL}, \text{ns}}^{(2)} + \widehat{D}_{\text{KL}, \text{s}}^{(1)} + \widehat{D}_{\text{KL}, \text{s}}^{(2)}. 
	\end{align*}
\end{enumerate}

In the above construction, the functions $h^{(1)}$ and $h^{(2)}$ are the bias-corrected plug-in estimators for the target function $-\log q$ and $p\log p$, respectively. In fact, the following part of the estimator $\widehat{D}_{\text{KL}, \text{ns}}^{(2)} + \widehat{D}_{\text{KL}, \text{s}}^{(2)}$ is already minimax rate-optimal for estimating the negative Shannon entropy without Poissonization, sample splitting, or explicit polynomial approximation. The estimation performance of the above estimator $\widehat{D}_{\text{KL}}$ is summarized in the following theorem. 

\begin{theorem}\label{thm:KL_upper}
	For integer $\alpha\ge 2$, constant $c_1>0$ large enough and $c_2\log n\ge 1$, it holds that
	\begin{align*}
	&\sup_{(P,Q) \in \calM_k(U)} \bE_{(P,Q)} |\widehat{D}_{\text{\rm KL}} - D_{\text{\rm KL}}(P\|Q) | \\
	&\lesssim_{\alpha,c_1,c_2} \frac{k}{m\log m} + \frac{kU}{n\log n} + (mn)^{cc_2}\left( \frac{\sqrt{k}}{m} + \frac{\sqrt{k}U}{n}\right) + \frac{\log U}{\sqrt{m}} + \sqrt{\frac{U}{n}},
	\end{align*}
	where $c>0$ is an absolute constant. 
\end{theorem}
Based on Theorem \ref{thm:KL_upper} and the assumption $\log k\gtrsim \log (m+n)$ in Theorem \ref{thm.KL}, we again conclude that by choosing the constant $c_2>0$ small enough, Theorem \ref{thm:KL_upper} implies the upper bound of Theorem \ref{thm.KL}. 

\section{Estimator Analysis}\label{sec:analysis}
In this section, we provide the roadmap of the error analysis of the estimator $\widehat{D}_\alpha$ and prove Theorem \ref{thm:alpha_upper}, relegating the similar analysis of the estimator $\widehat{D}_{\text{KL}}$ and the proofs of main lemmas to the Appendices. Clearly, a triangle inequality gives
\begin{align}\label{eq:triangle}
\bE|\widehat{D}_{\alpha}  - D_\alpha(P \| Q)| \le \frac{\bE|\widehat{D}_{\alpha,\text{ns}}  - D_{\alpha,\text{ns}}| + \bE|\widehat{D}_{\alpha,\text{s}}  - D_{\alpha,\text{s}}|}{\alpha(\alpha-1)},
\end{align}
and it suffices to upper bound the estimation error of the estimators $\widehat{D}_{\alpha,\text{ns}}, \widehat{D}_{\alpha,\text{s}}$ in \eqref{eq:nonsmooth_final}, \eqref{eq:smooth_final} in estimating the quantities $D_{\alpha,\text{ns}}, D_{\alpha,\text{s}}$ in \eqref{eq:nonsmooth_component}, \eqref{eq:smooth_component}, respectively. To this end, Section \ref{subsec:implicit} shows that for the plug-in estimator $\widehat{D}_{\alpha,\text{ns}}$ of the moment matching estimator, a deterministic upper bound of $|\widehat{D}_{\alpha,\text{ns}} - D_{\alpha,\text{ns}}|$ in terms of $L(\mu, \widehat{M})$ for a properly constructed measure $\mu$ is available via an \emph{implicit} polynomial approximation. Consequently, the estimation error of the non-smooth component $\bE|\widehat{D}_{\alpha,\text{ns}}  - D_{\alpha,\text{ns}}|$ is effectively controlled by $\bE[L(\mu,\widehat{M})]$, which further reduces to the bias-variance analysis of the smoothed moments estimator presented in Section \ref{subsec:nonsmooth}. Finally, Section \ref{subsec:smooth} deals with the estimation error (specifically, bias and variance) of the bias-corrected plug-in approach for the smooth component $D_{\alpha,\text{s}}$. 

\subsection{A deterministic inequality via implicit polynomial approximation}\label{subsec:implicit}
In this subsection, we prove a deterministic upper bound on $|\widehat{D}_{\alpha,\text{ns}} - D_{\alpha,\text{ns}}|$ for the plug-in approach of the measure $\widehat{\mu}$ in the linear program \eqref{eq:LP}. To this end, consider the following measure $\mu$ based on the perfect knowledge of $(P,Q)$: 
\begin{align}\label{eq:measure_true}
\mu = \sum_{i=1}^k \delta_{(p_i,q_i)}\cdot \bP\left(\mathsf{B}\left(\frac{n}{2},q_i\right)\le c_1\log n \right)\1\left(q_i \le \frac{3c_1\log n}{n} \right), 
\end{align}
where $\delta_{(p,q)}$ denotes the Dirac point measure on the single point $(p,q)$. Clearly the measure $\mu$ is supported on $\calS$ and satisfies $\mu(\bR^2)\le k$, thus it is a feasible solution to the linear program \eqref{eq:LP}. Moreover, a combination of \eqref{eq:nonsmooth_component} and \eqref{eq:measure_true} gives
\begin{align*}
\left| \int \frac{p^\alpha}{q^{\alpha-1}}\mu(dp,dq)  - D_{\alpha,\text{ns}}\right| &= \sum_{i=1}^k \frac{p_i^\alpha}{q_i^{\alpha-1}}\cdot \bP\left(\mathsf{B}\left(\frac{n}{2},q_i\right)\le c_1\log n \right)\1\left(q_i > \frac{3c_1\log n}{n} \right)\\
&\le \frac{1}{n^{5\alpha}}\sum_{i=1}^k \frac{p_i^\alpha}{q_i^{\alpha-1}} \le \frac{U^{\alpha-1}}{n^{5\alpha}}\sum_{i=1}^k p_i = \frac{U^{\alpha-1}}{n^{5\alpha}},
\end{align*}
where the inequalities follow from Lemma \ref{lemma.localization} and the bounded likelihood ratio $p_i\le Uq_i$. Hence, by a triangle inequality, the following deterministic inequality holds: 
\begin{align}\label{eq:integral_diff}
|\widehat{D}_{\alpha,\text{ns}} - D_{\alpha,\text{ns}}| \le \left| \int \frac{p^\alpha}{q^{\alpha-1}}(\widehat{\mu}(dp,dq) - \mu(dp,dq)) \right| + \frac{U^{\alpha-1}}{n^{5\alpha}}, 
\end{align}
and the quantity of interest is the integral difference of the function $p^\alpha/q^{\alpha-1}$ with respect to the measures $\widehat{\mu}$ and $\mu$. 

Next we introduce an approximating polynomial of the function $p^\alpha/q^{\alpha-1}$ on the set $\calS$, where it is implicit in the sense that it is only used in the estimator analysis but not in the estimator construction. By Lemma \ref{lemma:approximation} in the Appendix, there exists a polynomial $Q_0(x) = \sum_{d=\alpha}^{G+\alpha} a_dx^d$ such that
\begin{align*}
\sup_{x\in [0,1]}\left|x -  Q_0(x)\right| \lesssim_{\alpha} \frac{1}{G^2}. 
\end{align*}
Now define 
\begin{align*}
Q(q) = \frac{3c_1\log n}{nq^\alpha}\cdot Q_0\left(\frac{nq}{3c_1\log n}\right), \quad q\in \left[0, \frac{3c_1\log n}{n}\right].
\end{align*}
It is clear that $Q$ is a degree-$G$ polynomial, and 
\begin{align}\label{eq:approximation_error}
\sup_{(p,q)\in \calS} \left|\frac{p^\alpha}{q^{\alpha-1}} - p^\alpha Q(q) \right| \le U^\alpha \sup_{q\in [0,3c_1\log n/n]} \left|q - q^\alpha Q(q)\right| \lesssim_{\alpha,c_1,c_2} \frac{U^\alpha}{n\log n}. 
\end{align}
In other words, the bivariate polynomial $p^\alpha Q(q)$ is a good approximation of $p^\alpha/q^{\alpha-1}$ on $\calS$ with a uniform approximation error in \eqref{eq:approximation_error}. Moreover, as $|Q_0(x)|$ is upper bounded by $O(1)$ on the unit interval $[0,1]$, the even polynomial $P_0(t) = Q_0(t^2)$ on $[-1,1]$ is also bounded by a constant. Consequently, Lemma \ref{lemma:coefficient_bound} shows that $\max_{\alpha\le d\le \alpha+G} |a_d|\le (1+\sqrt{2})^{2(\alpha+G)} = O_\alpha(n^{cc_2})$ with some absolute constant $c>0$. As a further result, if we express $Q(q) = \sum_{d=0}^G b_d q^d$, the coefficients have the upper bound
\begin{align}\label{eq:coefficient_bound}
|b_d| = \left(\frac{3c_1\log n}{n} \right)^{1-d-\alpha}|a_{d+\alpha}| \le n^{cc_2}\cdot \left(\frac{3c_1\log n}{n} \right)^{1-d-\alpha}, \quad d=0,1,\cdots,G. 
\end{align}

Consequently, by \eqref{eq:approximation_error} and \eqref{eq:coefficient_bound}, the integral difference in \eqref{eq:integral_diff} can be upper bounded by
\begin{align*}
&\left| \int \frac{p^\alpha}{q^{\alpha-1}}(\widehat{\mu}(dp,dq) - \mu(dp,dq)) \right| \\
&\le \left| \int_\calS \left(\frac{p^\alpha}{q^{\alpha-1}} - p^\alpha Q(q)\right)(\widehat{\mu}(dp,dq) - \mu(dp,dq)) \right| + \sum_{d=0}^G |b_d|\cdot \left|\int p^\alpha q^d(\widehat{\mu}(dp,dq) - \mu(dp,dq)) \right| \\
&\lesssim_{\alpha,c_1,c_2}  \int_\calS \frac{U^\alpha}{n\log n} (\widehat{\mu}(dp,dq) + \mu(dp,dq))  \\
&\quad + n^{cc_2}\left(\frac{3c_1\log n}{n}\right)^{1-\alpha}\sum_{d=0}^G \left(\frac{3c_1\log n}{n}\right)^{-d}\left|\int p^\alpha q^d(\widehat{\mu}(dp,dq) - \mu(dp,dq)) \right| \\
&\lesssim_{\alpha,c_1,c_2} \frac{kU^\alpha}{n\log n} + n^{cc_2}\left(\frac{3c_1\log n}{n}\right)^{1-\alpha} L(\widehat{\mu}, \mu),
\end{align*}
where the last inequality is due to the constraint $\mu(\bR^2), \widehat{\mu}(\bR^2)\le k$, and we slightly abuse the notation $L(\widehat{\mu}, \mu)$ to denote $L(\widehat{\mu},M)$ in \eqref{eq:LP} with $M_{\alpha,d} = \int p^\alpha q^d\mu(dp,dq)$. Now by the triangle inequality and the definition of the minimizer $\widehat{\mu}$, we have
\begin{align}\label{eq:LP_triangle}
L(\widehat{\mu}, \mu) \le L(\widehat{\mu}, \widehat{M}) + L(\mu, \widehat{M}) \le 2 L(\mu, \widehat{M}),
\end{align}
and therefore the following final inequality holds for the difference $|\widehat{D}_{\alpha,\text{ns}} - D_{\alpha,\text{ns}}|$: 
\begin{align}\label{eq:deterministic_ineq}
|\widehat{D}_{\alpha,\text{ns}} - D_{\alpha,\text{ns}}| \lesssim_{\alpha,c_1,c_2} \frac{kU^\alpha}{n\log n} + n^{cc_2}\left(\frac{3c_1\log n}{n}\right)^{1-\alpha}L(\mu, \widehat{M}), 
\end{align}
where $c>0$ is an absolute constant.

\subsection{Bias-variance analysis in the non-smooth regime}\label{subsec:nonsmooth}
To complete the error analysis for the non-smooth component, the deterministic inequality \eqref{eq:deterministic_ineq} shows that it suffices to upper bound the quantity $\bE[L(\mu, \widehat{M})]$, or equivalently, to upper bound the estimation error $\bE|\widehat{M}_{\alpha,d} - M_{\alpha,d}^\star|$ of the smoothed moments for each $d=0,1,\cdots,G$, where
\begin{align*}
M_{\alpha,d}^\star \triangleq \int p^\alpha q^d\mu(dp,dq) = \sum_{i=1}^k p_i^\alpha q_i^d\cdot \bP\left(\mathsf{B}\left(\frac{n}{2},q_i\right)\le c_1\log n\right)\1\left(q_i\le \frac{3c_1\log n}{n}\right)
\end{align*}
is very close to the smoothed moments defined in \eqref{eq:smoothed_moments_true}. Based on the following bias-variance decomposition
\begin{align*}
\bE|\widehat{M}_{\alpha,d} - M_{\alpha,d}^\star| \le |\bE\widehat{M}_{\alpha,d} - M_{\alpha,d}^\star| + \sqrt{\var(\widehat{M}_{\alpha,d})}, 
\end{align*}
it suffices to analyze the bias and variance of the estimator $\widehat{M}_{\alpha,d}$, respectively. Specifically, the bias $|\bE\widehat{M}_{\alpha,d} - M_{\alpha,d}^\star|$ is expected to be very small, as the estimator $\widehat{M}_{\alpha,d}$ would be exactly unbiased if the modification $\widetilde{g}_d(x)$ in \eqref{eq:moment_est_trunc} were not applied and $M_{\alpha,d}^\star$ were equal to $M_{\alpha,d}$, and the expected differences incurred by the modification $\widetilde{g}_d$ as well as $|M_{\alpha,d} - M_{\alpha,d}^\star|$ are both small due to the concentration inequalities in Lemma \ref{lemma.localization}. As for the variance $\var(\widehat{M}_{\alpha,d})$, the truncation in $\widetilde{g}_d(x)$ ensures that changing one observation only results in a tiny change in the estimator $\widehat{M}_{\alpha,d}$, and thus Lemma \ref{lemma:ES_multinomial} in the appendix, which is a corollary and convenient form of the Efron-Stein-Steele inequality for bivariate functional estimation, gives a small variance. The following lemma summarizes the upper bounds on the bias and the variance. 
\begin{lemma}\label{lemma:nonsmooth_bias_variance}
	For integer $\alpha\ge 2$, constant $c_1>0$ large enough, $c_2\log n\ge 1$ and any $d=0,1,\cdots,G$, if $(P,Q)\in \calM_k(U)$, it holds that
	\begin{align*}
	|\bE\widehat{M}_{\alpha,d} - M_{\alpha,d}^\star| &\lesssim_{\alpha,c_1,c_2} \frac{kU^\alpha}{n^{4\alpha}}\left(\frac{3c_1\log n}{n}\right)^{\alpha+d}, \\
	\sqrt{\var(\widehat{M}_{\alpha,d})} &\lesssim_{\alpha,c_1,c_2} \sqrt{k}U^\alpha\cdot n^{c'c_2}\left(\frac{3c_1\log n}{n}\right)^{\alpha+d},
	\end{align*}
	where $c'>0$ is an absolute constant. 
\end{lemma}
Based on Lemma \ref{lemma:nonsmooth_bias_variance}, the quantity $\bE[L(\mu, \widehat{M})]$ can be upper bounded as
\begin{align}\label{eq:objective_upper_bound}
\bE[L(\mu, \widehat{M})] &\le \sum_{d=0}^G \left(\frac{3c_1\log n}{n}\right)^{-d}\left[|\bE\widehat{M}_{\alpha,d} - M_{\alpha,d}^\star| + \sqrt{\var(\widehat{M}_{\alpha,d})} \right] \nonumber \\
&\lesssim_{\alpha,c_1,c_2} \sum_{d=0}^G \left(\frac{3c_1\log n}{n}\right)^{-d}\cdot \left(\frac{3c_1\log n}{n}\right)^{\alpha+d}\left[\frac{kU^\alpha}{n^{4\alpha}} + \sqrt{k}U^\alpha\cdot n^{c'c_2} \right] \nonumber\\
&\lesssim_{\alpha,c_1,c_2} \left(\frac{3c_1\log n}{n}\cdot U\right)^{\alpha}\left[\frac{k\log n}{n^{4\alpha}} + \sqrt{k}n^{c'c_2}\log n\right],
\end{align}
which together with \eqref{eq:deterministic_ineq} completes the analysis of $\bE|\widehat{D}_{\alpha,\text{ns}} - D_{\alpha,\text{ns}}|$. 

\subsection{Bias-variance analysis in the smooth regime}\label{subsec:smooth}
Now the only remaining quantity of interest is $\bE|\widehat{D}_{\alpha,\text{s}} - D_{\alpha,\text{s}}|$, and we use the bias-variance decomposition again to write
\begin{align*}
\bE|\widehat{D}_{\alpha,\text{s}} - D_{\alpha,\text{s}}| \le |\bE\widehat{D}_{\alpha,\text{s}} - D_{\alpha,\text{s}}| + \sqrt{\var(\widehat{D}_{\alpha,\text{s}})}. 
\end{align*}
To handle the bias, we first need the following lemma on the performance of $h_\alpha$ in \eqref{eq:bias_correction}. 
\begin{lemma}\label{lemma:smooth_bias}
	Let $X\sim \mathsf{B}(n/2,q)$ with $q\ge c_1\log n/n$. Then for integer $\alpha\ge 2$ and constant $c_1>0$ large enough, we have
	\begin{align*}
	|\bE[h_\alpha(X)] - q^{1-\alpha}| \lesssim_{\alpha,c_1} \frac{1}{nq^\alpha\log n}. 
	\end{align*}
\end{lemma}

Let $X_i\sim \mathsf{B}(n/2,q_i)$ for all $i\in [k]$, then thanks to the identity
\begin{align*}
\bE\left[\prod_{\ell=0}^{\alpha-1}\frac{m\widehat{p}_i - \ell}{m-\ell}\sum_{s>c_1\log n}\bP\left(\mathsf{HG}\left(n,n\widehat{q}_i, \frac{n}{2}\right) = s\right) h_\alpha(n\widehat{q}_i - s)\right] = p_i^\alpha\bE[h_\alpha(X_i)]\cdot \bP\left(\mathsf{B}\left(\frac{n}{2},q_i\right) > c_1\log n\right),
\end{align*}
we have
\begin{align*}
|\bE\widehat{D}_{\alpha,\text{s}} - D_{\alpha,\text{s}}| \le \sum_{i=1}^k p_i^\alpha|\bE[h_\alpha(X_i)] - q_i^{1-\alpha}|\cdot \bP\left(\mathsf{B}\left(\frac{n}{2},q_i\right) > c_1\log n\right). 
\end{align*}
By Lemma \ref{lemma.localization}, the Binomial probability is at most $n^{-5\alpha}$ if $q_i\le c_1\log n/n$; since $\|h_\alpha\|_\infty = O_\alpha(n^{\alpha-1})$ and $p_i\le Uq_i$, the quantity $p_i^\alpha|\bE[h_\alpha(X_i)] - q_i^{1-\alpha}|$ is always $O_\alpha(p_i(U^{\alpha-1} + n^{\alpha-1}))$. Therefore, the total contribution of symbols $i\in [k]$ with small probability $q_i \le c_1\log n/n$ to the above quantity is at most $O_{\alpha,c_1}(n^{-5\alpha}(U^{\alpha-1} + n^{\alpha-1}))$. Now applying Lemma \ref{lemma:smooth_bias} for symbols with large probability gives the final bias bound
\begin{align}\label{eq:smooth_bias}
|\bE\widehat{D}_{\alpha,\text{s}} - D_{\alpha,\text{s}}| \lesssim_{\alpha,c_1} \frac{U^{\alpha-1} + n^{\alpha-1}}{n^{5\alpha}} + \sum_{i=1}^k \frac{p_i^\alpha}{nq_i^\alpha\log n} \lesssim_{\alpha,c_1} \frac{kU^\alpha}{n\log n}. 
\end{align}

The above bias analysis is similar to the prior work using explicit polynomial approximation, e.g. \cite{bu2018estimation}. The following variance analysis will be different, where in Poissonized models one can conveniently use the independence of frequency counts, while we use the independence of raw samples instead. Specifically, we again make use of Lemma \ref{lemma:ES_multinomial} to upper bound the variance and utilize the stability of the estimator $\widehat{D}_{\alpha,\text{\rm s}}$ in the local change of empirical frequencies. The following lemma presents the final variance bound. 
\begin{lemma}\label{lemma:smooth_variance}
	For integer $\alpha\ge 2$ and constant $c_1>0$ large enough, it holds that
	\begin{align*}
	\sqrt{\var(\widehat{D}_{\alpha,\text{\rm s}})} \lesssim_{\alpha,c_1,\varepsilon} \frac{U^{\alpha-1}}{\sqrt{m}} + \frac{U^{\alpha-1/2}}{\sqrt{n}} + \frac{\sqrt{k}U^\alpha}{n^{1-\varepsilon}}, 
	\end{align*}
	where $\varepsilon>0$ is any absolute constant. 
\end{lemma}

Hence, combining the inequalities \eqref{eq:triangle}, \eqref{eq:deterministic_ineq}, \eqref{eq:objective_upper_bound}, \eqref{eq:smooth_bias} and Lemma \ref{lemma:smooth_variance}, we conclude that
\begin{align*}
\bE_{(P,Q)} |\widehat{D}_\alpha - D_\alpha(P\|Q) | \lesssim_{\alpha,c_1,c_2} \frac{kU^\alpha}{n\log n} + \frac{\sqrt{k}U^\alpha}{n^{1-cc_2}} + \frac{U^{\alpha-1}}{\sqrt{m}} + \frac{U^{\alpha-1/2}}{\sqrt{n}}
\end{align*}
for some absolute constant $c>0$ and all $(P,Q)\in \calM_k(U)$, which is exactly the statement of Theorem \ref{thm:alpha_upper}.

\appendix
\section{Auxiliary Lemmas}\label{appendix:auxiliary}
\begin{lemma}[Lemma 17 of \cite{han2018local}]\label{lemma.localization}
	Fix any $\alpha>0$. There exists a constant $c = c(\alpha)>0$ such that for all $c_1>c$, the following inequality holds for each $p\in [0,1]$: 
	\begin{align*}
	\bP(\mathsf{B}(n/2,p)\le c_1\log n \mid p\ge 3c_1\log n/n) &\le n^{-5\alpha}, \\
	\bP(\mathsf{B}(n/2,p)\ge c_1\log n \mid p\le c_1\log n/n) & \le n^{-5\alpha}. 
	\end{align*}
\end{lemma}

\begin{lemma}\label{lemma:approximation}
	Fix any integer $\alpha \ge 2$. There exist absolute constants $c_\alpha,C_\alpha>0$ such that for any $n\ge \alpha$, it holds that
	\begin{align*}
	\frac{c_\alpha}{n^2} \le  \inf_{a_\alpha,\cdots,a_n\in \bR}\sup_{x\in [0,1]} \left|x - \sum_{d = \alpha}^n a_dx^d \right| \le \frac{C_\alpha}{n^2}. 
	\end{align*}
\end{lemma}

\begin{lemma}[\!\!\cite{markov1892functions}]\label{lemma:coefficient_bound}
	Let $p(x) = \sum_{d=0}^n a_dx^d$ be a polynomial of degree $n$ with $|p(x)|\le 1$ for all $x\in [-1,1]$. Then
	\begin{align*}
	\max_{0\le d\le n} |a_d| \le (1+\sqrt{2})^n. 
	\end{align*}
\end{lemma}

 \begin{lemma}\label{lemma:ES_multinomial}
	Let $(m\widehat{p}_1,\cdots,m\widehat{p}_k)\sim \mathsf{Multi}(m;p_1,\cdots,p_k)$ and $(n\widehat{q}_1,\cdots,n\widehat{q}_k)\sim \mathsf{Multi}(n;q_1,\cdots,q_k)$ be independent random vectors. For $S = \sum_{i=1}^k f_i(\widehat{p}_i, \widehat{q}_i)$ with any $f_i: [0,1]^2\to \bR$, we have
	\begin{align*}
	\var(S) &\le 2m\cdot \sum_{i=1}^k \bE\left[\widehat{p}_i \left(f_i(\widehat{p}_i, \widehat{q}_i) -f_i\left(\widehat{p}_i - \frac{1}{m}, \widehat{q}_i\right) \right)^2 \right] \\
	&\quad + 2n\cdot \sum_{i=1}^k \bE\left[\widehat{q}_i \left(f_i(\widehat{p}_i, \widehat{q}_i) -f_i\left(\widehat{p}_i, \widehat{q}_i - \frac{1}{n}\right) \right)^2 \right], 
	\end{align*}
	with the convention that whenever $\widehat{p}_i = 0$ or $\widehat{q}_i = 0$, the respective product inside the expectation is zero. 
\end{lemma}
\section{Upper Bound Analysis of the KL Divergence}\label{appendix:KL}
The analysis of the KL divergence estimator $\widehat{D}_{\text{KL}}$ follows similar lines to that of the $\alpha$-divergences. First, using a triangle inequality, we may express the error as
\begin{align}\label{eq:triangle_KL}
&\bE |\widehat{D}_{\text{KL}} - D_{\text{KL}}(P\|Q) | \nonumber\\
&\le \bE |\widehat{D}_{\text{KL}, \text{ns}}^{(1)} - {D}_{\text{KL}, \text{ns}}^{(1)}| + \bE |\widehat{D}_{\text{KL}, \text{ns}}^{(2)} - {D}_{\text{KL}, \text{ns}}^{(2)}| + \bE|\widehat{D}_{\text{KL}, \text{s}}^{(1)} + \widehat{D}_{\text{KL}, \text{s}}^{(2)} - ({D}_{\text{KL}, \text{s}}^{(1)} + {D}_{\text{KL}, \text{s}}^{(2)}) |,
\end{align}
where 
\begin{align*}
{D}_{\text{KL}, \text{ns}}^{(1)} &= \sum_{i=1}^k p_i\log \frac{1}{q_i} \cdot \bP\left(\mathsf{B}\left(\frac{n}{2},q_i\right) \le c_1\log n\right), \\
{D}_{\text{KL}, \text{ns}}^{(2)} &= \sum_{i=1}^k p_i\log p_i \cdot \bP\left(\mathsf{B}\left(\frac{m}{2},p_i\right) \le c_1\log m\right),\\
{D}_{\text{KL}, \text{s}}^{(1)} &= \sum_{i=1}^k p_i\log \frac{1}{q_i} \cdot \bP\left(\mathsf{B}\left(\frac{n}{2},q_i\right) > c_1\log n\right), \\
{D}_{\text{KL}, \text{s}}^{(2)} &= \sum_{i=1}^k p_i\log p_i \cdot \bP\left(\mathsf{B}\left(\frac{m}{2},p_i\right) > c_1\log m\right). 
\end{align*}
Based on \eqref{eq:triangle_KL}, we apply the duality between moment matching and best polynomial approximation to upper bound the first two terms, and the bias-variance analysis for the last term. Similar to Section \ref{sec:analysis}, the following subsections provide the roadmap of the analysis, with detailed proofs of the main lemmas postponed to Appendix \ref{appendix:main_proof}. 

\subsection{Deterministic inequalities}
Similar to Section \ref{subsec:implicit}, we first find a random upper bound which holds deterministically for the quantities $|\widehat{D}_{\text{KL}, \text{ns}}^{(1)} - {D}_{\text{KL}, \text{ns}}^{(1)}|$ and $|\widehat{D}_{\text{KL}, \text{ns}}^{(2)} - {D}_{\text{KL}, \text{ns}}^{(2)}|$. To start with, note that the measures
\begin{align*}
\mu^{(1)} &= \sum_{i=1}^k \delta_{(p_i,q_i)}\bP\left(\mathsf{B}\left(\frac{n}{2},q_i\right) \le c_1\log n\right)\cdot \1\left(q_i \le \frac{3c_1\log n}{n} \right), \\
\mu^{(2)} &= \sum_{i=1}^k \delta_{p_i}\bP\left(\mathsf{B}\left(\frac{m}{2},p_i\right) \le c_1\log m\right)\cdot \1\left(p_i \le \frac{3c_1\log m}{m} \right)
\end{align*}
are feasible solutions to the respective linear programs for the estimated measure $\widehat{\mu}$, and it remains to find good polynomial approximations of the functions $-p\log q$ and $p\log p$ over the respective domains. To this end, we recall the following result from \cite[Lemma 18]{Jiao--Venkat--Han--Weissman2015minimax}. 
\begin{lemma}\label{lemma:entropy_approx}
There is an absolute constant $C>0$ such that for all $n\in \NN$, there exists a polynomial $P_n(x)$ of degree at most $n$ such that $P_n(0) = 0$ and 
\begin{align*}
\sup_{x\in [0,1]} \left|x\log x - P_n(x) \right| \le \frac{C}{n^2}. 
\end{align*}
\end{lemma}
Let $P_0(x)$ and $Q_0(x)$ be the polynomials given by the Lemma \ref{lemma:entropy_approx} with degrees $G' = \lceil c_2\log m\rceil$ and $G = \lceil c_2\log n\rceil$, respectively. Then by a simple scaling, the polynomial
\begin{align*}
P(p) = \frac{3c_1\log m}{m} P_0\left(\frac{mp}{3c_1\log m} \right) + p\log\left( \frac{3c_1\log m}{m} \right)
\end{align*}
satisfies that $|P(p) - p\log p| \lesssim_{c_1,c_2} (m\log m)^{-1}$ for all $0\le p\le 3c_1\log m/m$. Similarly, another polynomial (note that $Q_0(0)=0$)
\begin{align*}
Q(q) = -\frac{1}{q}\left[\frac{3c_1\log n}{n} Q_0\left(\frac{nq}{3c_1\log n} \right) + q\log\left( \frac{3c_1\log n}{n} \right) \right]
\end{align*}
satisfies that $|Q(q) + \log q| \lesssim_{c_1,c_2} (nq\log n)^{-1}$ for every $0\le q\le 3c_1\log n/n$. Hence, the bivariate polynomial $pQ(q)$ satisfies that for every $(p,q)\in\calS$, it holds that
\begin{align*}
|pQ(q) + p\log q| \lesssim_{c_1,c_2} \frac{p}{nq\log n} \lesssim_{c_1,c_2} \frac{U}{n\log n}. 
\end{align*}
Hence, following the same lines of analysis of Section \ref{subsec:implicit}, the coefficient bound of $P_n(x)$ in Lemma \ref{lemma:entropy_approx} gives the following deterministic inequalities: 
\begin{equation}\label{eq:deterministic_KL}
\begin{split}
|\widehat{D}_{\text{KL}, \text{ns}}^{(1)} - {D}_{\text{KL}, \text{ns}}^{(1)}| &\lesssim_{c_1,c_2} \frac{kU}{n\log n} + n^{O(c_2)}\cdot L^{(1)}(\mu^{(1)}, \widehat{M}^{(1)}), \\
|\widehat{D}_{\text{KL}, \text{ns}}^{(2)} - {D}_{\text{KL}, \text{ns}}^{(2)}| &\lesssim_{c_1,c_2} \frac{k}{m\log m} + m^{O(c_2)-1}\cdot L^{(2)}(\mu^{(2)}, \widehat{M}^{(2)}). 
\end{split}
\end{equation}

\subsection{Bias-variance analysis in the non-smooth regime}
We follow the same lines of Section \ref{subsec:nonsmooth} to upper bound $\bE[L^{(1)}(\mu^{(1)}, \widehat{M}^{(1)})]$ and $\bE[L^{(2)}(\mu^{(2)}, \widehat{M}^{(2)})]$, or in other words, derive the estimation performance of the (smoothed) moment estimators $\widehat{M}_d^{(1)}$ and $\widehat{M}_d^{(2)}$. Specifically, let
\begin{align*}
M_{1,d}^\star &= \sum_{i=1}^k p_iq_i^d \bP\left(\mathsf{B}\left(\frac{n}{2},q_i\right) \le c_1\log n\right)\cdot \1\left(q_i \le \frac{3c_1\log n}{n} \right), \\
M_{2,d}^\star &= \sum_{i=1}^k p_i^d \bP\left(\mathsf{B}\left(\frac{m}{2},p_i\right) \le c_1\log m\right)\cdot \1\left(p_i \le \frac{3c_1\log m}{m} \right)
\end{align*}
be the true moments of $\mu^{(1)}$ and $\mu^{(2)}$ respectively, and we will show that the moment estimators are almost unbiased in estimating the above quantities. Moreover, these estimators enjoy small variance by a perturbation argument in the Efron--Stein--Steele inequality. The following lemma summarizes the bias-variance analysis of the moment estimators. 
\begin{lemma}\label{lemma:nonsmooth_KL}
Let $c_2\log n\ge 1$, and constant $c_1>0$ be large enough. Then for $0\le d\le \lceil c_2\log n\rceil$ and $(P,Q)\in \calM_k(U)$, it holds that
\begin{align*}
|\bE[\widehat{M}_d^{(1)} ] - M_{1,d}^\star | &\lesssim_{c_1,c_2} \frac{1}{n^5}\left(\frac{3c_1\log n}{n}\right)^d, \\
\sqrt{ \var(\widehat{M}_{d}^{(1)}) } &\lesssim_{c_1,c_2} n^{O(c_2)}\left( \frac{\sqrt{k}U}{n} + \frac{\sqrt{k}}{m} \right)\left(\frac{3c_1\log n}{n} \right)^d. 
\end{align*}
Similarly, for $c_2\log m\ge 1$ and $0\le d\le \lceil c_2\log m\rceil$, it holds that
\begin{align*}
|\bE[\widehat{M}_d^{(2)} ] - M_{2,d}^\star | &\lesssim_{c_1,c_2} \frac{1}{m^5}\left(\frac{3c_1\log m}{m}\right)^d, \\
\sqrt{ \var(\widehat{M}_{d}^{(2)}) } &\lesssim_{c_1,c_2} \sqrt{k}m^{O(c_2)}\cdot \left(\frac{3c_1\log m}{m} \right)^d. 
\end{align*}
\end{lemma}

Based on Lemma \ref{lemma:nonsmooth_KL} and the definition of the objective function in the linear program, it holds that
\begin{align*}
\bE[L^{(1)}(\mu^{(1)}, \widehat{M}^{(1)})] &= \sum_{d=0}^{\lceil c_2\log n\rceil} \left(\frac{3c_1\log n}{n} \right)^{-d} \cdot \bE|\widehat{M}_d^{(1)} - M_{1,d}^\star| \\
&\le \sum_{d=0}^{\lceil c_2\log n\rceil} \left(\frac{3c_1\log n}{n} \right)^{-d} \cdot \left(|\bE\widehat{M}_d^{(1)} - M_{1,d}^\star| + \sqrt{\var(\widehat{M}_d^{(1)})} \right) \\
&\lesssim_{c_1,c_2} \sum_{d=0}^{\lceil c_2\log n\rceil} \left( \frac{1}{n^5} + n^{O(c_2)}\left(\frac{\sqrt{k}U}{n} + \frac{\sqrt{k}}{m}\right) \right)\\
& \lesssim_{c_1,c_2} n^{O(c_2)}\left(\frac{\sqrt{k}}{m} + \frac{\sqrt{k}U}{n} \right). 
\end{align*}
Similarly, we also have
\begin{align*}
\bE[L^{(2)}(\mu^{(2)}, \widehat{M}^{(2)})] \lesssim_{c_1,c_2} m^{O(c_2)}\cdot \sqrt{k}.  
\end{align*}
Consequently, plugging the previous inequalities into the deterministic inequality \eqref{eq:deterministic_KL}, we conclude that
\begin{align}\label{eq:nonsmooth_final_KL}
\bE |\widehat{D}_{\text{KL}, \text{ns}}^{(1)} - {D}_{\text{KL}, \text{ns}}^{(1)}| + \bE |\widehat{D}_{\text{KL}, \text{ns}}^{(2)} - {D}_{\text{KL}, \text{ns}}^{(2)}| \lesssim_{c_1,c_2} \frac{kU}{n\log n} + \frac{k}{m\log m} + (mn)^{O(c_2)}\left(\frac{\sqrt{k}}{m} + \frac{\sqrt{k}U}{n} \right). 
\end{align}

\subsection{Bias-variance analysis in the smooth regime}
Using the usual bias-variance decomposition, we have
\begin{align}\label{eq:triangle_smooth}
&\bE|\widehat{D}_{\text{KL}, \text{s}}^{(1)} + \widehat{D}_{\text{KL}, \text{s}}^{(2)} - ({D}_{\text{KL}, \text{s}}^{(1)} + {D}_{\text{KL}, \text{s}}^{(2)}) | \nonumber \\
&\le |\bE[ \widehat{D}_{\text{KL}, \text{s}}^{(1)}] - {D}_{\text{KL}, \text{s}}^{(1)}| +  |\bE[ \widehat{D}_{\text{KL}, \text{s}}^{(2)}] - {D}_{\text{KL}, \text{s}}^{(2)}| + \sqrt{\var(\widehat{D}_{\text{KL}, \text{s}}^{(1)} + \widehat{D}_{\text{KL}, \text{s}}^{(2)}) }. 
\end{align}
Note that in \eqref{eq:triangle_smooth} we did not apply the triangle inequality to the variance part, as taking the sum inside the variance will lead to a significant reduction on the total variance\footnote{Specifically, the variance drops from $O_{c_1}(\log(kU)/\sqrt{m} + \sqrt{U/n})$ to $O_{c_1}(\log U/\sqrt{m} + \sqrt{U/n})$.}. To deal with the biases in \eqref{eq:triangle_smooth}, the first step is to analyze the bias-corrected plug-in estimators $h^{(1)}$ and $h^{(2)}$ in the estimator construction. 
\begin{lemma}\label{lemma:smooth_KL_bias}
Let $X\sim \mathsf{B}(n/2,q)$ with $q\ge c_1\log n/n$, for a constant $c_1>0$ large enough. Then 
\begin{align*}
| \bE[h^{(1)}(X) ] + \log q | \lesssim_{c_1} \frac{1}{nq\log n}. 
\end{align*}
Similarly, if $Y\sim \mathsf{B}(m/2,p)$ with $p\ge c_1\log m/m$, it holds that
\begin{align*}
| \bE[h^{(2)}(Y) ] - p\log p | \lesssim_{c_1} \frac{1}{m\log m}. 
\end{align*}
\end{lemma}
By \eqref{eq:unbiased_smoothed_moments} and the triangle inequality, we have
\begin{align*}
|\bE[ \widehat{D}_{\text{KL}, \text{s}}^{(1)}] - {D}_{\text{KL}, \text{s}}^{(1)}| \le \sum_{i=1}^k \bP\left(\mathsf{B}\left(\frac{n}{2},q_i\right)>c_1\log n \right)\cdot p_i|\bE[h^{(1)}(X_i)] + \log q_i|. 
\end{align*}
For each $i\in [k]$, if $q_i\le c_1\log n/n$, the above Binomial probability is at most $n^{-5}$ by Lemma \ref{lemma.localization}. Since $\| h^{(1)}\|_\infty \lesssim \log n$ and $p_i\le Uq_i$, the total contribution of symbols $i\in [k]$ with $q_i\le c_1\log n/n$ in the above sum is at most $O(kU/n^4)$. For symbols $i\in [k]$ with a large probability $q_i > c_1\log n/n$, we use Lemma \ref{lemma:smooth_KL_bias} to conclude that
\begin{align}\label{eq:smooth_bias_1}
|\bE[ \widehat{D}_{\text{KL}, \text{s}}^{(1)}] - {D}_{\text{KL}, \text{s}}^{(1)}| \lesssim_{c_1} \frac{kU}{n^4} + \sum_{i=1}^k \frac{p_i}{nq_i\log n} \lesssim_{c_1} \frac{kU}{n\log n}. 
\end{align}
Similarly, the following upper bound holds for the other bias: 
\begin{align}\label{eq:smooth_bias_2}
|\bE[ \widehat{D}_{\text{KL}, \text{s}}^{(2)}] - {D}_{\text{KL}, \text{s}}^{(2)}| \lesssim_{c_1} \frac{k}{m\log m}. 
\end{align}

Next we deal with the variance $\var(\widehat{D}_{\text{KL}, \text{s}}^{(1)} + \widehat{D}_{\text{KL}, \text{s}}^{(2)}) $ based on Lemma \ref{lemma:efron-stein}, and the upper bound is summarized in the following lemma. 
\begin{lemma}\label{lemma:smooth_KL_variance}
Let $c_1>0$ be a constant large enough and $\varepsilon>0$ be any constant, then
\begin{align*}
\sqrt{ \var(\widehat{D}_{\text{\rm KL}, \text{\rm s}}^{(1)} + \widehat{D}_{\text{\rm KL}, \text{\rm s}}^{(2)}) } \lesssim_{c_1,\varepsilon} \frac{\log U}{\sqrt{m}} + \sqrt{\frac{U}{n}} + (mn)^{\varepsilon}\left(\frac{\sqrt{k}U}{n} + \frac{\sqrt{k}}{m} \right). 
\end{align*}
\end{lemma}
Hence, by \eqref{eq:triangle_smooth}, \eqref{eq:smooth_bias_1}, \eqref{eq:smooth_bias_2}, and Lemma \ref{lemma:smooth_KL_variance}, we conclude that
\begin{align}\label{eq:smooth_final_KL}
& \bE|\widehat{D}_{\text{KL}, \text{s}}^{(1)} + \widehat{D}_{\text{KL}, \text{s}}^{(2)} - ({D}_{\text{KL}, \text{s}}^{(1)} + {D}_{\text{KL}, \text{s}}^{(2)}) | \nonumber\\
& \lesssim_{c_1,\varepsilon} \frac{k}{m\log m} + \frac{kU}{n\log n} + \frac{\log U}{\sqrt{m}} + \sqrt{\frac{U}{n}} + (mn)^{\varepsilon}\left(\frac{\sqrt{k}U}{n} + \frac{\sqrt{k}}{m} \right). 
\end{align}
Therefore, the desired Theorem \ref{thm:KL_upper} is a direct consequence of \eqref{eq:triangle_KL}, \eqref{eq:nonsmooth_final_KL}, and \eqref{eq:smooth_final_KL}.

\section{Minimax Lower Bounds}\label{appendix:lower_bound}
There are two main lemmas that we employ towards the proof of the minimax lower bounds. The first is the Le Cam's two-point method, which helps to prove the minimax lower bound corresponding to the variance, or equivalently, the classical asymptotics. Suppose we observe a random vector ${\bf Z} \in \mathcal{Z}$ with $\sigma$-algebra $\mathcal{A}$ and distribution $P_\theta$ where $\theta \in \Theta$. Let $\theta_0$ and $\theta_1$ be two elements of $\Theta$. Let $\widehat{T} = \widehat{T}({\bf Z})$ be an arbitrary estimator of a function $T(\theta)$ based on $\bf Z$. Le Cam's two-point method gives the following general minimax lower bound.
\begin{lemma}[Section 2.4.2 of \cite{Tsybakov2008}]\label{lemma:two-points}
 The following inequality holds:
	\begin{align*}
	\inf_{\widehat{T}} \sup_{\theta \in \Theta} \bP_\theta\left( |\widehat{T} - T(\theta)| \geq \frac{|T(\theta_1)-T(\theta_0)|}{2} \right) \geq
	\frac{1}{4}\exp\left(-D_{\text{\rm KL}}\left(P_{\theta_1}\|P_{\theta_0}\right)\right).
	\end{align*}
\end{lemma}

The second lemma is the generalized two-point method or method of two fuzzy hypotheses. Let $\sigma_0$ and $\sigma_1$ be two prior distributions supported on $\Theta$. Write $F_i$ for the marginal distribution of $\mathbf{Z}$ when the prior is $\sigma_i$ for $i = 0,1$, and let $\widehat{T} = \widehat{T}({\bf Z})$ be an arbitrary estimator of a function $T(\theta)$ based on $\bf Z$. We have the following general minimax lower bound.

\begin{lemma}[Theorem 2.15 of \cite{Tsybakov2008}] \label{lemma:fuzzy}
Suppose there exist $\zeta\in \mathbb{R}, s>0, 0\leq \beta_0,\beta_1 <1$ such that
	\begin{align*}
	\sigma_0(\theta: T(\theta) \leq \zeta -s) & \geq 1-\beta_0 \\
	\sigma_1(\theta: T(\theta) \geq \zeta + s) & \geq 1-\beta_1.
	\end{align*}
	If $\mathsf{TV}(F_1,F_0) \leq \eta<1$, then
	\begin{align*}
	\inf_{\hat{T}} \sup_{\theta \in \Theta} \bP_\theta\left( |\hat{T} - T(\theta)| \geq s \right) \geq \frac{1-\eta - \beta_0 - \beta_1}{2},
	\end{align*}
	where $\mathsf{TV}(P,Q) = \int|dP-dQ|/2$ is the total variation distance between two probability measures $P$ and $Q$. 
\end{lemma}

\subsection{Lower Bounds for $\alpha$-Divergences}\label{subsec:alpha_lower_proof}
In this section, we will make use of Lemma \ref{lemma:two-points} and Lemma \ref{lemma:fuzzy} to prove the following minimax lower bound in Theorem \ref{thm.alpha_div}. 
\begin{theorem}\label{thm:alpha_div_lower}
	Let $\alpha\ge 2$ be an integer. Then for any $m\gtrsim_\alpha U^{2(\alpha-1)}$, $n\gtrsim_\alpha kU^\alpha / \log k + U^{2\alpha-1}$, $U\gtrsim_\alpha (\log k)^2$, and $\log k\gtrsim_\alpha \log n$, 
	\begin{align*}
	\inf_{\widehat{D}}\sup_{(P,Q)\in \calM_k(U)} \bE_{(P,Q)} |\widehat{D} - D_\alpha(P\|Q)| \gtrsim_\alpha \frac{kU^{\alpha}}{n\log n} + \frac{U^{\alpha-1}}{\sqrt{m}} + \frac{U^{\alpha-1/2}}{\sqrt{n}}. 
	\end{align*}
\end{theorem}
Clearly Theorem \ref{thm:alpha_div_lower} implies the lower bound of Theorem \ref{thm.alpha_div}. We will use Lemma \ref{lemma:two-points} to prove the last two terms, and apply Lemma \ref{lemma:fuzzy} for the first term. 

First consider the following two-point construction. Fix $\varepsilon\in (0,1/2)$ to be chosen later, let 
\begin{align*}
P_1 &= \left(\frac{1-\varepsilon}{2(k-1)}, \cdots, \frac{1-\varepsilon}{2(k-1)}, \frac{1+\varepsilon}{2}\right), \\
P_2 &= \left(\frac{1+\varepsilon}{2(k-1)}, \cdots, \frac{1+\varepsilon}{2(k-1)}, \frac{1-\varepsilon}{2}\right), \\
Q &= \left(\frac{1}{U(k-1)}, \cdots \frac{1}{U(k-1)}, 1 - \frac{1}{U}\right). 
\end{align*}
Clearly, for $U\ge 3$ we have $(P_1,Q), (P_2,Q)\in \calM_k(U)$. Under the above construction, the functional value difference between these two points is 
\begin{align*}
|D_\alpha(P_1\|Q) - D_{\alpha}(P_2\|Q)| &= \frac{U^{\alpha-1}}{\alpha(\alpha-1)2^\alpha}\left(1 - \frac{1}{(U-1)^{\alpha-1}}\right)\left((1+\varepsilon)^\alpha - (1-\varepsilon)^\alpha\right) = \Omega_\alpha\left(U^{\alpha-1}\varepsilon\right). 
\end{align*}
Moreover, the KL divergence between the observations is
\begin{align*}
D_{\text{KL}}(P_1^{\otimes m} \| P_2^{\otimes m}) &= mD_{\text{KL}}(P_1 \| P_2) \\
&= m\left( \frac{1-\varepsilon}{2}\log \frac{1-\varepsilon}{1+\varepsilon} + \frac{1+\varepsilon}{2}\log \frac{1+\varepsilon}{1-\varepsilon} \right) \\
&= O(m\varepsilon^2). 
\end{align*}
Consequently, choosing $\varepsilon = 1/(2\sqrt{m})$ in Lemma \ref{lemma:two-points} gives
\begin{align}\label{eq:alpha_var_lower_1}
\inf_{\widehat{D}}\sup_{(P,Q)\in \calM_k(U)} \bE_{(P,Q)} |\widehat{D} - D_\alpha(P\|Q)| \gtrsim_\alpha \frac{U^{\alpha-1}}{\sqrt{m}}. 
\end{align}

Next consider another two-point construction. Fix $\varepsilon\in (0,1/2)$ to be chosen later, let 
\begin{align*}
P &= \left(\frac{1}{2(k-1)}, \cdots, \frac{1}{2(k-1)}, \frac{1}{2}\right), \\
Q_1 &= \left(\frac{1-\varepsilon}{4(k-1)U}, \cdots, \frac{1-\varepsilon}{4(k-1)U}, 1-\frac{1-\varepsilon}{4U}\right), \\
Q_2 &= \left(\frac{1+\varepsilon}{4(k-1)U}, \cdots, \frac{1+\varepsilon}{4(k-1)U}, 1-\frac{1+\varepsilon}{4U}\right). 
\end{align*}
Clearly we have $(P,Q_1), (P,Q_2)\in \calM_k(U)$. Moreover, the functional value difference is
\begin{align*}
&|D_\alpha(P\|Q_1) - D_{\alpha}(P\|Q_2)| \\
&= \frac{1}{\alpha(\alpha-1)}\left|(2U)^{\alpha-1}\left(\frac{1}{(1-\varepsilon)^{\alpha-1}} - \frac{1}{(1+\varepsilon)^{\alpha-1}}  \right) + \frac{1}{2^\alpha}\left(\left(1-\frac{1-\varepsilon}{4U}\right)^{\alpha-1} - \left(1-\frac{1+\varepsilon}{4U}\right)^{\alpha-1} \right) \right| \\
&= \Omega_\alpha(U^{\alpha-1}\varepsilon),
\end{align*}
and the KL divergence between observations is
\begin{align*}
D_{\text{KL}}(Q_1^{\otimes n} \| Q_2^{\otimes n}) &= nD_{\text{KL}}(Q_1 \| Q_2) \\
&= n\left(\frac{1-\varepsilon}{4U}\log\frac{1-\varepsilon}{1+\varepsilon} + \left(1-\frac{1-\varepsilon}{4U}\right)\log \frac{4U-1+\varepsilon}{4U-1-\varepsilon} \right) = O\left(\frac{n\varepsilon^2}{U}\right).
\end{align*}
Hence, choosing $\varepsilon = \sqrt{U/(4n)}$ in this two-point construction gives
\begin{align}\label{eq:alpha_var_lower_2}
\inf_{\widehat{D}}\sup_{(P,Q)\in \calM_k(U)} \bE_{(P,Q)} |\widehat{D} - D_\alpha(P\|Q)| \gtrsim_\alpha \frac{U^{\alpha-1/2}}{\sqrt{n}}. 
\end{align}

Finally we construct the fuzzy hypotheses $\sigma_0$ and $\sigma_1$ in Lemma \ref{lemma:fuzzy}. To this end, we first recall the following duality result between best polynomial approximation and moment matching. 
\begin{lemma}[Appendix E of \cite{wu2016minimax}]\label{lem:measure}
Given a compact interval $I=[a,b]$ with $a>0$, an integer $L>0$ and a continuous function $\phi$ on $I$, let 
\begin{align*}
E_{L}(\phi;I) \triangleq \inf_{\{a_i\}} \sup_{x\in I} \left| \sum_{i=0}^L a_i x^i
-\phi(x)\right|
\end{align*}
denote the best uniform approximation error of $\phi$ by degree-$L$ polynomials. Then
\begin{align*}
2 E_{L}(\phi;I) = \max & ~ \int \phi(t) \nu_1(dt) - \int \phi(t) \nu_0(dt)   \\
\text{\rm s.t.}     & ~ \int t^{l} \nu_1(dt) = \int t^{l} \nu_0(dt), \quad l=0,\ldots,L,
\end{align*}
where the maximum is taken over pairs of probability measures $\nu_0$ and $\nu_1$ supported on $I$.
\end{lemma}
To apply Lemma \ref{lem:measure}, we choose
\begin{align*}
I = \left[\frac{d_0}{n\log n}, \frac{d_1\log n}{n} \right], \qquad L = \lceil d_2\log n \rceil,
\end{align*}
with $d_0,d_1,d_2>0$ be constants specified later. By the proof of the lower bound in Lemma \ref{lemma:approximation} (cf. \eqref{eq:approximation_lower_target}) and proper scaling, for $d_0>0$ small enough (depending only on $\alpha,d_1,d_2$) it holds that
\begin{align*}
E_L(x^{1-\alpha}; I) \gtrsim_{\alpha} (n\log n)^{\alpha-1}. 
\end{align*}
Hence, by Lemma \ref{lem:measure}, there exist probability measures $\nu_0, \nu_1$ supported on $I$ such that they have matching first $L$ moments and $\Delta\triangleq \int x^{1-\alpha}(\nu_1(dx) - \nu_0(dx)) \gtrsim_\alpha (n\log n)^{\alpha-1}$. 

Now consider the following priors $\sigma_0$ and $\sigma_1$ on $(P,Q)$. First, under both priors, the probability vector $P$ is always
\begin{align*}
P = \left(\frac{cU}{n\log n}, \cdots, \frac{cU}{n\log n}, 1 - \frac{c(k-1)U}{n\log n} \right), 
\end{align*}
where $c>0$ is a small constant. Since $n\gtrsim kU^\alpha/\log k$, we have $cUk/(n\log n)<1$ for constant $c>0$ small enough, and $P$ is a probability vector. As for the prior distributions on the probability vector $Q=(q_1,\cdots,q_k)$, the prior $\mu_i$ with $i\in \{0,1\}$ assigns an i.i.d. prior $\nu_i$ to the first $(k-1)$ entries $q_1,\cdots,q_{k-1}$, and $q_k = 1 - (k-1)\bE_{X\sim \nu_0}[X] = 1 - (k-1)\bE_{X\sim \nu_1}[X]$ is a common deterministic scalar. Again, as $X\sim \nu_i$ is at most $d_1\log n/n$, the assumptions $U\gtrsim (\log k)^2$ and $n\gtrsim kU^\alpha/\log k$ ensure that $q_k>0$ for constant $d_1>0$ small enough. However, the above product measures $\mu_0, \mu_1$ may not make $Q$ a probability vector (i.e. sum into one), and thus we consider the following set of approximate probability vectors: 
\begin{align}\label{eq:approximate_probability}
\calM_k(U,\varepsilon) = \left\{(P,Q): P\in \calM_k, Q \in \bR_+^k, \left|\sum_{i=1}^k q_i - 1 \right|\le\varepsilon, p_i\le Uq_i, \forall i\in [k] \right\}. 
\end{align}
Finally, we define the priors $\sigma_i, i\in \{0,1\}$ to be used in Lemma \ref{lemma:fuzzy} to be the pushforward measure of the restriction of $\mu_i, i\in \{0,1\}$ to the following set: 
\begin{align}\label{eq:restriction}
E_i = \calM_k\left(U,\frac{k}{n\log n}\right) \cap \left\{(P,Q): | D_\alpha(P\|Q) - \bE_{(P,Q)\sim \mu_i}[D_{\alpha}(P\|Q)] | \le \frac{(k-1)\Delta}{4}\left(\frac{cU}{n\log n} \right)^{\alpha} \right\}. 
\end{align}

To arrive at the desired minimax lower bound based on the above construction, we define the following quantities. Let $R^\star(m,n,U)$ denote the minimax $L_1$ risk in estimating $D_\alpha(P\|Q)$ over $(P,Q)\in \calM_k(U)$, and $R_{\text{P}}^\star(m,n,U,\varepsilon)$ be the counterpart over $(P,Q)\in \calM_k(U,\varepsilon)$ under the \emph{Poissonized} model one observes the mutually independent histogram $h_i\sim \mathsf{Poi}(nq_i), i\in [k]$ for distribution $Q$ (we do not change the sampling scheme for $P$). The following lemma shows that lower bounds of $R_{\text{P}}^\star(m,n,U,\varepsilon)$ in the Poissonized model translate to those of $R^\star(m,n,U)$ under the original sampling model. 

\begin{lemma}\label{lemma:risk_reduction_alpha}
For $\varepsilon\in (0,1)$, the following inequality holds: 
\begin{align*}
R_{\text{\rm P}}^\star(m,n,U,\varepsilon) \lesssim_\alpha \left(\varepsilon + e^{-n/8} \right)U^{\alpha-1} + R^\star(m,n/2,(1+\varepsilon)U). 
\end{align*}
\end{lemma}
The proof of Lemma \ref{lemma:risk_reduction_alpha} is postponed to Appendix \ref{appendix:auxiliary_proof}. By Lemma \ref{lemma:risk_reduction_alpha}, it is clear that for the choice of $\varepsilon = k/(n\log n)$ in \eqref{eq:restriction}, the target minimax lower bound
\begin{align}\label{eq:alpha_bias_lower}
R^\star(m,n,U) = \inf_{\widehat{D}}\sup_{(P,Q)\in \calM_k(U)} \bE_{(P,Q)} |\widehat{D} - D_\alpha(P\|Q)| \gtrsim_\alpha \frac{kU^\alpha}{n\log n}
\end{align}
follows from $R_{\text{\rm P}}^\star(m,n,U,\varepsilon) \gtrsim kU^\alpha/(n\log n)$. To this end, we apply Lemma \ref{lemma:fuzzy} to the priors $\sigma_0, \sigma_1$ and parameters $\theta = (P,Q), T(\theta) = D_\alpha(P\|Q), \zeta = (\bE_{(P,Q)\sim \mu_0}[D_{\alpha}(P\|Q)] + \bE_{(P,Q)\sim \mu_1}[D_{\alpha}(P\|Q)])/2, s = (k-1)\Delta/4\cdot (cU/n\log n)^\alpha$, and $\Theta = \calM_k(U,\varepsilon)$. By the construction of the measures $\nu_i$ and $\mu_i$, we have
\begin{align*}
\bE_{(P,Q)\sim \mu_1}[D_{\alpha}(P\|Q)] - \bE_{(P,Q)\sim \mu_0}[D_{\alpha}(P\|Q)] = \left(\frac{cU}{n\log n}\right)^\alpha \cdot (k-1)\int x^{1-\alpha}(\nu_1(dx) - \nu_0(dx)) = 4s, 
\end{align*}
and consequently \eqref{eq:restriction} implies that $D_{\alpha}(P\|Q)\le \zeta - s$ almost surely for $(P,Q)\sim \mu_0$, and that $D_{\alpha}(P\|Q) \ge \zeta + s$ almost surely for $(P,Q)\sim \mu_1$. In other words, $\beta_0 = \beta_1 = 0$ in Lemma \ref{lemma:fuzzy}. Hence, it remains to upper bound the total variation distance $\mathsf{TV}(\sigma_0, \sigma_1)$. 

First we show that $\mu_i(E_i^c)$ is small for $i\in \{0,1\}$. In fact, choosing a constant $c>0$ small enough in the probability vector $P$ and using the support of $\nu_i$ shows that $p_i/q_i\le U$ always holds in the support of $\mu_0, \mu_1$. Hence, for $i\in \{0,1\}$, 
\begin{align}\label{eq:triangle_1}
\mu_i\left( \calM_k(U,\varepsilon)^c \right) &= \mu_i \left(\left|\sum_{i=1}^k q_i - 1 \right| > \varepsilon \right) \nonumber \\
&= \nu_i^{\otimes (k-1)}\left(\left|\sum_{i=1}^{k-1} (q_i - \bE_{\nu_i}[q_i] ) \right| > \varepsilon \right) \nonumber\\
&\le \frac{(k-1)\var_{\nu_i}(q)}{\varepsilon^2} \le \frac{(k-1)(n\log n)^2}{k^2}\cdot \left(\frac{d_1\log n}{n}\right)^2 \to 0, 
\end{align}
where we have used the Chebyshev's inequality and the assumption $\log k\gtrsim_\alpha \log n$. Similarly, using the support of $\nu_i$, applying the Chebyshev's inequality again gives
\begin{align}\label{eq:triangle_2}
&\mu_i\left\{(P,Q): | D_\alpha(P\|Q) - \bE_{(P,Q)\sim \mu_i}[D_{\alpha}(P\|Q)] | > \frac{(k-1)\Delta}{4}\left(\frac{cU}{n\log n} \right)^{\alpha} \right\} \nonumber\\
& \le \left[\frac{(k-1)\Delta}{4}\left(\frac{cU}{n\log n} \right)^{\alpha}\right]^{-2}\cdot (k-1)\var_{q\sim \nu_i}\left(\left(\frac{cU}{n\log n}\right)^{\alpha}q^{1-\alpha} \right) \nonumber \\
&\lesssim_\alpha \frac{(n\log n)^2}{(k-1)U^{2\alpha}}\cdot \left(\frac{U}{n\log n}\right)^{2\alpha}\cdot \left(\frac{1}{n\log n}\right)^{2(1-\alpha)}  \to 0.
\end{align}
Consequently, by \eqref{eq:triangle_1} and \eqref{eq:triangle_2}, a union bound gives $\mu_i(E_i^c)\to 0$ as $n\to\infty$. Furthermore, after choosing $d_2>0$ large enough and $d_1>0$ small enough, \cite[Lemma 3]{wu2016minimax} gives
\begin{align*}
\mathsf{TV}\left(\mu_0\circ (P^{\otimes m},Q^{\otimes n})^{-1}, \mu_1\circ (P^{\otimes m},Q^{\otimes n})^{-1} \right) \le (k-1)\mathsf{TV}\left(\bE_{X\sim \nu_0}[\mathsf{Poi}(X)] ,\bE_{X\sim \nu_1}[\mathsf{Poi}(X)] \right) \to 0. 
\end{align*}
Let $\mu_i' = \mu_i\circ (P^{\otimes m},Q^{\otimes n})^{-1}$ for $i\in \{0,1\}$, a triangle inequality gives
\begin{align*}
\mathsf{TV}(\sigma_0, \sigma_1) &\le \mathsf{TV}\left(\sigma_0, \mu_0' \right) + \mathsf{TV}\left(\sigma_1, \mu_1' \right) + \mathsf{TV}\left(\mu_0', \mu_1' \right) \\
&\le \mu_0(E_0^c) + \mu_1(E_1^c) + \mathsf{TV}\left(\mu_0', \mu_1' \right) \to 0, 
\end{align*}
and therefore $\eta \to 1/2$ in Lemma \ref{lemma:fuzzy}. Hence, as $s\gtrsim_\alpha kU^\alpha/(n\log n)$, Lemma \ref{lemma:fuzzy} and \ref{lemma:risk_reduction_alpha} lead to the desired inequality \eqref{eq:alpha_bias_lower}. 

Finally, a combination of \eqref{eq:alpha_var_lower_1}, \eqref{eq:alpha_var_lower_2} and \eqref{eq:alpha_bias_lower} completes the proof of Theorem \ref{thm:alpha_div_lower}. 

\subsection{Lower Bounds for KL Divergence}
The proof of minimax lower bounds for the KL divergence is similar to those of $\alpha$-divergences. In this section we prove the following theorem. 
\begin{theorem}\label{thm.KL_lower}
	If $m\gtrsim k/\log k, n\gtrsim kU/\log k, U\gtrsim (\log k)^2$ and $\log k\gtrsim \log(m+n)$, we have
	\begin{align*}
	\inf_{\widehat{D}}\sup_{(P,Q)\in \calM_k(U)} \bE_{(P,Q)} |\widehat{D} - D_{\text{\rm KL}}(P\|Q)| \gtrsim \frac{k}{m\log m} + \frac{kU}{n\log n} + \frac{\log U}{\sqrt{m}} + \sqrt{\frac{U}{n}}. 
	\end{align*}
\end{theorem}
Clearly Theorem \ref{thm.KL_lower} implies the lower bound of Theorem \ref{thm.KL}. Similarly, we will apply Lemma \ref{lemma:two-points} for the last two terms of Theorem \ref{thm.KL_lower}, and Lemma \ref{lemma:fuzzy} for the first two terms. 

Consider the same two-point constructions in Appendix \ref{subsec:alpha_lower_proof}. In the first construction, 
\begin{align*}
| D_{\text{KL}}(P_1\|Q) - D_{\text{KL}}(P_2\|Q) | = \left|\varepsilon\log \frac{U}{1-U^{-1}} + (1+\varepsilon)\log\frac{1+\varepsilon}{2} - (1-\varepsilon)\log\frac{1-\varepsilon}{2} \right| = \Omega(\varepsilon\log U),
\end{align*}
and Appendix \ref{subsec:alpha_lower_proof} shows that $D_{\text{KL}}(P_1^{\otimes m}\|P_2^{\otimes m})=O(m\varepsilon^2)$. Hence, by choosing $\varepsilon = 1/(2\sqrt{m})$, we arrive at
\begin{align}\label{eq:KL_lower_var_1}
\inf_{\widehat{D}}\sup_{(P,Q)\in \calM_k(U)} \bE_{(P,Q)} |\widehat{D} - D_{\text{\rm KL}}(P\|Q)| \gtrsim \frac{\log U}{\sqrt{m}}. 
\end{align}

In the second construction, we have
\begin{align*}
| D_{\text{KL}}(P\|Q_1) - D_{\text{KL}}(P\|Q_2) | = \left|\frac{1}{2}\log\frac{1+\varepsilon}{1-\varepsilon} - \frac{1}{2}\log\frac{4U-1+\varepsilon}{4U-1-\varepsilon} \right| = \Omega(\varepsilon),
\end{align*}
and Appendix \ref{subsec:alpha_lower_proof} shows that $D_{\text{KL}}(Q_1^{\otimes n}\|Q_2^{\otimes n})=O(n\varepsilon^2/U)$. Hence, choosing $\varepsilon = \sqrt{U/(4n)}$, we have
\begin{align}\label{eq:KL_lower_var_2}
\inf_{\widehat{D}}\sup_{(P,Q)\in \calM_k(U)} \bE_{(P,Q)} |\widehat{D} - D_{\text{\rm KL}}(P\|Q)| \gtrsim \sqrt{\frac{U}{n}}. 
\end{align}

Next we construct the fuzzy hypotheses of the first two terms of Theorem \ref{thm.KL_lower}. The first term $\Omega(k/(n\log n))$ essentially follows from the minimax lower bound of entropy estimation: \cite{wu2016minimax} constructed two priors on $P$ such that $p_i \in [\Omega(1/(n\log n)), O(\log n/n)]$ for all $i\in [k]$ under both priors. Hence, setting identical $q_i = \Theta(1/(n\log n))$ deterministically, the assumption $U\gtrsim (\log n)^2$ implies that the additional likelihood ratio condition is satisfied almost surely under both priors. Furthermore, the estimation of KL divergence reduces to the estimation of the Shannon entropy $H(P) = \sum_{i=1}^k -p_i\log p_i$ of $P$. Consequently, the lower bound in \cite{wu2016minimax} shows that
\begin{align}\label{eq:KL_lower_bias_1}
\inf_{\widehat{D}}\sup_{(P,Q)\in \calM_k(U)} \bE_{(P,Q)} |\widehat{D} - D_{\text{\rm KL}}(P\|Q)| \gtrsim \frac{k}{m\log m}. 
\end{align}

As for the second term, we use a similar construction to that in Section \ref{subsec:alpha_lower_proof}. First recall the following lemma, which is a counterpart of Lemma \ref{lemma:approximation} for $\alpha=1$. 
\begin{lemma}[Lemma 5 of \cite{wu2016minimax}]\label{lemma:approximation_log}
There exist absolute constants $c,c'>0$ such that for $n\in \NN$, 
\begin{align*}
\inf_{a_0,\cdots,a_n} \sup_{x\in [c/n^2, 1]} \left| \log x - \sum_{d=0}^n a_dx^d \right| \ge c'. 
\end{align*}
\end{lemma}
By Lemma \ref{lem:measure}, Lemma \ref{lemma:approximation_log} and a proper scaling, for the parameters
\begin{align*}
I = \left[\frac{d_0}{n\log n}, \frac{d_1\log n}{n} \right], \qquad L = \lceil d_2\log n \rceil,
\end{align*}
with $d_0,d_1,d_2>0$ be constants specified later, there exist two probability measures $\nu_0, \nu_1$ supported on $I$ with matching moments up to order $L$, and $\Delta \triangleq \int \log x(\mu_1(dx) - \mu_0(dx)) = \Omega(1)$. Next we choose te priors $\mu_0, \mu_1$ on $(P,Q)$ as follows. Again, set $P$ to be a deterministic vector
\begin{align*}
P = \left(\frac{cU}{n\log n}, \cdots, \frac{cU}{n\log n}, 1 - \frac{c(k-1)U}{n\log n} \right), 
\end{align*}
with $c>0$ small enough to ensure that $P$ is a valid probability vector. For $Q=(q_1,\cdots,q_k)$, the measure $\mu_i$ assigns $\nu_i$ independently to each coordinate $q_1,\cdots,q_{k-1}$, and sets $q_k = 1 - (k-1)\bE_{X\sim \nu_0}[X] = 1-(k-1)\bE_{X\sim\nu_1}[X]$ for the last coordinate. Similarly, we define the set of approximate probability vectors in \eqref{eq:approximate_probability}, and let the priors $\sigma_i$ be the pushforward measure of the restriction of $\mu_i$ to the following set: 
\begin{align}\label{eq:restriction_KL}
E_i = \calM_k\left(U,\frac{k}{n\log n}\right) \cap \left\{(P,Q): | D_{\text{KL}}(P\|Q) - \bE_{(P,Q)\sim \mu_i}[D_{\text{KL}}(P\|Q)] | \le \frac{(k-1)\Delta}{4}\left(\frac{cU}{n\log n} \right) \right\}. 
\end{align}

Again, let $R^\star(m,n,U)$ and $R_{\text{P}}^\star(m,n,U,\varepsilon)$ be the target minimax risk over $(P,Q)\in \calM_k(U)$ and the Poissonized minimax risk over $(P,Q)\in \calM_k(U,\varepsilon)$, respectively. The relationship between these two quantities is summarized in the following lemma. 
\begin{lemma}\label{lemma:risk_reduction_KL}
	For $\varepsilon\in (0,1)$, the following inequality holds: 
	\begin{align*}
	R_{\text{\rm P}}^\star(m,n,U,\varepsilon) \lesssim_\alpha \left(\varepsilon + e^{-n/8} \right)\log U+ R^\star(m,n/2,(1+\varepsilon)U). 
	\end{align*}
\end{lemma}
Consequently, choosing $\varepsilon = k/(n\log n)$ as in \eqref{eq:restriction_KL}, to show that the desired lower bound
\begin{align}\label{eq:KL_lower_bias_2}
\inf_{\widehat{D}}\sup_{(P,Q)\in \calM_k(U)} \bE_{(P,Q)} |\widehat{D} - D_{\text{\rm KL}}(P\|Q)| \gtrsim \frac{kU}{n\log n}, 
\end{align}
Lemma \ref{lemma:risk_reduction_KL} implies that it suffices to show that $R_{\text{P}}^\star(m,n,U,\varepsilon) = \Omega(kU/(n\log n))$. To this end, similar arguments lead to $\beta_0 = \beta_1 = 0$ in Lemma \ref{lemma:fuzzy} with 
\begin{align*}
\zeta = \frac{ \bE_{(P,Q)\sim \mu_0}[D_{\text{KL}}(P\|Q)] +  \bE_{(P,Q)\sim \mu_1}[D_{\text{KL}}(P\|Q)] }{2}, \quad s = \frac{(k-1)\Delta}{4}\left(\frac{cU}{n\log n} \right)\gtrsim \frac{kU}{n\log n}. 
\end{align*}
Moreover, the same arguments give $\mu_i(\calM_k(U,\varepsilon)^c)\to 0$ for both $i \in \{0,1\}$, and 
\begin{align*}
&\mu_i\left\{(P,Q): | D_{\text{KL}}(P\|Q) - \bE_{(P,Q)\sim \mu_i}[D_{\text{KL}}(P\|Q)] | > \frac{(k-1)\Delta}{4}\left(\frac{cU}{n\log n} \right) \right\} \nonumber\\
& \le \left[\frac{(k-1)\Delta}{4}\left(\frac{cU}{n\log n} \right)\right]^{-2}\cdot (k-1)\var_{q\sim \nu_i}\left( \frac{cU}{n\log n}\log \frac{1}{q} \right) \nonumber \\
&\lesssim \frac{(n\log n)^2}{(k-1)U^{2}}\cdot \left(\frac{U}{n\log n}\right)^{2}\cdot (\log n)^2 \to 0.
\end{align*}
Hence, the union bound gives $\mu_i(E_i^c)\to 0$ for $i\in \{0,1\}$ as $n\to\infty$, and the triangle inequality for the total variation distance gives that $\TV(\sigma_0, \sigma_1) \to 0$ and thus $\eta \to 1/2$. Therefore, \eqref{eq:KL_lower_bias_2} is a direct consequence of Lemma \ref{lemma:fuzzy}, and a combination of \eqref{eq:KL_lower_var_1}, \eqref{eq:KL_lower_var_2}, \eqref{eq:KL_lower_bias_1}, and \eqref{eq:KL_lower_bias_2} completes the proof of Theorem \ref{thm.KL_lower}.

\section{Proof of Main Lemmas}\label{appendix:main_proof}
\subsection{Proof of Lemma \ref{lemma:nonsmooth_bias_variance}}
For the bias bound, recall the quantity $M_{m,d}$ in \eqref{eq:smoothed_moments_true} and define the following quantity: 
\begin{align*}
\widehat{M}_{\alpha,d}^\star =  \sum_{i=1}^k \left(\prod_{\ell=0}^{\alpha-1}\frac{m\widehat{p}_i - \ell}{m-\ell}\cdot \sum_{0\le s\le c_1\log n} \bP\left(\mathsf{HG}\left(n,n\widehat{q}_i, \frac{n}{2}\right) = s\right) g_d(n\widehat{q}_i - s)\right). 
\end{align*}
As similar arguments of \eqref{eq:unbiased_smoothed_moments} lead to $\bE[\widehat{M}_{\alpha,d}^\star] = M_{\alpha,d}$, a triangle inequality gives
\begin{align}\label{eq:alpha_bias_nonsmooth}
|\bE[\widehat{M}_{\alpha,d}] - M_{\alpha,d}^\star| \le |M_{\alpha,d} - M_{\alpha,d}^\star| + | \bE[\widehat{M}_{\alpha,d} - \widehat{M}_{\alpha,d}^\star] |. 
\end{align}
We upper bound the terms of \eqref{eq:alpha_bias_nonsmooth} separately. First, 
\begin{align}\label{eq:alpha_bias_nonsmooth_1}
|M_{\alpha,d} - M_{\alpha,d}^\star| &= \sum_{i=1}^k p_i^\alpha q_i^d\cdot \bP\left(\mathsf{B}\left(\frac{n}{2},q_i\right)\le c_1\log n \right)\1\left(q_i > \frac{3c_1\log n}{n}\right)\nonumber \\
&\stepa{\le} \sum_{i=1}^k U^\alpha q_i^{\alpha+d}\cdot \exp\left(-\Omega\left(\frac{(nq_i - 2c_1\log n)^2}{nq_i}\wedge |nq_i - 2c_1\log n| \right) \right)\1\left(q_i > \frac{3c_1\log n}{n}\right) \nonumber\\
&\stepb{\lesssim}_{\alpha,c_1,c_2} \sum_{i=1}^k U^\alpha \left(\frac{3c_1\log n}{n}\right)^{\alpha+d}\cdot \frac{1}{n^{4\alpha}} = \frac{kU^{\alpha}}{n^{4\alpha}}\left(\frac{3c_1\log n}{n}\right)^{\alpha+d},
\end{align}
where (a) follows from the Chernoff bound, and (b) follows from the substitution $q_i = 3c_1\log n/n \cdot x$ with $x\ge 1$ and $\sup_{x\ge 1} x^{\alpha+d}n^{-c_1(x-1)} = 1$ as long as $c_1\log n\ge \alpha+d$, which is satisfied for all $0\le d\le G=\lceil c_2\log n\rceil$ if we choose $c_1$ much larger than $c_2$. Second, for random variables $X_i\sim \mathsf{B}(n/2,q_i)$, we have
\begin{align}\label{eq:alpha_bias_nonsmooth_2}
|\bE[\widehat{M}_{\alpha,d} - \widehat{M}_{\alpha,d}^\star]| &= \sum_{i=1}^k p_i^\alpha \cdot \bP\left(\mathsf{B}\left(\frac{n}{2},q_i\right)\le c_1\log n \right)\bE[g_d(X_i) - \widetilde{g}_d(X_i) ]\nonumber\\
&\le \sum_{i=1}^k p_i^\alpha \cdot \bP\left(\mathsf{B}\left(\frac{n}{2},q_i\right)\le c_1\log n \right)\bE\left[g_d(X_i)\1(X_i> 2c_1\log n)\right] \nonumber\\
&\stepc{\le} U^\alpha\sum_{i: q_i \le 3c_1\log n/n} q_i^\alpha\cdot \bE[(2X_i/n)^d\1(X_i>2c_1\log n) ] \nonumber \\
&\qquad + U^\alpha \sum_{i: q_i > 3c_1\log n/n} q_i^{\alpha+d}\cdot \bP\left(\mathsf{B}\left(\frac{n}{2},q_i\right)\le c_1\log n \right) \nonumber \\
&\stepd{\lesssim}_{\alpha,c_1,c_2} \frac{kU^{\alpha}}{n^{4\alpha}}\left(\frac{3c_1\log n}{n}\right)^{\alpha+d},
\end{align}
where in (c) we have used that $0\le g_d(x) \le (2x/n)^d$ for $x\in \NN$ and $\bE[g_d(X_i)] = q_i^d$, and (d) follows from the similar tail comparisons as in \eqref{eq:alpha_bias_nonsmooth_1}. Hence, the desired bias bound in Lemma \ref{lemma:nonsmooth_bias_variance} follows from \eqref{eq:alpha_bias_nonsmooth}, \eqref{eq:alpha_bias_nonsmooth_1}, and \eqref{eq:alpha_bias_nonsmooth_2}. 

We apply Lemma \ref{lemma:ES_multinomial} to upper bound the variance, where the function $f_i$ is
\begin{align*}
f_i(\widehat{p}_i, \widehat{q}_i) \equiv f(\widehat{p}_i, \widehat{q}_i) \triangleq u(\widehat{p}_i)v(\widehat{q}_i) \triangleq \prod_{\ell=0}^{\alpha-1}\frac{m\widehat{p}_i - \ell}{m-\ell}\cdot \sum_{0\le s\le c_1\log n} \bP\left(\mathsf{HG}\left(n,n\widehat{q}_i, \frac{n}{2}\right) = s\right) \widetilde{g}_d(n\widehat{q}_i - s). 
\end{align*}
The following lemma summarizes some properties of functions $u$ and $v$. 
\begin{lemma}\label{lemma:uv}
Let $m\widehat{p}\sim \mathsf{B}(n,p)$ and $n\widehat{q}\sim \mathsf{B}(n,q)$, then the following inequalities hold: 
\begin{align*}
\bE[u(\widehat{p})^2] &\lesssim_\alpha p^{2\alpha},\qquad \bE\left[\widehat{p}\left(u(\widehat{p}) - u\left(\widehat{p} - \frac{1}{m} \right) \right)^2 \right] \lesssim_{\alpha} \frac{p^{2\alpha-1}}{m^2}, \\
\bE[v(\widehat{q})^2] &\lesssim  \left(\frac{4c_1\log n}{n}\right)^{2d}\exp\left(-c(nq - 3c_1\log n)_+ \right), \\
\bE\left[\widehat{q}\left(v(\widehat{q}) - v\left(\widehat{q} - \frac{1}{n} \right) \right)^2 \right] &\lesssim \left(\frac{4c_1\log n}{n}\right)^{2d+1}\exp\left(-c(nq - 3c_1\log n)_+ \right), 
\end{align*}
where $c>0$ is an absolute constant. 
\end{lemma}
Based on Lemma \ref{lemma:ES_multinomial} and Lemma \ref{lemma:uv}, we conclude that 
\begin{align*}
\var(\widehat{M}_{\alpha,d}) &\lesssim m\sum_{i=1}^k \frac{p_i^{2\alpha-1}}{m^2}\left(\frac{4c_1\log n}{n}\right)^{2d}\exp\left(-c(nq_i - 3c_1\log n)_+ \right) \\
&\qquad + n\sum_{i=1}^k p_i^{2\alpha} \cdot \left(\frac{4c_1\log n}{n}\right)^{2d+1}\exp\left(-c(nq_i - 3c_1\log n)_+ \right). 
\end{align*}
Hence, with a large constant $c_1>0$, the total contribution of symbols $i\in [k]$ with $q_i > 4c_1\log n/n$ to the above variance is at most $kn^{-5\alpha}\cdot (4c_1\log n/n)^{2d}$, which is negligible compared to the claimed result. Restricting to symbols with $q_i \le 4c_1\log n/n$, using $p_i\le Uq_i$ and $d\le c_2\log n+1$ in the above variance bound gives
\begin{align*}
\var(\widehat{M}_{m,d}) \lesssim_{\alpha,c_1} kU^{2\alpha}\log n\cdot \left(\frac{4c_1\log n}{n}\right)^{2(d+\alpha)} \lesssim_{\alpha,c_1,c_2} kU^{2\alpha}\left(\frac{3c_1\log n}{n}\right)^{2(d+\alpha)}\cdot n^{O(c_2)}, 
\end{align*}
as claimed. 

\subsection{Proof of Lemma \ref{lemma:smooth_bias}}
First, we recall from the Chernoff bound (cf. Lemma \ref{lemma.localization}) that if $X\sim \mathsf{B}(n/2,q)$ and $q\ge c_1\log n/n$, for constant $c_1>0$ large enough we have
\begin{align}\label{eq:concentration}
\bP\left(\frac{nq}{4} \le X \le nq \right) \ge 1 - \frac{1}{n^{5\alpha}}. 
\end{align}
Consequently, as $\|h_\alpha\|_\infty \lesssim_\alpha n^{\alpha-1}$, we have
\begin{align*}
\left|\bE[h_\alpha(X)] - \bE\left[h_\alpha(X)\1\left(\frac{nq}{4}\le X\le nq\right) \right] \right| \lesssim \frac{1}{n^{4\alpha+1}}. 
\end{align*}

For $nq/4 \le X\le nq$, the Taylor expansion of $x^{1-\alpha}$ around $x=q$ gives
\begin{align*}
\frac{1}{(2X/n)^{\alpha-1}} &= \frac{1}{q^{\alpha-1}} - \frac{\alpha-1}{q^\alpha}\left(\frac{2X}{n} - q \right) + \frac{\alpha(\alpha-1)}{2q^{\alpha+1}}\left(\frac{2X}{n} - q \right)^2 \\
&\qquad - \frac{\alpha(\alpha-1)(\alpha-2)}{6q^{\alpha+2}}\left(\frac{2X}{n} - q\right)^3 + \frac{\alpha(\alpha-1)(\alpha-2)(\alpha-3)}{24\xi^{\alpha+3}}\left(\frac{2X}{n}-q\right)^4,
\end{align*}
with $\xi\in [q/2,2q]$. Using the central moments of the Binomial distribution
\begin{equation}\label{eq:binomial_central_moments}
\begin{split}
&\bE\left[\left(\frac{2X}{n} - q \right) \right] =0, \qquad\bE\left[\left(\frac{2X}{n} - q \right)^2 \right] = \frac{2q(1-q)}{n}, \\
&\bE\left[\left(\frac{2X}{n} - q \right)^3 \right] \lesssim \frac{q}{n^2}, \qquad \bE\left[\left(\frac{2X}{n} - q \right)^4 \right] \lesssim \frac{q^2}{n^2} + \frac{q}{n^3}, 
\end{split}
\end{equation}
taking expectation at both sides of the Taylor expansion together with \eqref{eq:concentration} gives
\begin{align}\label{eq:bias_bound_taylor_1}
&\left| \bE\left[ \left(\frac{1}{(2X/n)^{\alpha-1}} - \frac{1}{q^{\alpha-1}} - \frac{\alpha(\alpha-1)(1-q)}{nq^\alpha}\right) \1\left(\frac{nq}{4}\le X\le nq\right) \right] \right| \nonumber \\
&\lesssim_\alpha \frac{1}{q^{\alpha+2}}\cdot \frac{q}{n^2} + \frac{1}{q^{\alpha+3}}\cdot \left(\frac{q^2}{n^2} + \frac{q}{n^3}\right) + \frac{1}{n^{3\alpha}} \nonumber \\
&\lesssim_\alpha \frac{1}{n^2q^{\alpha+1}} + \frac{1}{n^3q^\alpha} + \frac{1}{n^{4\alpha}} \lesssim_{\alpha,c_1} \frac{1}{nq^\alpha \log n}, 
\end{align}
where in the last inequality we have used the assumption $q\ge c_1\log n/n$. Hence, it now remains to show that the bias-correction term $(1-(2X/n))/(2X/n)^\alpha$ in the estimator $h_\alpha$ is a good estimate of the target quantity $(1-q)/q^\alpha$. To this end, the Taylor expansion of $(1-x)/x^\alpha$ around $x=q$ again gives
\begin{align*}
\frac{1 - (2X/n)}{(2X/n)^\alpha} = \frac{1-q}{q^\alpha} - \frac{\alpha - (\alpha-1)q}{q^{\alpha+1}}\left(\frac{2X}{n}-q\right) + \frac{\alpha(\alpha+1) - \alpha(\alpha-1)\xi}{2\xi^{\alpha+2}}\left(\frac{2X}{n}-q\right)^2, 
\end{align*}
with $\xi \in [q/2,2q]$ for $nq/4\le X\le nq$. Using the central moments \eqref{eq:binomial_central_moments} again, we conclude that
\begin{align}\label{eq:bias_bound_taylor_2}
\left| \bE\left[\left( \frac{1 - (2X/n)}{(2X/n)^\alpha} - \frac{1-q}{q^\alpha}  \right)\1\left(\frac{nq}{4}\le X\le nq\right) \right] \right| \lesssim_\alpha \frac{1}{n^{3\alpha}} + \frac{1}{nq^{\alpha+1}} \lesssim_{\alpha,c_1} \frac{1}{q^\alpha\log n}. 
\end{align}
Hence, the desired bias bound follows from \eqref{eq:bias_bound_taylor_1}, \eqref{eq:bias_bound_taylor_2} and the concentration inequality \eqref{eq:concentration}. 

\subsection{Proof of Lemma \ref{lemma:smooth_variance}}
We use Lemma \ref{lemma:ES_multinomial} to upper bound the variance of $\widehat{D}_{\alpha,\text{s}}$, where we deal with the perturbations of $X^m\sim P$ and $Y^n\sim Q$, respectively. Specifically, the function $f_i$ in Lemma \ref{lemma:ES_multinomial} is
\begin{align*}
f_i(\widehat{p}_i, \widehat{q}_i) \equiv f(\widehat{p}_i, \widehat{q}_i) \triangleq u(\widehat{p}_i)w(\widehat{q}_i) \triangleq \prod_{\ell=0}^{\alpha-1}\frac{m\widehat{p}_i - \ell}{m-\ell}\cdot \sum_{s>c_1\log n}\bP\left(\mathsf{HG}\left(n,n\widehat{q}_i, \frac{n}{2}\right) = s\right) h_\alpha(n\widehat{q}_i - s). 
\end{align*}
The properties of $u$ are summarized in Lemma \ref{lemma:uv}, so it remains to consider the properties of the function $w$. First, for the second moment of $w(\widehat{q})$ with $n\widehat{q}\sim \mathsf{B}(n,q)$, Cauchy--Schwartz inequality gives
\begin{align*}
\bE[w(\widehat{q})^2] &\le \bE\left[ \sum_{s>c_1\log n}\bP\left(\mathsf{HG}\left(n,n\widehat{q}_i, \frac{n}{2}\right) = s\right) h_\alpha(n\widehat{q}_i - s)^2 \right] \\
&= \bP\left(\mathsf{B}\left(\frac{n}{2},q\right) > c_1\log n\right)\cdot \bE[h_\alpha(X)^2],
\end{align*}
with $X\sim \mathsf{B}(n/2,q)$, where the last identity uses the unbiased/sample splitting property in \eqref{eq:unbiased_smoothed_moments}. The following lemma gives an upper bound of the second moment of $h_\alpha(X)$. 
\begin{lemma}\label{lemma:h_alpha_moment}
For $q\ge c_1\log n/n$ with constant $c_1>0$ large enough, it holds that
\begin{align*}
\bE[h_\alpha(X)^2] \lesssim_{\alpha,c_1} \frac{1}{q^{2(\alpha-1)}}. 
\end{align*}
\end{lemma}
Consequently, by Lemma \ref{lemma.localization} and Lemma \ref{lemma:h_alpha_moment}, we conclude that
\begin{align*}
\bE[w(\widehat{q})^2] \lesssim_{\alpha,c_1} \frac{1}{n^{3\alpha}}\cdot \1\left(q< \frac{c_1\log n}{n}\right) + \frac{1}{q^{2(\alpha-1)}}\cdot \1\left(q\ge \frac{c_1\log n}{n}\right). 
\end{align*}
Hence, the variance upper bound $S_m$ corresponding to the perturbation in $X^m\sim P$ in Lemma \ref{lemma:ES_multinomial} is 
\begin{align}\label{eq:variance_upper_1}
S_m &\triangleq 2m\cdot \sum_{i=1}^k \bE\left[\widehat{p}_i\left(u(\widehat{p}_i) - u\left(\widehat{p}_i - \frac{1}{m}\right) \right)^2 \right]\cdot \bE[w(\widehat{q}_i)^2] \nonumber \\
&\lesssim_{\alpha,c_1} \sum_{i\in [k]: q_i < c_1\log n/n}\frac{p_i^{2\alpha-1}}{m}\cdot \frac{1}{n^{3\alpha}} + \sum_{i\in [k]: q_i \ge c_1\log n/n}\frac{p_i^{2\alpha-1}}{m}\cdot \frac{1}{q_i^{2(\alpha-1)}} \nonumber \\
&\lesssim_{\alpha,c_1} \sum_{i=1}^k \frac{U^{2(\alpha-1)}p_i}{m} = \frac{U^{2(\alpha-1)}}{m},
\end{align}
where in the last inequality we have used that $p_i\le Uq_i$ for all $i\in [k]$. 

Next we deal with the perturbation in $Y^n\sim Q$, where a central quantity is to upper bound the difference $w(\widehat{q}) - w(\widehat{q} - 1/n)$. To do so, we distinguish into three cases: 
\begin{itemize}
	\item Case I: $q\le c_1\log n/n$. In this case, the above upper bound on the second moment of $w(\widehat{q})$ gives
	\begin{align*}
	\bE\left[\widehat{q}\left(w(\widehat{q}) - w\left(\widehat{q} - \frac{1}{n} \right) \right)^2 \right] \le 2\bE[w(\widehat{q})^2] + 2\bE[w(\widehat{q} - 1/n)^2] \lesssim_{\alpha,c_1} \frac{1}{n^{3\alpha}}. 
	\end{align*}
	Consequently, by Lemma \ref{lemma:uv} we have
	\begin{align}\label{eq:variance_upper_2_small}
	&2n\cdot \sum_{i\in [k]: q_i \le {c_1\log n}/{n}} \bE[u(\widehat{p}_i)^2]\cdot \bE\left[\widehat{q}_i\left(w(\widehat{q}_i) - w\left(\widehat{q}_i - \frac{1}{n} \right) \right)^2 \right] \nonumber \\
	&\lesssim_{\alpha,c_1} \frac{1}{n^{3\alpha-1}}\sum_{i\in [k]: q_i \le {c_1\log n}/{n}} p_i^{2\alpha} \nonumber \\
	&\lesssim_{\alpha,c_1} \frac{1}{n^{3\alpha-1}}\sum_{i\in [k]: q_i \le {c_1\log n}/{n}} p_i\left(\frac{U\log n}{n}\right)^{2\alpha-1} \lesssim_{\alpha,c_1} \frac{U^{2\alpha-1}}{n}. 
	\end{align}
	\item Case II: $c_1\log n/n < q \le 3c_1\log n/n$. In this case, it is clear that $| w(\widehat{q}) | \le \|h_\alpha\|_\infty \lesssim_{\alpha} n^{\alpha-1}$. Hence, 
		\begin{align*}
	\bE\left[\widehat{q}\left(w(\widehat{q}) - w\left(\widehat{q} - \frac{1}{n} \right) \right)^2 \right] \lesssim_{\alpha} n^{2(\alpha-1)}\cdot \bE[\widehat{q}] = n^{2(\alpha-1)}q. 
	\end{align*}
	Consequently, it holds that
	\begin{align}\label{eq:variance_upper_2_medium}
&2n\cdot \sum_{i\in [k]: {c_1\log n}/{n} < q_i \le 3c_1\log n/n} \bE[u(\widehat{p}_i)^2]\cdot \bE\left[\widehat{q}_i\left(w(\widehat{q}_i) - w\left(\widehat{q}_i - \frac{1}{n} \right) \right)^2 \right] \nonumber \\
&\lesssim_{\alpha,c_1} n^{2\alpha-1}\sum_{i\in [k]: q_i \le {3c_1\log n}/{n}} p_i^{2\alpha}q_i \nonumber \\
&\lesssim_{\alpha,c_1} n^{2\alpha-1}\sum_{i\in [k]} U^{2\alpha}\left(\frac{3c_1\log n}{n}\right)^{2\alpha+1} \lesssim_{\alpha,c_1,\varepsilon} \frac{kU^{2\alpha}}{n^{2-\varepsilon}}
\end{align}
for any $\varepsilon>0$. 
\item Case III: $q>3c_1\log n/n$. We claim that in this case $w(\widehat{q})$ is close to the following function $w^\star(\widehat{q})$ under the $L_2$ norm: 
\begin{align*}
w^\star(\widehat{q}) \triangleq  \sum_{s\ge 0}\bP\left(\mathsf{HG}\left(n,n\widehat{q}, \frac{n}{2}\right) = s\right) h_\alpha(n\widehat{q} - s). 
\end{align*}
In fact, the Cauchy--Schwartz inequality gives
\begin{align*}
\bE[(w(\widehat{q}) - w^\star(\widehat{q}))^2 ] &= \bE\left[\left(\sum_{s\le c_1\log n}\bP\left(\mathsf{HG}\left(n,n\widehat{q}, \frac{n}{2}\right) = s\right) h_\alpha(n\widehat{q} - s) \right)^2 \right] \\
&\le \bE\left[\sum_{s\le c_1\log n}\bP\left(\mathsf{HG}\left(n,n\widehat{q}, \frac{n}{2}\right) = s\right) h_\alpha(n\widehat{q} - s)^2 \right] \\
&\le \bP\left(\mathsf{B}\left(\frac{n}{2},q\right)\le c_1\log n \right)\cdot \|h_\alpha\|_\infty^2 \lesssim_{c_1,\alpha} \frac{1}{n^{3\alpha}},
\end{align*}
where in the last inequality we have used Lemma \ref{lemma.localization}. Hence, by triangle inequality it suffices to work with the function $w^\star(\widehat{q})$. To this end, note that an alternative definition of $w^\star(\widehat{q})$ is $w^\star(\widehat{q}) = \bE[h_\alpha(n\widehat{q} - Y)]$ with $Y\sim \mathsf{HG}(n,n\widehat{q},n/2)$. Similarly, $w^\star(\widehat{q} - 1/n) = \bE[h_\alpha(n\widehat{q} - 1 - Z)]$ with $Z\sim \mathsf{HG}(n,n\widehat{q}-1,n/2)$. By definition of the hypergeometric distribution, it is clear that there exists a coupling between random variables $Y$ and $Z$ such that $Y-1\le Z\le Y$ almost surely, and consequently
\begin{align*}
\left| w^\star(\widehat{q}) - w^\star\left(\widehat{q} - \frac{1}{n} \right) \right| \le \bE \left|h_\alpha(n\widehat{q} - Y) - h_\alpha(n\widehat{q} - 1 - Z) \right| \le \bE [\Delta h_\alpha(n\widehat{q} - Y)],
\end{align*}
with $\Delta h_\alpha(x) \triangleq |h_\alpha(x) - h_\alpha(x-1)|$. The key property of the function $\Delta h_\alpha$ is summarized in the following lemma. 
\begin{lemma}\label{lemma:delta_h_alpha_moment}
For $X\sim \mathsf{B}(n/2,q)$ and $q\ge 3c_1\log n/n$ with constant $c_1>0$ large enough, it holds that
\begin{align*}
\bE[\Delta h_\alpha(X)^2] \lesssim_{\alpha,c_1} \frac{1}{n^2q^{2\alpha}}, \qquad \bE[X\Delta h_\alpha(X)^2] \lesssim_{\alpha,c_1} \frac{1}{nq^{2\alpha-1}}. 
\end{align*}
\end{lemma}
Based on Lemma \ref{lemma:delta_h_alpha_moment}, we have the following chain of inequalities: 
\begin{align*}
&\bE\left[\widehat{q}\left(w^\star (\widehat{q}) - w^\star \left(\widehat{q} - \frac{1}{n} \right) \right)^2 \right] \\
&\stepa{\le} \bE\left[\widehat{q}\cdot \sum_{s\ge 0} \bP\left(\mathsf{HG}\left(n,n\widehat{q}, \frac{n}{2}\right) = s\right) \Delta h_\alpha(n\widehat{q} - s)^2 \right] \\
&= \bE\left[\sum_{s\ge 0} \frac{s}{n}\bP\left(\mathsf{HG}\left(n,n\widehat{q}, \frac{n}{2}\right) = s\right) \cdot \Delta h_\alpha(n\widehat{q} - s)^2 \right] \\
&\qquad + \bE\left[\sum_{s\ge 0} \left(\widehat{q}-\frac{s}{n}\right)\bP\left(\mathsf{HG}\left(n,n\widehat{q}, \frac{n}{2}\right) = s\right) \cdot \Delta h_\alpha(n\widehat{q} - s)^2 \right] \\
&\stepb{=} \bE\left[\frac{X_1}{n} \right]\cdot \bE[\Delta h_\alpha(X_2)^2] + \frac{ \bE [X_2\Delta h_\alpha(X_2)^2]}{n} \\
&\stepc{\lesssim}_{\alpha,c_1} \frac{1}{n^2q^{2\alpha-1}}, 
\end{align*}
where (a) is again due to the Cauchy--Schwartz inequality applied to $\bE^2[\Delta h_\alpha(n\widehat{q}-Y)]$, (b) is due to the unbiased property similar to \eqref{eq:unbiased_smoothed_moments} with i.i.d. random variables $X_1, X_2 \sim \mathsf{B}(n/2,q)$, and (c) follows from Lemma \ref{lemma:delta_h_alpha_moment}. In summary, it holds that
	\begin{align}\label{eq:variance_upper_2_large}
&2n\cdot \sum_{i\in [k]: q_i > 3c_1\log n/n} \bE[u(\widehat{p}_i)^2]\cdot \bE\left[\widehat{q}_i\left(w(\widehat{q}_i) - w\left(\widehat{q}_i - \frac{1}{n} \right) \right)^2 \right] \nonumber \\
&\lesssim_{\alpha,c_1} \sum_{i\in [k]: q_i > {3c_1\log n}/{n}} \left(\frac{p_i^{2\alpha}}{n^{3\alpha-1}} + \frac{p_i^{2\alpha}}{nq_i^{2\alpha-1}} \right) \nonumber \\
&\lesssim_{\alpha,c_1} \sum_{i=1}^k \left(\frac{p_i}{n^{3\alpha-1}} + \frac{U^{2\alpha-1}p_i}{n} \right) \lesssim_{\alpha,c_1} \frac{U^{2\alpha-1}}{n}. 
\end{align}
\end{itemize}

Combining the above scenarios and inequalities \eqref{eq:variance_upper_2_small}, \eqref{eq:variance_upper_2_medium}, \eqref{eq:variance_upper_2_large}, the variance upper bound $S_n$ corresponding to the perturbation in $Y^n\sim Q$ in Lemma \ref{lemma:ES_multinomial} is 
\begin{align}\label{eq:variance_upper_2}
S_n \triangleq 2n\cdot \sum_{i=1}^k \bE[u(\widehat{p}_i)^2]\cdot \bE\left[\widehat{q}_i\left(w(\widehat{q}_i) - w\left(\widehat{q}_i - \frac{1}{n} \right) \right)^2 \right] \lesssim_{\alpha,c_1,\varepsilon} \frac{U^{2\alpha-1}}{n} + \frac{kU^{2\alpha}}{n^{2-\varepsilon}}. 
\end{align}
Therefore, a combination of \eqref{eq:variance_upper_1}, \eqref{eq:variance_upper_2} and Lemma \ref{lemma:ES_multinomial} completes the proof of Lemma \ref{lemma:smooth_variance}. 

\subsection{Proof of Lemma \ref{lemma:nonsmooth_KL}}
We shall only prove the statements on $\widehat{M}_d^{(1)}$, as the analysis is entirely similar for $\widehat{M}_d^{(2)}$. Adopting the same idea of the proof of Lemma \ref{lemma:nonsmooth_bias_variance}, the bias can be expressed as
\begin{align}\label{eq:triangle_bias}
|\bE[\widehat{M}_{d}^{(1)}] - M_{1,d}^\star | &\le \underbrace{\sum_{i=1}^k p_iq_i^d\cdot \bP\left(\mathsf{B}\left(\frac{n}{2},q_i\right) \le c_1\log n\right)\cdot \1\left(q_i > \frac{3c_1\log n}{n}\right)}_{\triangleq A_1} \nonumber \\
&\qquad + \underbrace{\sum_{i=1}^k p_i\bP\left(\mathsf{B}\left(\frac{n}{2},q_i\right)\le c_1\log n\right)\cdot \bE|g_{d}(X_i) - \widetilde{g}_d(X_i) | }_{\triangleq A_2}
\end{align}
for $X_i \sim \mathsf{B}(n/2,q_i)$. By the Chernoff bound and tail comparison, for constant $c_1>0$ large enough and $c_2>0$ small enough we have
\begin{align*}
\sup_{ q \ge 3c_1\log n/n} q^d\cdot \bP\left(\mathsf{B}\left(\frac{n}{2},q\right) \le c_1\log n\right) \lesssim_{c_1,c_2} \frac{1}{n^5}\left(\frac{3c_1\log n}{n}\right)^d. 
\end{align*}
Consequently, we have
\begin{align}\label{eq:A_1}
A_1 \lesssim_{c_1,c_2} \frac{1}{n^5}\left(\frac{3c_1\log n}{n}\right)^d\sum_{i=1}^k p_i = \frac{1}{n^5}\left(\frac{3c_1\log n}{n}\right)^d. 
\end{align}

As for $A_2$, by definition of the modification $\widetilde{g}_d$ in \eqref{eq:moment_est_trunc}, we have
\begin{align*}
\bE|g_{d}(X_i) - \widetilde{g}_d(X_i) | &\le \bE\left[ g_d(X_i) \1(X_i \ge 2c_1\log n) \right] \\
&\lesssim_{c_1,c_2} \frac{1}{n^5}\left(\frac{3c_1\log n}{n}\right)^d\cdot \1\left(q_i \le \frac{3c_1\log n}{n}\right) + q_i^d\cdot \1\left(q_i > \frac{3c_1\log n}{n}\right), 
\end{align*}
where the second inequality follows from similar tail comparison and $g_d(x) \le (2x/n)^d$. Hence, 
\begin{align}\label{eq:A_2}
A_2 \lesssim_{c_1,c_2} A_1 + \frac{1}{n^5}\left(\frac{3c_1\log n}{n}\right)^d\sum_{i=1}^k p_i  \lesssim_{c_1,c_2} \frac{1}{n^5}\left(\frac{3c_1\log n}{n}\right)^d. 
\end{align}
Therefore, the claimed bias bound follows from a combination of \eqref{eq:triangle_bias}, \eqref{eq:A_1}, and \eqref{eq:A_2}. 

As for the variance of $\widehat{M}_d^{(1)}$, we apply Lemma \ref{lemma:ES_multinomial} to the function
\begin{align*}
f_i(\widehat{p}_i,\widehat{q}_i) \equiv f(\widehat{p}_i,\widehat{q}_i) \triangleq \widehat{p}_i v(\widehat{q}_i) \triangleq \widehat{p}_i\cdot \sum_{s\le c_1\log n}\bP\left(\mathsf{HG}\left(n,n\widehat{q}_i,\frac{n}{2}\right) = s \right)\widetilde{g}_d(n\widehat{q}_i-s). 
\end{align*}
Since Lemma \ref{lemma:uv} summarizes some properties of the function $v$, Lemma \ref{lemma:ES_multinomial} leads to
\begin{align*}
\var(\widehat{M}_d^{(1)}) &\le 2m \cdot \sum_{i=1}^k \bE\left[\frac{\widehat{p}_i}{m^2}\right]\bE\left[v(\widehat{q}_i)^2 \right] + 2n\cdot \sum_{i=1}^k \bE[\widehat{p}_i^2]\cdot \bE\left[\widehat{q}_i\left(v(\widehat{q}_i) - v\left(\widehat{q}_i - \frac{1}{n}\right) \right)^2 \right] \\
&\lesssim_{c_1,c_2} \frac{1}{m}\sum_{i=1}^k p_i\left(\frac{4c_1\log n}{n}\right)^{2d}\exp(-c(nq_i - 3c_1\log n)_+) \\
&\qquad + \sum_{i=1}^k \left(p_i^2 + \frac{p_i}{m}\right)\cdot n\left(\frac{4c_1\log n}{n}\right)^{2d+1}\exp(-c(nq_i - 3c_1\log n)_+). 
\end{align*}
Using $d\le \lceil c_2\log n\rceil$ and distinguishing into two cases $q_i \le 4c_1\log n/n$ and $q_i>4c_1\log n/n$, we conclude that
\begin{align*}
\var(\widehat{M}_d^{(1)}) &\lesssim_{c_1,c_2} n^{O(c_2)}\cdot \left( \frac{kU}{mn} + \frac{kU^2}{n^2} \right)\left(\frac{3c_1\log n}{n}\right)^{2d} \\
&\lesssim_{c_1,c_2} n^{O(c_2)}\left(\frac{\sqrt{k}}{m} + \frac{\sqrt{k}U}{n} \right)^2\left(\frac{3c_1\log n}{n}\right)^{2d},
\end{align*}
establishing the desired variance bound. 

\subsection{Proof of Lemma \ref{lemma:smooth_KL_bias}}
As the second inequality essentially follows from the analysis of \cite[Lemma 3]{Jiao--Venkat--Han--Weissman2015minimax}, we solely focus on the first inequality. Recall that the concentration inequality \eqref{eq:concentration} gives $nq/4\le X\le nq$ with probability at least $1 - n^{-5}$, we may primarily work on the regime $X\in [nq/4, nq]$. In this regime, the Taylor expansion of $-\log x$ around $x=q$ gives that
\begin{align*}
-\log\frac{2X}{n} = -\log q - \frac{1}{q}\left(\frac{2X}{n}-q\right) + \frac{1}{2q^2}\left(\frac{2X}{n}-q\right)^2 - \frac{1}{3q^3}\left(\frac{2X}{n}-q\right)^3 + \frac{1}{4\xi^4}\left(\frac{2X}{n}-q\right)^4, 
\end{align*}
with $\xi\in [q/2,2q]$ if $nq/4\le X\le nq$. Using the central moments of Binomial random variable in \eqref{eq:binomial_central_moments}, we have
\begin{align}\label{eq:Taylor_1}
\left|\bE\left[\left(-\log \frac{2X}{n} - \frac{1-q}{nq} + \log q\right)\1\left(\frac{nq}{4}\le X\le nq\right) \right] \right| \lesssim_{c_1} \frac{1}{n^2q^2} + \frac{1}{n^3q^3} + \frac{1}{n^5} \lesssim_{c_1} \frac{1}{nq\log n}, 
\end{align}
where the last step is thanks to the assumption $q\ge c_1\log n/n$. As for the bias correction term, the Taylor expansion of $(1-x)/(nx)$ around $x=q$ gives
\begin{align*}
\frac{1-2X/n}{2X} = \frac{1-q}{nq} - \frac{1}{nq^2}\left(\frac{2X}{n} - q\right) + \frac{1}{n\xi^3}\left(\frac{2X}{n} - q\right)^2
\end{align*}
for some $\xi \in [q/2,2q]$ if $nq/4\le X\le nq$. Hence, using the central moments \eqref{eq:binomial_central_moments} again yields to
\begin{align}\label{eq:Taylor_2}
\left|\bE\left[\left(\log \frac{2X}{n} - \frac{1-q}{nq}\right)\1\left(\frac{nq}{4}\le X\le nq\right) \right] \right| \lesssim_{c_1} \frac{1}{n^2q^2} + \frac{1}{n^5}\lesssim_{c_1} \frac{1}{nq\log n}.  
\end{align}
Now the desired bias upper bound follows from the inequalities \eqref{eq:Taylor_1}, \eqref{eq:Taylor_2}, and the concentration bound \eqref{eq:concentration}. 

\subsection{Proof of Lemma \ref{lemma:smooth_KL_variance}}
By the estimator construction for the KL divergence, it is clear that $\widehat{D}_{\text{KL}, \text{s}}^{(1)} + \widehat{D}_{\text{KL}, \text{s}}^{(2)} = \sum_{i=1}^k f(\widehat{p}_i,\widehat{q}_i)$, with the bivariate function $f$ given by
\begin{align*}
f(\widehat{p},\widehat{q}) &= \widehat{p}\sum_{s>c_1\log n}\bP\left(\mathsf{HG}\left(n,n\widehat{q},\frac{n}{2}\right) = s \right)h^{(1)}(n\widehat{q}-s) \\
&\qquad + \sum_{s>c_1\log m} \bP\left(\mathsf{HG}\left(m,m\widehat{p},\frac{m}{2}\right) = s \right)h^{(2)}(m\widehat{p}-s). 
\end{align*}
For notational simplicity, we write $f(\widehat{p},\widehat{q}) = \widehat{p}\cdot u(\widehat{q}) + v(\widehat{p})$. We begin with some properties of the functions $u$ and $v$ which are summarized in the following lemma. 
\begin{lemma}\label{lemma:uv_new}
For $m\widehat{p}\sim\mathsf{B}(m,p)$, $n\widehat{q}\sim\mathsf{B}(n,q)$, the following inequalities hold: 
\begin{align*}
\bE[u(\widehat{q})^2] &\lesssim_{c_1} [1 + (\log q)^2]\cdot \1\left(q>\frac{3c_1\log n}{n} \right) + (\log n)^2\cdot \1\left(q\le \frac{3c_1\log n}{n} \right), \\
\bE\left[\widehat{p}\left(v(\widehat{p}) - v\left(\widehat{p} - \frac{1}{m}\right) \right)^2 \right] &\lesssim_{c_1} \frac{p[1+(\log p)^2]}{m^2}\cdot \1\left(p>\frac{3c_1\log m}{m}\right) + \frac{(\log m)^5}{m^3}\cdot \1\left(p\le \frac{3c_1\log m}{m} \right), \\
\bE\left[\widehat{q}\left(u(\widehat{q}) - u\left(\widehat{q} - \frac{1}{n}\right) \right)^2 \right] &\lesssim_{c_1} \frac{1}{n^2q}\cdot \1\left(q>\frac{3c_1\log n}{n}\right) + \frac{(\log n)^3}{n}\cdot \1\left(q\le \frac{3c_1\log n}{n} \right). 
\end{align*}
\end{lemma}

Based on Lemma \ref{lemma:uv_new}, the perturbation of $\widehat{q}$ in Lemma \ref{lemma:ES_multinomial} can be directly controlled. To see this, note that
\begin{align*}
f(\widehat{p}, \widehat{q}) - f\left(\widehat{p}, \widehat{q} - \frac{1}{n}\right) = \widehat{p}\left(u(\widehat{q}) - u\left(\widehat{q} - \frac{1}{n}\right) \right). 
\end{align*}
Hence, by Lemma \ref{lemma:uv_new}, 
\begin{align}\label{eq:perturbation_q}
& 2n\cdot \sum_{i=1}^k \bE\left[ \widehat{q}_i \left( f(\widehat{p}_i, \widehat{q}_i) - f\left(\widehat{p}_i, \widehat{q}_i - \frac{1}{n}\right) \right)^2 \right] \nonumber \\
&= 2n\sum_{i=1}^k \bE[\widehat{p}_i^2]\cdot \bE\left[\widehat{q}_i\left(u(\widehat{q}_i) - u\left(\widehat{q}_i - \frac{1}{n}\right) \right)^2 \right] \nonumber\\
&\lesssim_{c_1} \sum_{i\in [k]: q_i \le 3c_1\log n/n} \left(p_i^2 + \frac{p_i}{m}\right)\cdot (\log n)^3 + \sum_{i\in [k]: q_i>3c_1\log n/n} \left(p_i^2 + \frac{p_i}{m}\right)\cdot \frac{1}{nq_i} \nonumber\\
&\lesssim_{c_1,\varepsilon} \frac{kU^2}{n^{2-2\varepsilon}} + \frac{kU}{n^{1-\varepsilon}m} + \frac{U}{n} + \frac{k}{nm} \lesssim_{c_1,\varepsilon} \frac{kU^2}{n^{2-2\varepsilon}} + \frac{k}{m^2} + \frac{U}{n}, 
\end{align}
where the last step is due to the inequality $kU/(n^{1-\varepsilon}m)\le kU^2/n^{2-2\varepsilon} + k/m^2$. 

The perturbation of $\widehat{p}$ is slightly more involved to deal with. We write that 
\begin{align*}
f(\widehat{p},\widehat{q}) - f\left(\widehat{p} - \frac{1}{m}, \widehat{q}\right) = \frac{u(\widehat{q})}{m} + v(\widehat{p}) - v\left(\widehat{p} - \frac{1}{m}\right). 
\end{align*}
We distinguish into three cases: 
\begin{itemize}
	\item Case I: $p \le 3c_1\log m/m$. By Lemma \ref{lemma:uv_new}, we see that
	\begin{align*}
	\bE\left[ \widehat{p}\left(f(\widehat{p},\widehat{q}) - f\left(\widehat{p} - \frac{1}{m}, \widehat{q}\right)\right)^2 \right] &\lesssim \frac{\bE[\widehat{p}]\cdot \bE[u(\widehat{q})^2]}{m^2} + \bE\left[\widehat{p}\left(v(\widehat{p}) - v\left(\widehat{p} - \frac{1}{m}\right) \right)^2 \right] \\
	&\lesssim_{c_1} \frac{p(\log n)^2}{m^2} + \frac{(\log m)^5}{m^3}. 
	\end{align*}
	Consequently, the total contribution of such symbols is 
	\begin{align}\label{eq:case_1}
	&2m\cdot \sum_{i\in [k]: p_i\le 3c_1\log m/m} \bE\left[ \widehat{p}_i \left( f(\widehat{p}_i, \widehat{q}_i) - f\left(\widehat{p}_i - \frac{1}{m}, \widehat{q}_i\right) \right)^2 \right] \nonumber\\
	&\lesssim_{c_1}  \sum_{i\in [k]: p_i\le 3c_1\log m/m} \left( \frac{p_i(\log n)^2}{m} + \frac{(\log m)^5}{m^2} \right) \nonumber \\
	&\lesssim_{c_1,\varepsilon} (mn)^\varepsilon \left(\frac{kU}{mn} + \frac{k}{m^2}\right) \lesssim_{c_1,\varepsilon} (mn)^\varepsilon \left(\frac{kU^2}{n^2} + \frac{k}{m^2}\right). 
	\end{align}
	\item Case II: $q\le 3c_1\log n/n$. In this case, by Lemma \ref{lemma:uv_new} again we have
	\begin{align*}
\bE\left[ \widehat{p}\left(f(\widehat{p},\widehat{q}) - f\left(\widehat{p} - \frac{1}{m}, \widehat{q}\right)\right)^2 \right] &\lesssim \frac{\bE[\widehat{p}]\cdot \bE[u(\widehat{q})^2]}{m^2} + \bE\left[\widehat{p}\left(v(\widehat{p}) - v\left(\widehat{p} - \frac{1}{m}\right) \right)^2 \right] \\
&\lesssim_{c_1} \frac{p(\log n)^2}{m^2} + \frac{(\log m)^5}{m^3} + \frac{p[1+(\log p)^2]}{m^2} \\
&\lesssim_{c_1} \frac{U(\log n)^3}{m^2n} + \frac{(\log m)^5}{m^3}, 
\end{align*}
where in the last step we have used that $p\le Uq$. Consequently, the total contribution of such symbols is
	\begin{align}\label{eq:case_2}
&2m\cdot \sum_{i\in [k]: q_i\le 3c_1\log n/n} \bE\left[ \widehat{p}_i \left( f(\widehat{p}_i, \widehat{q}_i) - f\left(\widehat{p}_i - \frac{1}{m}, \widehat{q}_i\right) \right)^2 \right] \nonumber\\
&\lesssim_{c_1}  \sum_{i\in [k]: q_i\le 3c_1\log n/n} \left( \frac{U(\log n)^2}{mn} + \frac{(\log m)^5}{m^2} \right) \nonumber \\
&\lesssim_{c_1,\varepsilon} (mn)^\varepsilon \left(\frac{kU}{mn} + \frac{k}{m^2}\right) \lesssim_{c_1,\varepsilon} (mn)^\varepsilon \left(\frac{kU^2}{n^2} + \frac{k}{m^2}\right). 
\end{align}
\item Case III: $p>3c_1\log m/m$ and $q>3c_1\log n/n$. We consider the following identity: 
\begin{align*}
f(\widehat{p},\widehat{q}) - f\left(\widehat{p} - \frac{1}{m}, \widehat{q}\right)
&= \frac{u(\widehat{q})}{m} + v(\widehat{p}) - v\left(\widehat{p} - \frac{1}{m}\right) \\
&= \frac{u(\widehat{q}) + \log q}{m} + \left(v(\widehat{p}) - v\left(\widehat{p}-\frac{1}{m}\right) - \frac{\log p}{m} \right) + \frac{1}{m}\log\frac{p}{q}. 
\end{align*}
The usefulness of the above decomposition lies in the following key result that each term now enjoys a small second moment. 
\begin{lemma}\label{lemma:second_order_diff}
For $m\widehat{p}\sim \mathsf{B}(m,p)$, $n\widehat{q}\sim \mathsf{B}(n,q)$ with $p>3c_1\log m/m$ and $q>3c_1\log n/n$, it holds that
\begin{align*}
\bE[ (u(\widehat{q}) + \log q)^2 ] &\lesssim_{c_1} 1, \\
\bE\left[\left(v(\widehat{p}) - v\left(\widehat{p}-\frac{1}{m}\right) - \frac{\log p}{m} \right)^2\right] &\lesssim_{c_1} \frac{1}{m^2}.
\end{align*}
\end{lemma}
Based on Lemma \ref{lemma:second_order_diff} and the fact \eqref{eq:concentration} that $\widehat{p}\le 2p$ with probability at least $1-m^{-5}$, we arrive at
\begin{align*}
&\bE\left[ \widehat{p}\left(f(\widehat{p},\widehat{q}) - f\left(\widehat{p} - \frac{1}{m}, \widehat{q}\right)\right)^2 \right] \\
&\lesssim_{c_1} \frac{1}{m^5} + p\cdot \left(\frac{\bE[ (u(\widehat{q}) + \log q)^2 ]  }{m^2} + \bE\left[\left(v(\widehat{p}) - v\left(\widehat{p}-\frac{1}{m}\right) - \frac{\log p}{m} \right)^2\right] + \frac{1}{m^2}\log^2 \frac{p}{q}\right) \\
&\lesssim_{c_1} \frac{p}{m^2} + \frac{p\log^2(p/q)}{m^2} \lesssim_{c_1} \frac{p(\log U)^2 + q}{m^2}, 
\end{align*}
where the last step follows from the elementary inequality $(\log x)^2 \lesssim (\log U)^2 + x^{-1}$ for all $0<x\le U$. Consequently, the total contribution of such symbols to the final variance is
	\begin{align}\label{eq:case_3}
&2m\cdot \sum_{i\in [k]: p_i>3c_1\log m/m, q_i> 3c_1\log n/n} \bE\left[ \widehat{p}_i \left( f(\widehat{p}_i, \widehat{q}_i) - f\left(\widehat{p}_i - \frac{1}{m}, \widehat{q}_i\right) \right)^2 \right] \nonumber\\
&\lesssim_{c_1}  \sum_{i\in [k]} \frac{p_i(\log U)^2 + q_i}{m}\lesssim_{c_1} \frac{(\log U)^2}{m}. 
\end{align}
\end{itemize}
In summary, a combination of the inequalities \eqref{eq:case_1}, \eqref{eq:case_2}, and \eqref{eq:case_3} leads to
\begin{align}\label{eq:perturbation_p}
&2m\cdot \sum_{i\in [k]} \bE\left[ \widehat{p}_i \left( f(\widehat{p}_i, \widehat{q}_i) - f\left(\widehat{p}_i - \frac{1}{m}, \widehat{q}_i\right) \right)^2 \right]\lesssim_{c_1,\varepsilon} \frac{(\log U)^2}{m} + (mn)^{\varepsilon}\left(\frac{kU^2}{n^2} + \frac{k}{m^2} \right),
\end{align}
and the final result follows from Lemma \ref{lemma:ES_multinomial} and \eqref{eq:perturbation_q}, \eqref{eq:perturbation_p}. 

\section{Proof of Auxiliary Lemmas}\label{appendix:auxiliary_proof}
\subsection{Proof of Lemma \ref{lemma:approximation}}
We prove the upper and lower approximation error bounds separately. The upper bound makes use of the M\"{u}ntz polynomial: it was shown in \cite[Page 169]{cheney1966introduction} that for real numbers $m,p_1,\cdots,p_n > -1/2$, the distance of $x^m$ to the linear span of $\{x^{p_1},\cdots, x^{p_n} \}$ in the $L_2[0,1]$ metric is
\begin{align*}
d_n(m; p_1,\cdots,p_n) = \frac{1}{\sqrt{2m+1}}\prod_{i=1}^n \frac{|m-p_i|}{m+p_i+1}. 
\end{align*}
Specializing to the case where $m=1/2$ and $\{p_1,\cdots,p_{n-\alpha+1}\} = \{\alpha-1/2,\cdots,n-1/2 \}$, the above result implies that there exists some rational series $P(x) = \sum_{d = \alpha-1}^{n-1} b_dx^{d+1/2}$ such that
\begin{align*}
\left\|\sqrt{x} - P(x) \right\|_{L_2[0,1]} \le \frac{1}{\sqrt{2}}\prod_{i=1}^{n-\alpha+1} \frac{\alpha+i-1/2}{\alpha+i+1/2} = O_\alpha\left(\frac{1}{n^2}\right). 
\end{align*}
Hence, for $Q(x) = \int_0^x P(t)dt$, the Cauchy--Schwartz inequality gives
\begin{align*}
\left|\frac{2}{3}x^{\frac{3}{2}} - Q(x) \right| \le \int_0^x |\sqrt{t} - P(t)|dt \le \sqrt{\left(\int_0^x dt \right)\left(\int_0^x |\sqrt{t} - P(t)|^2dt \right) } = O_\alpha\left(\frac{\sqrt{x}}{n^2}\right)
\end{align*}
for all $x\in [0,1]$. After dividing $2\sqrt{x}/3$ at both sides, we conclude that the polynomial $R(x) = 3Q(x)/(2\sqrt{x})$ satisfies the claimed upper bound. 

As for the lower bound, note that for any $c\in (0,1)$, 
\begin{align*}
\inf_{a_\alpha,\cdots,a_n\in \bR}\sup_{x\in [0,1]} \left|x - \sum_{d = \alpha}^n a_dx^d \right| &\ge \inf_{a_0,\cdots,a_n\in \bR}\sup_{x\in [0,1]} x^\alpha \left|x^{1-\alpha} - \sum_{d = 0}^n a_dx^d \right|  \\
&\ge \inf_{a_0,\cdots,a_n\in \bR}\sup_{x\in [(c/n)^2,1]} x^\alpha \left|x^{1-\alpha} - \sum_{d = 0}^n a_dx^d \right| \\
&\ge \left(\frac{c}{n}\right)^{2\alpha} \inf_{a_0,\cdots,a_n\in \bR}\sup_{x\in [(c/n)^2,1]}\left|x^{1-\alpha} - \sum_{d = 0}^n a_dx^d \right|. 
\end{align*}
Hence, it suffices to show that for some constant $c\in (0,1)$, it holds that
\begin{align}\label{eq:approximation_lower_target}
\inf_{a_0,\cdots,a_n\in \bR}\sup_{x\in [(c/n)^2,1]}\left|x^{1-\alpha} - \sum_{d = 0}^n a_dx^d \right| = \Omega_\alpha\left(n^{2(\alpha-1)}\right). 
\end{align}
To this end, we introduce some necessary definitions and results from approximation theory. For functions defined on $[0,1]$, define the $r$-th order Ditzian--Totik modulus of smoothness by \cite{Ditzian--Totik1987}
\begin{align}\label{eq.DT_modulus}
\omega_\varphi^r(f,t)_\infty \triangleq \sup_{0<h\le t} \|\Delta_{h\varphi(x)}^r f(x)\|_\infty =  \sup_{0<h\le t} \left\|\sum_{\ell=0}^r \binom{r}{\ell}(-1)^{\ell} f\left(x + \left(\ell-\frac{r}{2}\right)h\varphi(x) \right)\right\|_\infty,
\end{align}
where $\varphi(x)\triangleq \sqrt{x(1-x)}$. This quantity is related to the polynomial approximation error via the following lemma. 
\begin{lemma}[\!\!{\cite{Ditzian--Totik1987}}]\label{lem.DT_modulus}
	For any integer $u>0$ and $n>u$, there exists some constant $M_u$ depending only on $u$, such that for all $t\in (0,1)$ and $f$,
	\begin{align*}
	E_{n}(f;[0,1]) &\le M_u\omega_\varphi^u(f,1/n)_\infty\\
	\frac{M_u}{n^u}\sum_{\ell=0}^n (\ell+1)^{u-1}E_{\ell}(f;[0,1]) &\ge \omega_\varphi^u(f,1/n)_\infty.
	\end{align*}
	where $E_{n}(f;[0,1])$ is the best degree-$n$ polynomial approximation error of $f$ on $[0,1]$. 
\end{lemma}
For function $f(x) = [(c/n)^2 + (1-(c/n)^2)x]^{1-\alpha}$ with $\alpha\ge 2$, by translation the approximation error $E_n(f;[0,1])$ is the LHS of \eqref{eq:approximation_lower_target}. It is straightforward to see that for any $1\le m\le n/c$, 
\begin{align*}
\omega_\varphi^1(f,1/m) &\le 2\|f\|_\infty \le 2(n/c)^{2(\alpha-1)}, \\
\omega_\varphi^1(f,1/m) &\ge \left|f(0) - f\left(\frac{1}{8m^2} \right)\right| = \Omega_\alpha\left( (n/c)^{2(\alpha-1)}\right). 
\end{align*}
Hence, there exist constants $c_\alpha, C_\alpha>0$ such that $c_\alpha (n/c)^{2(\alpha-1)}\le \omega_\varphi^1(f,1/m)\le C_\alpha (n/c)^{2(\alpha-1)}$ for $1\le m\le n/c$. Choosing $u=1$ and $D=1/c$ (w.l.o.g. we assume that $D>1$ is an integer), Lemma \ref{lem.DT_modulus} gives
\begin{align*}
c_\alpha (n/c)^{2(\alpha-1)} &\le \omega_\varphi^1\left(f,\frac{1}{Dn}\right) \\
&\le \frac{M_1}{Dn}\sum_{\ell=0}^{Dn} (\ell+1)^{u-1}E_{\ell}(f;[0,1]) \\
&\le \frac{M_1}{Dn}\left( \left(\frac{n}{c}\right)^{2(\alpha-1)} + M_1\sum_{\ell=1}^{n} C_\alpha \left(\frac{n}{c}\right)^{2(\alpha-1)}  + (Dn-n)E_n(f;[0,1]) \right),
\end{align*}
which gives
\begin{align*}
E_n(f;[0,1]) \ge \frac{1}{n(1/c-1)}\left(\frac{c_\alpha}{M_1}\left(\frac{n}{c}\right)^{2\alpha} - \left(\frac{n}{c}\right)^{2(\alpha-1)} - nM_1C_\alpha\left(\frac{n}{c}\right)^{2(\alpha-1)}   \right). 
\end{align*}
Consequently, choosing the constant $c>0$ small enough gives the claimed lower bound \eqref{eq:approximation_lower_target}. 

\subsection{Proof of Lemma \ref{lemma:ES_multinomial}}
We first prove the following result: for $(m\widehat{p}_1,\cdots,m\widehat{p}_k) \sim \mathsf{Multi}(m;p_1,\cdots,p_k)$ and $T = \sum_{i=1}^k f_i(\widehat{p}_i)$, it holds that
\begin{align}\label{eq:ES_1d}
\var(T) \le 2m\cdot \sum_{i=1}^k \bE\left[\widehat{p}_i \left(f(\widehat{p}_i) - f\left(\widehat{p}_i - \frac{1}{m} \right) \right)^2 \right]. 
\end{align}

We recall the following Efron--Stein--Steele inequality: 
\begin{lemma}[Efron--Stein--Steele Inequality \cite{steele1986efron}]\label{lemma:efron-stein}
	Let $X_1,X_2,\cdots,X_n$ be independent random variables, and for $i=1,\cdots,n$, let $X_i'$ be an independent copy of $X_i$. Then for any $f$,
	\begin{align*}
	\var(f(X_1,\cdots,X_n)) \le \frac{1}{2}\sum_{i=1}^n \bE(f(X_1,\cdots,X_n) - f(X_1,\cdots,X_{i-1},X_i',X_{i+1},\cdots,X_n))^2.
	\end{align*}
\end{lemma}

To prove \eqref{eq:ES_1d}, let $X^m\sim (p_1,\cdots,p_k)$ be $m$ i.i.d. observations and write $T$ as a function of $X^m$. By the Efron--Stein--Steele inequality, we have 
\begin{align}\label{eq:ES_original}
\var(T) \le \frac{m}{2}\cdot \bE\left[ \left( T(X_1,\cdots,X_m) - T(X_1',\cdots,X_m)\right)^2 \right], 
\end{align}
where $X_1'$ is an independent copy of $X_1$ mutually independent of $(X_2,\cdots,X_m)$. By the definition of $T$, one may express the difference as
\begin{align*}
T(X_1',\cdots,X_m) - T(X_1,\cdots,X_n) = D_- + D_+, 
\end{align*}
where
\begin{align*}
D_- &= f_{X_1}\left(\widehat{p}_{X_1} \right) - f_{X_1}\left(\widehat{p}_{X_1} - \frac{1}{m} \right), \\
D_+ &= \begin{cases}
f_{X_1'}\left(\widehat{p}_{X_1'} + \frac{1}{m} \right) - f_{X_1'}\left(\widehat{p}_{X_1'} \right) &\text{if } X_1' \neq X_1, \\
f_{X_1'}\left(\widehat{p}_{X_1'}\right) - f_{X_1'}\left(\widehat{p}_{X_1'} - \frac{1}{m} \right) &\text{if } X_1' = X_1. 
\end{cases}
\end{align*}
Note that conditioning on the empirical probabilities $(\widehat{p}_1,\cdots,\widehat{p}_k)$, the random variable $X_1$ follows the distribution $(\widehat{p}_1,\cdots,\widehat{p}_k)$ while the new observation $X_1'$ follows the distribution $(p_1,\cdots,p_k)$, and they are mutually independent. Hence, 
\begin{align*}
\bE[D_-^2] = \bE[ \bE[D_-^2 \mid \widehat{p}_1,\cdots, \widehat{p}_k ]] = \sum_{i=1}^k \bE\left[\widehat{p}_i \left(f_i(\widehat{p}_i) - f_i\left(\widehat{p}_i - \frac{1}{m} \right) \right)^2 \right], 
\end{align*}
and
\begin{align*}
\bE[D_+^2] &= \bE[ \bE[D_+^2 \mid \widehat{p}_1,\cdots, \widehat{p}_k ]] \\
&= \sum_{i=1}^k p_i\cdot \bE\left[(1-\widehat{p}_i)\left(f_i\left(\widehat{p}_i + \frac{1}{m}\right) - f_i\left(\widehat{p}_i \right) \right)^2 + \widehat{p}_i \left(f_i(\widehat{p}_i) - f_i\left(\widehat{p}_i - \frac{1}{m} \right) \right)^2 \right] \\
&= \sum_{i=1}^k\sum_{j=1}^m p_i\left(f_i\left(\frac{j}{m}\right) - f_i\left(\frac{j-1}{m}\right) \right)^2\times \\
&\qquad \left[\frac{m-j+1}{m}\cdot \binom{m}{j-1}p_i^{j-1}(1-p_i)^{m-j+1} + \frac{j}{m}\cdot \binom{m}{j}p_i^j(1-p_i)^{m-j} \right] \\
&= \sum_{i=1}^k\sum_{j=1}^m \left(f_i\left(\frac{j}{m}\right) - f_i\left(\frac{j-1}{m}\right) \right)^2\times \left[ \frac{j}{m}\cdot \binom{m}{j}p_i^j(1-p_i)^{m-j} \right] \\
&= \sum_{i=1}^k \bE\left[\widehat{p}_i \left(f_i(\widehat{p}_i) - f_i\left(\widehat{p}_i - \frac{1}{m} \right) \right)^2 \right] = \bE[D_-^2]. 
\end{align*}
Hence, by \eqref{eq:ES_original} and the triangle inequality, we have 
\begin{align*}
\var(T) \le \frac{m}{2}\bE[(D_- + D_+)^2] \le m\left(\bE[D_-^2] + \bE[D_+^2] \right) = 2m\cdot \sum_{i=1}^k \bE\left[\widehat{p}_i \left(f_i(\widehat{p}_i) - f_i\left(\widehat{p}_i - \frac{1}{m} \right) \right)^2 \right], 
\end{align*}
establishing \eqref{eq:ES_1d}. For the original statement, write the sum $S$ as a function of $X^m\sim (p_1,\cdots,p_k)$ and $Y^n\sim (q_1,\cdots,q_k)$, and deal with the single changes in $X^m$ and $Y^n$ separately as \eqref{eq:ES_1d} in the Efron--Stein--Steele inequality. 

\subsection{Proof of Lemma \ref{lemma:risk_reduction_alpha}}
For each $n\in \NN$, let $\widehat{D}_n$ be an estimator achieving the minimax risk $R^\star(m,n,(1+\varepsilon)U)$ under the original sampling model with $n$ samples from $Q$. By sufficiency arguments we may assume that $\widehat{D}_n$ only depends on the histograms. Then we construct an estimator $\widehat{D}$ for the Poissonized model as follows: let $N=\sum_{i=1}^k h_i$ from the observed Poisson histograms, then the estimator $\widehat{D}$ is chosen to be $\widehat{D}_{N}$. Clearly, $N\sim \mathsf{Poi}(n\sum_{i=1}^kq_i)$, and conditioning on the realization of $N$, the Poisson sampling model is equivalent to sampling $N$ independent samples from the discrete distribution $Q/\|Q\|_1$. Hence, let $Q_0 = Q/\|Q\|_1$ for $(P,Q)\in \calM_k(U,\varepsilon)$, it holds that $(P,Q_0)\in \calM_k((1+\varepsilon)U)$. Hence, the estimation performance of the estimator $\widehat{D}$ under the Poissonized model is 
\begin{align*}
\bE_{(P,Q)} |\widehat{D} - D_\alpha(P\|Q) | &\le  |D_\alpha(P\| Q_0) - D_\alpha(P\| Q)| + \bE_{(P,Q)} |\widehat{D} - D_\alpha(P\|Q_0) | \\
&= |D_\alpha(P\| Q_0) - D_\alpha(P\| Q)| + \sum_{\ell=0}^\infty \bE_{(P,Q_0)} |\widehat{D}_\ell - D_\alpha(P\| Q_0) | \cdot \bP(\mathsf{Poi}(n) = \ell) \\
&\stepa{\lesssim}_{\alpha} \varepsilon U^{\alpha-1} + \sum_{\ell=0}^\infty R^\star(m,\ell,(1+\varepsilon)U)  \cdot \bP(\mathsf{Poi}(n) = \ell) \\
&\stepb{\le} \varepsilon U^{\alpha-1} + \frac{U^{\alpha-1}-1}{\alpha(\alpha-1)}\cdot \bP(\mathsf{Poi}(n) < n/2) + R^\star(m,n/2,(1+\varepsilon)U) \\
&\stepc{\le} \varepsilon U^{\alpha-1} + \frac{U^{\alpha-1}-1}{\alpha(\alpha-1)}e^{-n/8} + R^\star(m,n/2,(1+\varepsilon)U), 
\end{align*}
where (a) is due to the fact that $(P,Q_0)\in \calM_k((1+\varepsilon)U)$ and for $(P,Q)\in \calM_k(U,\varepsilon)$, 
\begin{align*}
|D_\alpha(P\| Q_0) - D_\alpha(P\| Q)| = \frac{| \|Q\|_1^{\alpha-1} -1|}{\alpha(\alpha-1)}\cdot \sum_{i=1}^k \frac{p_i^\alpha}{q_i^{\alpha-1}} \lesssim_\alpha \varepsilon\sum_{i=1}^k \frac{p_i^\alpha}{q_i^{\alpha-1}}= O_\alpha\left( \varepsilon U^{\alpha-1} \right), 
\end{align*}
(b) follows from the decreasing property of the map $\ell \mapsto R^\star(m,\ell,(1+\varepsilon)U)$ and $R^\star(m,0,(1+\varepsilon)U) \le (U^{\alpha-1}-1)/(\alpha(\alpha-1))$ by considering an estimator which always outputs zero, and (c) is due to the Chernoff bound. Finally, taking the supremum over $(P,Q)\in \calM_k(U,\varepsilon)$ on the LHS completes the proof of the lemma. 

\subsection{Proof of Lemma \ref{lemma:risk_reduction_KL}}
The proof of Lemma \ref{lemma:risk_reduction_KL} is similar to that of Lemma \ref{lemma:risk_reduction_alpha}. Define the estimators $(\widehat{D}_n, n\in \NN)$ and $\widehat{D}$ as in the proof of Lemma \ref{lemma:risk_reduction_alpha}, and for $(P,Q)\in \calM_k(U,\varepsilon)$, define $Q_0 = Q/\|Q\|_1$ which satisfies $(P,Q_0)\in \calM_k((1+\varepsilon)U)$. Consequently, the estimation performance of the estimator $\widehat{D}$ under the Poissonized model is 
\begin{align*}
\bE_{(P,Q)} |\widehat{D} - D_{\text{KL}}(P\|Q) | &\le  |D_{\text{KL}}(P\| Q_0) - D_{\text{KL}}(P\| Q)| + \bE_{(P,Q)} |\widehat{D} - D_{\text{KL}}(P\|Q_0) | \\
&= |D_{\text{KL}}(P\| Q_0) - D_{\text{KL}}(P\| Q)| + \sum_{\ell=0}^\infty \bE_{(P,Q_0)} |\widehat{D}_\ell - D_{\text{KL}}(P\| Q_0) | \cdot \bP(\mathsf{Poi}(n) = \ell) \\
&\stepa{\lesssim} \varepsilon\log U + \sum_{\ell=0}^\infty R^\star(m,\ell,(1+\varepsilon)U)  \cdot \bP(\mathsf{Poi}(n) = \ell) \\
&\stepb{\le} \varepsilon\log U + \log U\cdot \bP(\mathsf{Poi}(n) < n/2) + R^\star(m,n/2,(1+\varepsilon)U) \\
&\stepc{\le} \varepsilon\log U + \log U\cdot e^{-n/8} + R^\star(m,n/2,(1+\varepsilon)U), 
\end{align*}
where (a) is due to the fact that $(P,Q_0)\in \calM_k((1+\varepsilon)U)$ and for $(P,Q)\in \calM_k(U,\varepsilon)$, 
\begin{align*}
|D_{\text{KL}}(P\| Q_0) - D_{\text{KL}}(P\| Q)| = | \log\|Q\|_1 -1| \cdot \sum_{i=1}^k p_i\log \frac{p_i}{q_i} \lesssim_\alpha \varepsilon\sum_{i=1}^k p_i\log \frac{p_i}{q_i}= O_\alpha\left( \varepsilon \log U \right), 
\end{align*}
step (b) follows from the decreasing property of the map $\ell \mapsto R^\star(m,\ell,(1+\varepsilon)U)$ and $R^\star(m,0,(1+\varepsilon)U) \le \log U$ by considering an estimator which always outputs zero, and (c) is due to the Chernoff bound. Finally, taking the supremum over $(P,Q)\in \calM_k(U,\varepsilon)$ on the LHS completes the proof of the lemma. 

\subsection{Proof of Lemma \ref{lemma:uv}}
To deal with the function $u$, recall that the moment generating function of Binomial distribution gives that for $m\widehat{p}\sim \mathsf{B}(m,p)$, 
\begin{align*}
\bE[e^{\lambda \widehat{p}}] = \left(1 - p + pe^{\lambda/m}\right)^m = \left(1 + p\sum_{i=1}^\infty \frac{(\lambda/m)^i}{i!} \right)^m, \qquad \forall \lambda\in \bR. 
\end{align*}
Hence, comparing the coefficients of $\lambda^k$ at both sides gives
\begin{align}\label{eq:binomial_moment}
\bE[\widehat{p}^k] \lesssim_k \sum_{\ell=1}^k \frac{p^{\ell}}{m^{k-\ell}}, \qquad \forall k\in \NN. 
\end{align}
Consequently, as $u(\widehat{p}) = \sum_{d=1}^\alpha a_d\widehat{p}^{d}$ with coefficient bound $|a_d| = O_\alpha(m^{d-\alpha})$, the Binomial moment bound \eqref{eq:binomial_moment} gives
\begin{align*}
\bE[u(\widehat{p})^2] \lesssim_\alpha \sum_{d=1}^\alpha \sum_{\ell=1}^{2d} m^{2d-2\alpha}\cdot \frac{p^{\ell}}{m^{2d-\ell}} \lesssim_\alpha p^{2\alpha}. 
\end{align*}
Similarly, simple algebra shows that $u(\widehat{p}) - u(\widehat{p}-1/m) = \sum_{d=1}^{\alpha-1} b_d\widehat{p}^d$ with $|b_d| = O_\alpha(m^{d-\alpha})$, and therefore \eqref{eq:binomial_moment} again leads to
\begin{align*}
\bE\left[\widehat{p}\left( u(\widehat{p}) - u\left(\widehat{p}-\frac{1}{m} \right) \right)^2 \right] \lesssim_\alpha \sum_{d=3}^{2\alpha-1}\sum_{\ell=1}^{d} m^{d-2\alpha-1}\cdot \frac{p^{\ell}}{m^{d-\ell}} \lesssim_\alpha \frac{p^{2\alpha-1}}{m^2}. 
\end{align*}

As for the function $v$, as $\|\widetilde{g}_d \|_\infty \le (4c_1\log n/n)^d$, it holds that
\begin{align*}
|v(\widehat{q})| \le \left(\frac{4c_1\log n}{n}\right)^d\cdot \bP\left(\mathsf{HG}\left(n, n\widehat{q}, \frac{n}{2} \right) \le  c_1\log n \right). 
\end{align*}
Hence, 
\begin{align*}
\bE[v(\widehat{q})^2] &\le \left(\frac{4c_1\log n}{n}\right)^{2d}\bE\left[ \bP\left(\mathsf{HG}\left(n, n\widehat{q}, \frac{n}{2} \right) \le  c_1\log n \right) \right] \\
&= \left(\frac{4c_1\log n}{n}\right)^{2d} \bP\left(\mathsf{B}\left(\frac{n}{2}, q \right) \le  c_1\log n \right) \\
&\le \left(\frac{4c_1\log n}{n}\right)^{2d} \exp(-\Omega(nq - 3c_1\log n)_+)), 
\end{align*}
where the last step follows from the Chernoff bound. Similarly, 
\begin{align*}
\bE\left[\widehat{q}\left( u(\widehat{q}) - u\left(\widehat{q}-\frac{1}{n} \right) \right)^2 \right] &\lesssim \left(\frac{4c_1\log n}{n}\right)^{2d}\bE\left[\widehat{q}\cdot \bP\left(\mathsf{HG}\left(n, n\widehat{q}, \frac{n}{2} \right) \le  c_1\log n \right) \right] \\
&\le \left(\frac{4c_1\log n}{n}\right)^{2d}\sqrt{\bE[\widehat{q}^2]\cdot \bE\left[\bP\left(\mathsf{HG}\left(n, n\widehat{q}, \frac{n}{2} \right) \le  c_1\log n \right) \right]}\\
&\le \left(\frac{4c_1\log n}{n}\right)^{2d} \cdot \sqrt{q^2 + \frac{q}{n}}\cdot \exp(-\Omega((nq - 3c_1\log n)_+) ) \\
&\lesssim \left(\frac{4c_1\log n}{n}\right)^{2d+1}\exp(-\Omega((nq - 3c_1\log n)_+)),
\end{align*}
where the last step follows from proper tail comparison. 

\subsection{Proof of Lemma \ref{lemma:h_alpha_moment}}
By definition of $h_\alpha$ in \eqref{eq:bias_correction}, it holds that $\|h_\alpha\|_\infty \lesssim_\alpha n^{\alpha-1}$. Moreover, for any $nq/4\le X\le nq$ with $q\ge c_1\log n/n$, it holds that
\begin{align*}
|h_\alpha(X)| \lesssim_\alpha \frac{1}{q^{\alpha-1}} + \frac{1}{nq^\alpha} \lesssim_{\alpha,c_1} \frac{1}{q^{\alpha-1}}. 
\end{align*}
Hence, the concentration bound \eqref{eq:concentration} leads to
\begin{align*}
\bE[h_\alpha(X)^2] &= \bE\left[h_\alpha(X)^2\1\left(\frac{nq}{4}\le X\le nq\right) \right] + \bE\left[h_\alpha(X)^2\1\left(X\notin \left[\frac{nq}{4},nq\right] \right) \right] \\
&\lesssim_{\alpha,c_1} \bE\left[\frac{1}{q^{2(\alpha-1)}} \1\left(\frac{nq}{4}\le X\le nq\right) \right] + n^{2(\alpha-1)}\bP\left(X\notin \left[\frac{nq}{4},nq\right] \right) \\
&\lesssim_{\alpha,c_1}  \frac{1}{q^{2(\alpha-1)}} + \frac{1}{n^{3\alpha}} \lesssim_{\alpha,c_1} \frac{1}{q^{2(\alpha-1)}}, 
\end{align*}
where in the last step we have again used that $q\ge c_1\log n/n$. 

\subsection{Proof of Lemma \ref{lemma:delta_h_alpha_moment}}
This proof is entirely similar to that of Lemma \ref{lemma:h_alpha_moment}. Specifically, it is easy to verify that $\Delta h_\alpha(x)\le 2\|h_\alpha\|_\infty \lesssim n^{\alpha-1}$ for all $x\in \NN$, and $\Delta h_\alpha(x) \lesssim_{\alpha,c_1} 1/(nq^\alpha)$ whenever $x\in [nq/4,nq]$. Hence, the concentration bound \eqref{eq:concentration} leads to
\begin{align*}
\bE[\Delta h_\alpha(X)^2] &= \bE\left[\Delta h_\alpha(X)^2\1\left(\frac{nq}{4}\le X\le nq\right) \right] + \bE\left[\Delta h_\alpha(X)^2\1\left(X\notin \left[\frac{nq}{4},nq\right] \right) \right] \\
&\lesssim_{\alpha,c_1} \bE\left[\frac{1}{n^2q^{2\alpha}} \1\left(\frac{nq}{4}\le X\le nq\right) \right] + n^{2(\alpha-1)}\bP\left(X\notin \left[\frac{nq}{4},nq\right] \right) \\
&\lesssim_{\alpha,c_1}  \frac{1}{n^2q^{2\alpha}} + \frac{1}{n^{3\alpha}} \lesssim_{\alpha,c_1} \frac{1}{n^2q^{2\alpha}}, 
\end{align*}
where in the last step we have again used that $q\ge 3c_1\log n/n$. The other inequality follows from the same arguments line by line. 

\subsection{Proof of Lemma \ref{lemma:uv_new}}
First we upper bound the second moment of $u(\widehat{q})$. By Cauchy--Schwartz, 
\begin{align*}
u(\widehat{q})^2 \le \sum_{s>c_1\log n} \bP\left(\mathsf{HG}\left(n,n\widehat{q},\frac{n}{2}\right)=s \right) h^{(1)}(n\widehat{q} - s)^2. 
\end{align*}
Consequently, \eqref{eq:unbiased_smoothed_moments} leads to
\begin{align*}
\bE[u(\widehat{q})^2] \le \bP\left(\mathsf{B}\left(\frac{n}{2},q\right)> c_1\log n \right)\cdot \bE[h^{(1)}(X)^2]
\end{align*}
for $X\sim \mathsf{B}(n/2,q)$. We distinguish into two scenarios. First, if $q\le c_1\log n/n$, Lemma \ref{lemma.localization} together with the norm bound $\|h^{(1)}\|_\infty \lesssim \log n$ gives an upper bound $O_{c_1}(n^{-5}\log^2n)$ on the above quantity. Second, if $q>c_1\log n/n$, following the same lines of the proofs of Lemma \ref{lemma:h_alpha_moment} and Lemma \ref{lemma:delta_h_alpha_moment} leads to $\bE[h^{(1)}(X)^2]\lesssim_{c_1} 1+(\log q)^2$. Then after some algebra, combining these two scenarios implies the first statement of the lemma (despite that the threshold becomes different). 

To deal with the function $v$, we again distinguish into two cases. First, when $p\le 3c_1\log m/m$, H\"{o}lder's inequality gives
\begin{align*}
\bE[v(\widehat{p})^4] \le \bP\left(\mathsf{B}\left(\frac{m}{2},p\right)> c_1\log m \right)\cdot \bE[h^{(2)}(X)^4]
\end{align*}
for $X\sim \mathsf{B}(m/2,p)$. Further distinguishing into two cases $p\le c_1\log m/m$ and $c_1\log m/m<p\le 3c_1\log m/m$ and using the above analysis for $u$, we conclude that for all $p\le 3c_1\log m/m$ it holds that
\begin{align*}
\bE[v(\widehat{p})^4] \lesssim_{c_1} \left(\frac{\log^2 m}{m} \right)^4. 
\end{align*}
Consequently, in this case the target quantity can be upper bounded as
\begin{align*}
\bE\left[\widehat{p}\left(v(\widehat{p}) - v\left(\widehat{p} - \frac{1}{m}\right) \right)^2 \right] &\lesssim \sqrt{\bE[\widehat{p}^2]}\cdot \sqrt{\bE[v(\widehat{p})^4] + \bE\left[v(\widehat{p} - 1/m)^4 \right] } \\
&\lesssim_{c_1} \sqrt{p^2 + \frac{p}{m}} \cdot \left(\frac{\log^2 m}{m}\right)^2 \lesssim_{c_1} \frac{(\log m)^5}{m^3}. 
\end{align*}

As for $p>3c_1\log m/m$, we use the coupling arguments as in the proof of Lemma \ref{lemma:smooth_variance}. First, by Lemma \ref{lemma.localization} and the assumption $p>3c_1\log m/m$, it suffices to replace $v$ by the function $v^\star$ defined as
\begin{align*}
v^\star(\widehat{p}) = \sum_{s\ge 0} \bP\left(\mathsf{HG}\left(m,m\widehat{p},\frac{m}{2}\right) =s \right)\cdot h^{(2)}(m\widehat{p}-s), 
\end{align*}
with an additive error at most $O_{c_1}(n^{-5})$. Second, we may express $v^\star(\widehat{p}) = \bE[h^{(2)}(m\widehat{p}-Y) ]$ with $Y\sim \mathsf{HG}(m,m\widehat{p},m/2)$, and $v^\star(\widehat{p}-1/m) = \bE[h^{(2)}(m\widehat{p}-1-Z) ]$ with $Z\sim \mathsf{HG}(m,m\widehat{p}-1,m/2)$. Since there exists a coupling between $(Y,Z)$ such that $Y-1\le Z\le Y$ holds almost surely, we conclude that
\begin{align*}
\left(v^\star(\widehat{p}) - v^\star\left(\widehat{p} - \frac{1}{m}\right)\right)^2 &= \left(\bE[h^{(2)}(m\widehat{p}-Y) - h^{(2)}(m\widehat{p}-1-Z) ]\right)^2  \\
&\le \left(\bE[\Delta h^{(2)}(m\widehat{p}-Y)]\right)^2\\
&\le \sum_{s\ge 0}\bP\left(\mathsf{HG}\left(m,m\widehat{p},\frac{m}{2}\right) =s \right)\cdot \Delta h^{(2)}(m\widehat{p}-s)^2,
\end{align*}
where $\Delta h^{(2)}(x) \triangleq |h^{(2)}(x) - h^{(2)}(x-1)|$. Similar to Lemma \ref{lemma:delta_h_alpha_moment}, for random variable $X\sim \mathsf{B}(m/2,p)$ with $p\ge 3c_1\log m/m$, the following inequalities hold: 
\begin{align*}
\bE\left[\Delta h^{(2)}(X)^2 \right] &\lesssim_{c_1} \frac{1+(\log p)^2}{m^2}, \qquad \bE\left[X\cdot \Delta h^{(2)}(X)^2 \right] \lesssim_{c_1} \frac{p[1+(\log p)^2]}{m}.
\end{align*}
Hence, for $X_1, X_2\sim \mathsf{B}(m/2,p)$, we have
\begin{align*}
\bE\left[\widehat{p}\left(v^\star(\widehat{p}) - v^\star\left(\widehat{p} - \frac{1}{m}\right) \right)^2 \right] & \le \bE\left[ \widehat{p}\cdot \sum_{s\ge 0} \bP\left(\mathsf{HG}\left(m,m\widehat{p},\frac{m}{2}\right) =s \right)\cdot \Delta h^{(2)}(m\widehat{p}-s)^2 \right]\\
&= \bE\left[ \sum_{s\ge 0} \frac{s}{m}\bP\left(\mathsf{HG}\left(m,m\widehat{p},\frac{m}{2}\right) =s \right)\cdot \Delta h^{(2)}(m\widehat{p}-s)^2 \right] \\
&\quad +  \bE\left[ \sum_{s\ge 0} \bP\left(\mathsf{HG}\left(m,m\widehat{p},\frac{m}{2}\right) =s \right)\cdot \left(\widehat{p} - \frac{s}{m}\right)\Delta h^{(2)}(m\widehat{p}-s)^2 \right]\\
&= \bE\left[\frac{X_1}{m}\right]\cdot \bE\left[\Delta h^{(2)}(X)^2 \right] + \frac{\bE\left[X\cdot \Delta h^{(2)}(X)^2 \right]}{m}\\
&\lesssim_{c_1} \frac{p[1+(\log p)^2]}{m^2}, 
\end{align*}
establishing the second inequality for $p\ge 3c_1\log m/m$. The third inequality follows from similar lines to the proof of the second inequality and is thus omitted. 

\subsection{Proof of Lemma \ref{lemma:second_order_diff}}
For the first inequality, consider the following bias-variance decomposition: 
\begin{align*}
\bE[(u(\widehat{q}) + \log q)^2 ] = |\bE[u(\widehat{q})] + \log q|^2 + \var(u(\widehat{q})). 
\end{align*}
First, the bias can be upper bounded as
\begin{align*}
|\bE[u(\widehat{q})] + \log q| &\stepa{=} \left|\bP\left(\mathsf{B}\left(\frac{n}{2},q\right) > c_1\log n \right)\cdot \bE[h^{(1)}(X)] + \log q \right| \\
&\stepb{\le} \left|\bE[h^{(1)}(X)] + \log q \right| + \frac{|\log q|}{n^5} \\
&\stepc{\lesssim}_{c_1} \frac{1}{nq\log n} + \frac{|\log q|}{n^5} \lesssim_{c_1} 1, 
\end{align*}
with $X\sim \mathsf{B}(n/2,q)$, where (a) follows from \eqref{eq:unbiased_smoothed_moments}, (b) is due to the triangle inequality and Lemma \ref{lemma.localization}, and (c) follows from Lemma \ref{lemma:smooth_KL_bias} and the assumption $q\ge 3c_1\log n/n$. Second, applying Lemma \ref{lemma:ES_multinomial} to the Binomial model (which is a special case of the Multinomial model), we have
\begin{align*}
\var(u(\widehat{q})) \le 2n\cdot \bE\left[\widehat{q}\left(u(\widehat{q}) - u\left(\widehat{q} - \frac{1}{n} \right) \right)^2 \right] \lesssim_{c_1} \frac{1}{nq} \lesssim_{c_1} 1, 
\end{align*}
where the second inequality is due to Lemma \ref{lemma:uv_new}. Hence, combining the above inequalities completes the proof of the first inequality. 

As for the second inequality, we again deal with the bias and the variance separately. For the bias part, following the above steps involving \eqref{eq:unbiased_smoothed_moments}, Lemma \ref{lemma.localization} and Lemma \ref{lemma:smooth_KL_bias} leads to
\begin{align*}
|\bE[v(\widehat{p})] - p\log p | &\lesssim_{c_1} \frac{1}{m\log m}, \\
\left|\bE\left[v\left(\widehat{p} - \frac{1}{m}\right)\right] - \left(p - \frac{1}{m}\right)\log \left(p-\frac{1}{m}\right) \right| &\lesssim_{c_1} \frac{1}{m\log m}. 
\end{align*}
Combining the above inequalities and observing that
\begin{align*}
\left|p\log p - \left(p - \frac{1}{m}\right)\log \left(p-\frac{1}{m}\right) - \frac{1}{m}\log p  \right| = \left(p-\frac{1}{m}\right)\log \frac{mp}{mp-1} \le \frac{1}{m}, 
\end{align*}
a triangle inequality leads to the squared bias bound
\begin{align}\label{eq:squared_bias}
\left|\bE\left[ v(\widehat{p}) - v\left(\widehat{p} - \frac{1}{m}\right)\right] - \frac{\log p}{m}  \right|^2 \lesssim_{c_1} \frac{1}{m^2}. 
\end{align}

The most challenging part is to upper bound the variance of the difference $v(\widehat{p}) - v(\widehat{p} - 1/m)$. To this end, by the assumption $p\ge 3c_1\log m/m$, the similar techniques in the proof of Lemma \ref{lemma:uv_new} show that it suffices to work with the new function $v^\star$ with
\begin{align*}
v^\star(\widehat{p}) = \sum_{s\ge 0}\bP\left(\mathsf{B}\left(m,m\widehat{p},\frac{m}{2}\right) =s \right)\cdot h^{(2)}(m\widehat{p}-s). 
\end{align*}
Consequently, applying the Efron--Stein--Steele inequality to the Binomial model again, Lemma \ref{lemma:ES_multinomial} yields to
\begin{align}\label{eq:ES_variance}
\var\left(v^\star(\widehat{p}) - v^\star\left(\widehat{p} - \frac{1}{m}\right) \right) \le 2m\cdot \bE\left[\widehat{p}\left(v^\star(\widehat{p}) - 2v^\star\left(\widehat{p} - \frac{1}{m}\right) +v^\star\left(\widehat{p} - \frac{2}{m}\right) \right)^2 \right]. 
\end{align}
To upper bound the finite difference in \eqref{eq:ES_variance}, we may write $v^\star(\widehat{p}) = \bE[h^{(2)}(m\widehat{p} - X)]$ with $X\sim \mathsf{HG}(m,m\widehat{p},m/2)$, $v^\star(\widehat{p}-1/m) = \bE[h^{(2)}(m\widehat{p} - 1 - Y)]$ with $Y\sim \mathsf{HG}(m,m\widehat{p}-1,m/2)$, and $v^\star(\widehat{p}-2/m) = \bE[h^{(2)}(m\widehat{p} - 2- Z)]$ with $Z\sim \mathsf{HG}(m,m\widehat{p}-2,m/2)$. Note that there exists a coupling of $(X,Y,Z)$ such that $X-1\le Y\le X$ and $Y-1\le Z\le Y$ hold almost surely, and $\bE[X+Z|Y] = 2Y$. Hence, a taylor expansion of $h^{(2)}(x)$ around $x=m\widehat{p} - Y-1$ gives
\begin{align*}
&\left| v^\star(\widehat{p}) - 2v^\star\left(\widehat{p} - \frac{1}{m}\right) +v^\star\left(\widehat{p} - \frac{2}{m}\right) \right| \\
&= \left|\bE[h^{(2)}(m\widehat{p}-X) - 2h^{(2)}(m\widehat{p}-Y-1) + h^{(2)}(m\widehat{p}-Z-2)]  \right| \\
&= \left|\bE\left[ Dh^{(2)}(m\widehat{p}-Y-1)(2Y-X-Z) + \frac{D^2h^{(2)}(\xi_1)}{2}(Y+1-X)^2 + \frac{D^2h^{(2)}(\xi_2)}{2}(Y-1-Z)^2 \right] \right|\\
&\le \bE\left[\sup_{m\widehat{p} -X-2 \le \xi \le m\widehat{p}-X}|D^2h^{(2)}(\xi)|\right] \triangleq \bE[h(m\widehat{p}-X)], 
\end{align*}
where we use the notations $Dh^{(2)}$ and $D^2h^{(2)}$ to denote the first- and second-order derivatives of $h^{(2)}$, respectively, and the first-order term in the Taylor expansion is zero via conditioning on $Y$. Consequently, the Cauchy--Schwartz inequality gives
\begin{align*}
\left(v^\star(\widehat{p}) - 2v^\star\left(\widehat{p} - \frac{1}{m}\right) +v^\star\left(\widehat{p} - \frac{2}{m}\right) \right)^2 \le \sum_{s\ge 0}\bP\left(\mathsf{B}\left(m,m\widehat{p},\frac{m}{2}\right) =s \right)\cdot h(m\widehat{p}-s)^2, 
\end{align*}
and therefore for $X_1, X_2\sim \mathsf{B}(m/2,p)$, 
\begin{align*}
&\bE\left[\widehat{p}\left(v^\star(\widehat{p}) - 2v^\star\left(\widehat{p} - \frac{1}{m}\right) +v^\star\left(\widehat{p} - \frac{2}{m}\right) \right)^2 \right] \\
&\le \bE\left[ \sum_{s\ge 0} \widehat{p} \cdot\bP\left(\mathsf{B}\left(m,m\widehat{p},\frac{m}{2}\right) =s \right)\cdot h(m\widehat{p}-s)^2 \right]\\
&= \bE\left[ \sum_{s\ge 0} \frac{s}{m}\cdot\bP\left(\mathsf{B}\left(m,m\widehat{p},\frac{m}{2}\right) =s \right)\cdot h(m\widehat{p}-s)^2 \right] + \bE\left[ \sum_{s\ge 0} \bP\left(\mathsf{B}\left(m,m\widehat{p},\frac{m}{2}\right) =s \right)\cdot \left(\widehat{p} - \frac{s}{m}\right) h(m\widehat{p}-s)^2 \right]\\
&= \bE\left[\frac{X_1}{m} \right]\cdot \bE[h(X_2)^2] + \frac{1}{m}\bE[X_2\cdot h(X_2)^2]. 
\end{align*}
Finally, following the same lines as the proof of Lemma \ref{lemma:h_alpha_moment} and Lemma \ref{lemma:delta_h_alpha_moment}, and using $h(X_2)=\Theta(1/(m^2p))$ whenever $mp/4\le X_2\le mp$, for $p\ge 3c_1\log m/m$ it holds that
\begin{align*}
\bE[h(X_2)^2] \lesssim_{c_1} \frac{1}{m^4p^2}, \qquad \bE[X_2\cdot h(X_2)^2] \lesssim_{c_1} \frac{1}{m^3p}. 
\end{align*}
Based on the above inequalities, \eqref{eq:ES_variance} gives the variance upper bound
\begin{align}\label{eq:variance}
\var\left(v^\star(\widehat{p}) - v^\star\left(\widehat{p} - \frac{1}{m}\right) \right) \lesssim_{c_1} \frac{1}{m^3p} \lesssim_{c_1} \frac{1}{m^2}. 
\end{align}
Hence, a combination of \eqref{eq:squared_bias} and \eqref{eq:variance} completes the proof of the second inequality.

\bibliographystyle{alpha}
\bibliography{di}

\newcommand{\etalchar}[1]{$^{#1}$}
\newcommand{\noopsort}[1]{}
\begin{thebibliography}{HJWW20}

\bibitem[Ach18]{acharya2018profile}
Jayadev Acharya.
\newblock Profile maximum likelihood is optimal for estimating kl divergence.
\newblock In {\em 2018 IEEE International Symposium on Information Theory
  (ISIT)}, pages 1400--1404. IEEE, 2018.

\bibitem[ACSS20]{ACSS20}
Nima Anari, Moses Charikar, Kirankumar Shiragur, and Aaron Sidford.
\newblock The {Bethe} and {Sinkhorn} permanents of low rank matrices and
  implications for profile maximum likelihood.
\newblock {\em arXiv preprint arXiv: 2004.02425}, 2020.

\bibitem[ACT20]{acharya2020inference}
Jayadev Acharya, Cl{\'e}ment~L Canonne, and Himanshu Tyagi.
\newblock Inference under information constraints i: Lower bounds from
  chi-square contraction.
\newblock {\em IEEE Transactions on Information Theory}, 66(12):7835--7855,
  2020.

\bibitem[ADOS17]{acharya2017unified}
Jayadev Acharya, Hirakendu Das, Alon Orlitsky, and Ananda~Theertha Suresh.
\newblock A unified maximum likelihood approach for estimating symmetric
  properties of discrete distributions.
\newblock In {\em International Conference on Machine Learning}, pages 11--21,
  2017.

\bibitem[Ama12]{amari2012differential}
Shun-ichi Amari.
\newblock {\em Differential-geometrical methods in statistics}, volume~28.
\newblock Springer Science \& Business Media, 2012.

\bibitem[Bir83]{birge1983approximation}
Lucien Birg{\'e}.
\newblock Approximation dans les espaces m{\'e}triques et th{\'e}orie de
  l'estimation.
\newblock {\em Zeitschrift f{\"u}r Wahrscheinlichkeitstheorie und verwandte
  Gebiete}, 65(2):181--237, 1983.

\bibitem[Bis06]{bishop2006pattern}
Christopher~M Bishop.
\newblock Pattern recognition.
\newblock {\em Machine Learning}, 128, 2006.

\bibitem[BZLV18]{bu2018estimation}
Yuheng Bu, Shaofeng Zou, Yingbin Liang, and Venugopal~V Veeravalli.
\newblock Estimation of kl divergence: Optimal minimax rate.
\newblock {\em IEEE Transactions on Information Theory}, 64(4):2648--2674,
  2018.

\bibitem[Che66]{cheney1966introduction}
Elliott~Ward Cheney.
\newblock Introduction to approximation theory.
\newblock 1966.

\bibitem[CK11]{csiszar2011information}
Imre Csiszar and J{\'a}nos K{\"o}rner.
\newblock {\em Information theory: coding theorems for discrete memoryless
  systems}.
\newblock Cambridge University Press, 2011.

\bibitem[CKV06]{Cai--Kulkarni--Verdu2006universal}
Haixiao Cai, Sanjeev~R Kulkarni, and Sergio Verd{\'u}.
\newblock Universal divergence estimation for finite-alphabet sources.
\newblock {\em IEEE Transactions on Information Theory}, 52(8):3456--3475,
  2006.

\bibitem[CL11]{Cai--Low2011}
T~Tony Cai and Mark~G Low.
\newblock Testing composite hypotheses, {H}ermite polynomials and optimal
  estimation of a nonsmooth functional.
\newblock {\em The Annals of Statistics}, 39(2):1012--1041, 2011.

\bibitem[CP04]{catoni2004statistical}
Olivier Catoni and Jean Picard.
\newblock {\em Statistical learning theory and stochastic optimization: Ecole
  d'Et{\'e} de Probabilit{\'e}s de Saint-Flour, XXXI-2001}, volume~31.
\newblock Springer Science \& Business Media, 2004.

\bibitem[Csi67]{csiszar1967information}
Imre Csisz{\'a}r.
\newblock Information-type measures of difference of probability distributions
  and indirect observation.
\newblock {\em studia scientiarum Mathematicarum Hungarica}, 2:229--318, 1967.

\bibitem[CSS19a]{charikar2019efficient}
Moses Charikar, Kirankumar Shiragur, and Aaron Sidford.
\newblock Efficient profile maximum likelihood for universal symmetric property
  estimation.
\newblock In {\em Proceedings of the 51st Annual ACM SIGACT Symposium on Theory
  of Computing}, pages 780--791, 2019.

\bibitem[CSS19b]{charikar2019general}
Moses Charikar, Kirankumar Shiragur, and Aaron Sidford.
\newblock A general framework for symmetric property estimation.
\newblock In {\em Advances in Neural Information Processing Systems}, pages
  12426--12436, 2019.

\bibitem[DLR77]{dempster1977maximum}
Arthur~P Dempster, Nan~M Laird, and Donald~B Rubin.
\newblock Maximum likelihood from incomplete data via the em algorithm.
\newblock {\em Journal of the royal statistical society. Series B
  (methodological)}, pages 1--38, 1977.

\bibitem[DT87]{Ditzian--Totik1987}
Zeev Ditzian and Vilmos Totik.
\newblock {\em Moduli of smoothness}.
\newblock Springer, 1987.

\bibitem[GBR{\etalchar{+}}06]{Gretton--Borgwardt--Rasch--Scholkopf--Smola2006kernel}
Arthur Gretton, Karsten~M Borgwardt, Malte Rasch, Bernhard Sch{\"o}lkopf, and
  Alex~J Smola.
\newblock A kernel method for the two-sample-problem.
\newblock In {\em Advances in neural information processing systems}, pages
  513--520, 2006.

\bibitem[H{\'a}j70]{Hajek1970characterization}
Jaroslav H{\'a}jek.
\newblock A characterization of limiting distributions of regular estimates.
\newblock {\em Zeitschrift f{\"u}r Wahrscheinlichkeitstheorie und verwandte
  Gebiete}, 14(4):323--330, 1970.

\bibitem[H{\'a}j72]{Hajek1972local}
Jaroslav H{\'a}jek.
\newblock Local asymptotic minimax and admissibility in estimation.
\newblock In {\em Proceedings of the sixth Berkeley symposium on mathematical
  statistics and probability}, volume~1, pages 175--194, 1972.

\bibitem[Han21]{han2020high}
Yanjun Han.
\newblock On the high accuracy limitation of adaptive property estimation.
\newblock {\em To appear in International Conference on Artificial Intelligence
  and Statistics (AISTATS)}, 2021.

\bibitem[Hel09]{hellinger1909neue}
Ernst Hellinger.
\newblock Neue begr{\"u}ndung der theorie quadratischer formen von
  unendlichvielen ver{\"a}nderlichen.
\newblock {\em Journal f{\"u}r die reine und angewandte Mathematik (Crelles
  Journal)}, 1909(136):210--271, 1909.

\bibitem[HJM20]{han2020estimation}
Yanjun Han, Jiantao Jiao, and Rajarshi Mukherjee.
\newblock On estimation of $l_r$-norms in gaussian white noise models.
\newblock {\em Probability Theory and Related Fields}, 177(3):1243--1294, 2020.

\bibitem[HJW16a]{han2016minimax}
Yanjun Han, Jiantao Jiao, and Tsachy Weissman.
\newblock Minimax rate-optimal estimation of divergences between discrete
  distributions.
\newblock {\em arXiv preprint arXiv:1605.09124v2}, 2016.

\bibitem[HJW16b]{han2016minimax_isita}
Yanjun Han, Jiantao Jiao, and Tsachy Weissman.
\newblock Minimax rate-optimal estimation of kl divergence between discrete
  distributions.
\newblock In {\em 2016 International Symposium on Information Theory and Its
  Applications (ISITA)}, pages 256--260. IEEE, 2016.

\bibitem[HJW18]{han2018local}
Yanjun Han, Jiantao Jiao, and Tsachy Weissman.
\newblock Local moment matching: A unified methodology for symmetric functional
  estimation and distribution estimation under wasserstein distance.
\newblock In {\em Conference On Learning Theory}, pages 3189--3221, 2018.

\bibitem[HJWW20]{han2020optimal}
Yanjun Han, Jiantao Jiao, Tsachy Weissman, and Yihong Wu.
\newblock Optimal rates of entropy estimation over lipschitz balls.
\newblock {\em Annals of Statistics}, 48(6):3228--3250, 2020.

\bibitem[HO19a]{hao2019broad}
Yi~Hao and Alon Orlitsky.
\newblock The broad optimality of profile maximum likelihood.
\newblock In {\em Advances in Neural Information Processing Systems}, pages
  10989--11001, 2019.

\bibitem[HO19b]{hao2019unified}
Yi~Hao and Alon Orlitsky.
\newblock Unified sample-optimal property estimation in near-linear time.
\newblock In {\em Advances in Neural Information Processing Systems}, pages
  11104--11114, 2019.

\bibitem[HP15]{hardt2015tight}
Moritz Hardt and Eric Price.
\newblock Tight bounds for learning a mixture of two gaussians.
\newblock In {\em Proceedings of the forty-seventh annual ACM symposium on
  Theory of computing}, pages 753--760, 2015.

\bibitem[HS21]{han2021competitive}
Yanjun Han and Kirankumar Shiragur.
\newblock On the competitive analysis and high accuracy optimality of profile
  maximum likelihood.
\newblock In {\em Proceedings of the 2021 ACM-SIAM Symposium on Discrete
  Algorithms (SODA)}, pages 1317--1336. SIAM, 2021.

\bibitem[IS12]{ingster2012nonparametric}
Yuri Ingster and Irina~A Suslina.
\newblock {\em Nonparametric goodness-of-fit testing under Gaussian models},
  volume 169.
\newblock Springer Science \& Business Media, 2012.

\bibitem[JHW18]{jiao2018minimax}
Jiantao Jiao, Yanjun Han, and Tsachy Weissman.
\newblock Minimax estimation of the $l_1$ distance.
\newblock {\em IEEE Transactions on Information Theory}, 64(10):6672--6706,
  2018.

\bibitem[JVHW15]{Jiao--Venkat--Han--Weissman2015minimax}
Jiantao Jiao, Kartik Venkat, Yanjun Han, and Tsachy Weissman.
\newblock Minimax estimation of functionals of discrete distributions.
\newblock {\em IEEE Transactions on Information Theory}, 61(5):2835--2885,
  2015.

\bibitem[KL51]{kullback1951information}
Solomon Kullback and Richard~A Leibler.
\newblock On information and sufficiency.
\newblock {\em The annals of mathematical statistics}, 22(1):79--86, 1951.

\bibitem[Kul97]{kullback1997information}
Solomon Kullback.
\newblock {\em Information theory and statistics}.
\newblock Courier Corporation, 1997.

\bibitem[KV17]{kong2017spectrum}
Weihao Kong and Gregory Valiant.
\newblock Spectrum estimation from samples.
\newblock {\em The Annals of Statistics}, 45(5):2218--2247, 2017.

\bibitem[KW13]{kingma2013auto}
Diederik~P Kingma and Max Welling.
\newblock Auto-encoding variational bayes.
\newblock {\em arXiv preprint arXiv:1312.6114}, 2013.

\bibitem[LC86]{LeCam1986asymptotic}
Lucien Le~Cam.
\newblock {\em Asymptotic methods in statistical decision theory}.
\newblock Springer, 1986.

\bibitem[LNS99]{Lepski--Nemirovski--Spokoiny1999estimation}
Oleg Lepski, Arkady Nemirovski, and Vladimir Spokoiny.
\newblock On estimation of the ${L}_r$ norm of a regression function.
\newblock {\em Probability theory and related fields}, 113(2):221--253, 1999.

\bibitem[LP06]{Lee--Park2006estimation}
Young~Kyung Lee and Byeong~U Park.
\newblock Estimation of {K}ullback--{L}eibler divergence by local likelihood.
\newblock {\em Annals of the Institute of Statistical Mathematics},
  58(2):327--340, 2006.

\bibitem[Mar92]{markov1892functions}
VA~Markov.
\newblock On functions deviating least from zero in a given interval.
\newblock {\em Izdat. Imp. Akad. Nauk, St. Petersburg}, pages 218--258, 1892.

\bibitem[NWJ10]{Nguyen--Wainwright--Jordan2010estimating}
XuanLong Nguyen, Martin~J Wainwright, and Michael~I Jordan.
\newblock Estimating divergence functionals and the likelihood ratio by convex
  risk minimization.
\newblock {\em Information Theory, IEEE Transactions on}, 56(11):5847--5861,
  2010.

\bibitem[OSVZ04]{orlitsky2004modeling}
Alon Orlitsky, Narayana~P Santhanam, Krishnamurthy Viswanathan, and Junan
  Zhang.
\newblock On modeling profiles instead of values.
\newblock In {\em Proceedings of the 20th conference on Uncertainty in
  artificial intelligence}, pages 426--435. AUAI Press, 2004.

\bibitem[OSW16]{orlitsky2016optimal}
Alon Orlitsky, Ananda~Theertha Suresh, and Yihong Wu.
\newblock Optimal prediction of the number of unseen species.
\newblock {\em Proceedings of the National Academy of Sciences},
  113(47):13283--13288, 2016.

\bibitem[Pan03]{Paninski2003}
Liam Paninski.
\newblock Estimation of entropy and mutual information.
\newblock {\em Neural Computation}, 15(6):1191--1253, 2003.

\bibitem[Pan04]{Paninski2004}
Liam Paninski.
\newblock Estimating entropy on $m$ bins given fewer than $m$ samples.
\newblock {\em Information Theory, IEEE Transactions on}, 50(9):2200--2203,
  2004.

\bibitem[PC08]{Perez2008kullback}
Fernando P{\'e}rez-Cruz.
\newblock {K}ullback-{L}eibler divergence estimation of continuous
  distributions.
\newblock In {\em IEEE International Symposium on Information Theory (ISIT)},
  pages 1666--1670. IEEE, 2008.

\bibitem[Pea94]{pearson1894contributions}
Karl Pearson.
\newblock Contributions to the mathematical theory of evolution.
\newblock {\em Philosophical Transactions of the Royal Society of London. A},
  185:71--110, 1894.

\bibitem[RW19]{rigollet2019uncoupled}
Philippe Rigollet and Jonathan Weed.
\newblock Uncoupled isotonic regression via minimum wasserstein deconvolution.
\newblock {\em Information and Inference: A Journal of the IMA}, 8(4):691--717,
  2019.

\bibitem[San58]{sanov1958probability}
IN~Sanov.
\newblock {\em On the probability of large deviations of random variables}.
\newblock United States Air Force, Office of Scientific Research, 1958.

\bibitem[Ste86]{steele1986efron}
J~Michael Steele.
\newblock An efron-stein inequality for nonsymmetric statistics.
\newblock {\em The Annals of Statistics}, 14(2):753--758, 1986.

\bibitem[TKV17]{tian2017learning}
Kevin Tian, Weihao Kong, and Gregory Valiant.
\newblock Learning populations of parameters.
\newblock In {\em Advances in neural information processing systems}, pages
  5778--5787, 2017.

\bibitem[Tsy09]{Tsybakov2008}
A.~Tsybakov.
\newblock {\em Introduction to Nonparametric Estimation}.
\newblock Springer-Verlag, 2009.

\bibitem[Ver19]{verdu2019empirical}
Sergio Verd{\'u}.
\newblock Empirical estimation of information measures: A literature guide.
\newblock {\em Entropy}, 21(8):720, 2019.

\bibitem[VV11a]{Valiant--Valiant2011}
Gregory Valiant and Paul Valiant.
\newblock Estimating the unseen: an $n/\log n$-sample estimator for entropy and
  support size, shown optimal via new {CLT}s.
\newblock In {\em Proceedings of the 43rd annual ACM symposium on Theory of
  computing}, pages 685--694. ACM, 2011.

\bibitem[VV11b]{Valiant--Valiant2011power}
Gregory Valiant and Paul Valiant.
\newblock The power of linear estimators.
\newblock In {\em Foundations of Computer Science (FOCS), 2011 IEEE 52nd Annual
  Symposium on}, pages 403--412. IEEE, 2011.

\bibitem[VV13]{Valiant--Valiant2013estimating}
Paul Valiant and Gregory Valiant.
\newblock Estimating the unseen: improved estimators for entropy and other
  properties.
\newblock In {\em Advances in Neural Information Processing Systems}, pages
  2157--2165, 2013.

\bibitem[WKV05]{Wang--Kulkarni--Verdu2005divergence}
Qing Wang, Sanjeev~R Kulkarni, and Sergio Verd{\'u}.
\newblock Divergence estimation of continuous distributions based on
  data-dependent partitions.
\newblock {\em IEEE Transactions on Information Theory}, 51(9):3064--3074,
  2005.

\bibitem[WKV09]{Wang--Kulkarni--Verdu2009divergence}
Qing Wang, Sanjeev~R Kulkarni, and Sergio Verd{\'u}.
\newblock Divergence estimation for multidimensional densities
  via-nearest-neighbor distances.
\newblock {\em IEEE Transactions on Information Theory}, 55(5):2392--2405,
  2009.

\bibitem[WY16]{wu2016minimax}
Yihong Wu and Pengkun Yang.
\newblock Minimax rates of entropy estimation on large alphabets via best
  polynomial approximation.
\newblock {\em IEEE Transactions on Information Theory}, 62(6):3702--3720,
  2016.

\bibitem[WY18]{wu2018optimal}
Yihong Wu and Pengkun Yang.
\newblock Optimal estimation of gaussian mixtures via denoised method of
  moments.
\newblock {\em arXiv preprint arXiv:1807.07237}, 2018.

\bibitem[WY19]{wu2019chebyshev}
Yihong Wu and Pengkun Yang.
\newblock Chebyshev polynomials, moment matching, and optimal estimation of the
  unseen.
\newblock {\em The Annals of Statistics}, 47(2):857--883, 2019.

\bibitem[ZG14]{Zhang--Grabchak2014nonparametric}
Zhiyi Zhang and Michael Grabchak.
\newblock Nonparametric estimation of {K}{\"u}llback-{L}eibler divergence.
\newblock {\em Neural computation}, 26(11):2570--2593, 2014.

\bibitem[ZVV{\etalchar{+}}16]{ZVVKCSLSDM16}
James Zou, Gregory Valiant, Paul Valiant, Konrad Karczewski, Siu~On Chan,
  Kaitlin Samocha, Monkol Lek, Shamil Sunyaev, Mark Daly, and Daniel~G.
  MacArthur.
\newblock Quantifying unobserved protein-coding variants in human populations
  provides a roadmap for large-scale sequencing projects.
\newblock {\em Nature Communications}, 7, 2016.

\end{thebibliography}

\end{document}